\documentstyle[12pt,fleqn]{article}
\catcode`\@=11
\long\def\@makefntext#1{ 
\protect\noindent \hbox to 3.2pt {\hskip-.9pt  
$^{{\ninerm\@thefnmark}}$\hfil}#1\hfill} 
\def\thefootnote{\fnsymbol{footnote}}
\def\@makefnmark{\hbox to 0pt{$^{\@thefnmark}$\hss}}  
\def\ps@myheadings{\let\@mkboth\@gobbletwo
\def\@oddhead{\hbox{} 
\rightmark\hfil\ninerm\thepage}   
\def\@oddfoot{}\def\@evenhead{\ninerm\thepage\hfil 
\leftmark\hbox{}}\def\@evenfoot{}
\def\sectionmark##1{}\def\subsectionmark##1{}}
\def\CVector#1#2#3{
\left\vert\begin{array}{c}
                 #1 \\ 
                 #2 
                \end{array}, #3\right>}
\def\AVector#1#2#3{\left|  #1, #2; #3 \right>}
\begin{document}
\newcommand{\symbolfootnote}{\renewcommand{\thefootnote}
        {\fnsymbol{footnote}}}
\renewcommand{\thefootnote}{\fnsymbol{footnote}}
\newcommand{\alphfootnote}
        {\setcounter{footnote}{0}
         \renewcommand{\thefootnote}{\sevenrm\alph{footnote}}}
\newcounter{sectionc}\newcounter{subsectionc}\newcounter{subsubsectionc}
\renewcommand{\section}[1] {\vspace{0.6cm}\addtocounter{sectionc}{1} 
\setcounter{subsectionc}{0}\setcounter{subsubsectionc}{0}\noindent 
        {\bf\thesectionc. #1}\par\vspace{0.4cm}}
\renewcommand{\subsection}[1] {\vspace{0.6cm}\addtocounter{subsectionc}{1} 
        \setcounter{subsubsectionc}{0}\noindent 
        {\it\thesectionc.\thesubsectionc. #1}\par\vspace{0.4cm}}
\renewcommand{\subsubsection}[1]
{\vspace{0.6cm}\addtocounter{subsubsectionc}{1}
        \noindent {\rm\thesectionc.\thesubsectionc.\thesubsubsectionc. 
        #1}\par\vspace{0.4cm}}
\newcommand{\nonumsection}[1] {\vspace{0.6cm}\noindent{\bf #1}
        \par\vspace{0.4cm}}
                                                 
\newcounter{appendixc}
\newcounter{subappendixc}[appendixc]
\newcounter{subsubappendixc}[subappendixc]
\renewcommand{\thesubappendixc}{\Alph{appendixc}.\arabic{subappendixc}}
\renewcommand{\thesubsubappendixc}
        {\Alph{appendixc}.\arabic{subappendixc}.\arabic{subsubappendixc}}

\renewcommand{\appendix}[1] {\vspace{0.6cm}
        \refstepcounter{appendixc}
        \setcounter{figure}{0}
        \setcounter{table}{0}
        \setcounter{equation}{0}
        \renewcommand{\thefigure}{\Alph{appendixc}.\arabic{figure}}
        \renewcommand{\thetable}{\Alph{appendixc}.\arabic{table}}
        \renewcommand{\theappendixc}{\Alph{appendixc}}
        \renewcommand{\theequation}{\Alph{appendixc}.\arabic{equation}}
        \noindent{\bf Appendix \theappendixc #1}\par\vspace{0.4cm}}
\newcommand{\subappendix}[1] {\vspace{0.6cm}
        \refstepcounter{subappendixc}
        \noindent{\bf Appendix \thesubappendixc. #1}\par\vspace{0.4cm}}
\newcommand{\subsubappendix}[1] {\vspace{0.6cm}
        \refstepcounter{subsubappendixc}
        \noindent{\it Appendix \thesubsubappendixc. #1}
        \par\vspace{0.4cm}}

\def\abstracts#1{{
        \centering{\begin{minipage}{30pc}\tenrm\baselineskip=12pt\noindent
        \centerline{\tenrm ABSTRACT}\vspace{0.3cm}
        \parindent=0pt #1
        \end{minipage} }\par}} 

\newcommand{\bibit}{\it}
\newcommand{\bibbf}{\bf}
\renewenvironment{thebibliography}[1]
        {\begin{list}{\arabic{enumi}.}
        {\usecounter{enumi}\setlength{\parsep}{0pt}
\setlength{\leftmargin 1.25cm}{\rightmargin 0pt}
         \setlength{\itemsep}{0pt} \settowidth
        {\labelwidth}{#1.}\sloppy}}{\end{list}}

\topsep=0in\parsep=0in\itemsep=0in
\parindent=1.5pc

\newcounter{itemlistc}
\newcounter{romanlistc}
\newcounter{alphlistc}
\newcounter{arabiclistc}
\newenvironment{itemlist}
        {\setcounter{itemlistc}{0}
         \begin{list}{$\bullet$}
        {\usecounter{itemlistc}
         \setlength{\parsep}{0pt}
         \setlength{\itemsep}{0pt}}}{\end{list}}

\newenvironment{romanlist}
        {\setcounter{romanlistc}{0}
         \begin{list}{$($\roman{romanlistc}$)$}
        {\usecounter{romanlistc}
         \setlength{\parsep}{0pt}
         \setlength{\itemsep}{0pt}}}{\end{list}}

\newenvironment{alphlist}
        {\setcounter{alphlistc}{0}
         \begin{list}{$($\alph{alphlistc}$)$}
        {\usecounter{alphlistc}
         \setlength{\parsep}{0pt}
         \setlength{\itemsep}{0pt}}}{\end{list}}

\newenvironment{arabiclist}
        {\setcounter{arabiclistc}{0}
         \begin{list}{\arabic{arabiclistc}}
        {\usecounter{arabiclistc}
         \setlength{\parsep}{0pt}
         \setlength{\itemsep}{0pt}}}{\end{list}}

\newcommand{\fcaption}[1]{
        \refstepcounter{figure}
        \setbox\@tempboxa = \hbox{\tenrm Fig.~\thefigure. #1}
        \ifdim \wd\@tempboxa > 6in
           {\begin{center}
        \parbox{6in}{\tenrm\baselineskip=12pt Fig.~\thefigure. #1 }
            \end{center}}
        \else
             {\begin{center}
             {\tenrm Fig.~\thefigure. #1}
              \end{center}}
        \fi}

\newcommand{\tcaption}[1]{
        \refstepcounter{table}
        \setbox\@tempboxa = \hbox{\tenrm Table~\thetable. #1}
        \ifdim \wd\@tempboxa > 6in
           {\begin{center}
        \parbox{6in}{\tenrm\baselineskip=12pt Table~\thetable. #1 }
            \end{center}}
        \else
             {\begin{center}
             {\tenrm Table~\thetable. #1}
              \end{center}}
        \fi}

\def\@citex[#1]#2{\if@filesw\immediate\write\@auxout
        {\string\citation{#2}}\fi
\def\@citea{}\@cite{\@for\@citeb:=#2\do
        {\@citea\def\@citea{,}\@ifundefined
        {b@\@citeb}{{\bf ?}\@warning
        {Citation `\@citeb' on page \thepage \space undefined}}
        {\csname b@\@citeb\endcsname}}}{#1}}

\newif\if@cghi
\def\cite{\@cghitrue\@ifnextchar [{\@tempswatrue
        \@citex}{\@tempswafalse\@citex[]}}
\def\citelow{\@cghifalse\@ifnextchar [{\@tempswatrue
        \@citex}{\@tempswafalse\@citex[]}}
\def\@cite#1#2{{$\null^{#1}$\if@tempswa\typeout
        {IJCGA warning: optional citation argument 
        ignored: `#2'} \fi}}
\newcommand{\citeup}{\cite}

\def\fnm#1{$^{\mbox{\scriptsize #1}}$}
\def\fnt#1#2{\footnotetext{\kern-.3em
        {$^{\mbox{\sevenrm #1}}$}{#2}}}

\font\twelvebf=cmbx10 scaled\magstep 1
\font\twelverm=cmr10 scaled\magstep 1
\font\twelveit=cmti10 scaled\magstep 1
\font\elevenbfit=cmbxti10 scaled\magstephalf
\font\elevenbf=cmbx10 scaled\magstephalf
\font\elevenrm=cmr10 scaled\magstephalf
\font\elevenit=cmti10 scaled\magstephalf
\font\bfit=cmbxti10
\font\tenbf=cmbx10
\font\tenrm=cmr10
\font\tenit=cmti10
\font\ninebf=cmbx9
\font\ninerm=cmr9
\font\nineit=cmti9
\font\eightbf=cmbx8
\font\eightrm=cmr8
\font\eightit=cmti8


\newdimen\mathindent\mathindent=6mm
\newcommand{\fl}{\hspace*{-\mathindent}}%
\newcommand{\tqs}{\hspace*{25pt}}%
\newcommand{\case}[2]{{\textstyle\frac{#1}{#2}}}
\newif\ifnumbysec
\newcommand{\eref}[1]{(\ref{#1})}%
\def\eqalign#1{\null\vcenter{\def\\{\cr}\openup\jot
  \ialign{\strut$\displaystyle{##}$\hfil&$\displaystyle{{}##}$\hfil
      \crcr#1\crcr}}\,}
\def\cases#1{%
     \left\{\,\vcenter{\def\\{\cr}\normalbaselines\openup1\jot
     \ialign{\strut$\displaystyle{##}\hfil$&\tqs
     \rm##\hfil\crcr#1\crcr}}\right.}%
\renewcommand{\thesection}{\normalsize\bf\arabic{section}}
\def\ack{%
{\bigskip\noindent\normalsize\bf~{Acknowledgments}\raggedright\par\medskip}}%

\centerline{\tenbf QUANTUM GROUPS AND THEIR APPLICATIONS IN NUCLEAR PHYSICS}
\vspace{0.8cm}
\centerline{\tenrm Dennis BONATSOS}
\baselineskip=13pt
\centerline{\tenit Institute of Nuclear Physics, N.C.S.R. ``Demokritos''}
\baselineskip=12pt
\centerline{\tenit GR-15310 Aghia Paraskevi, Attiki, Greece }
\vspace{0.3cm}
\centerline{\tenrm C. DASKALOYANNIS}
\baselineskip=13pt
\centerline{\tenit Department of Physics, Aristotle University of Thessaloniki}
\baselineskip=12pt 
\centerline{\tenit GR-54006 Thessaloniki, Greece}
\vspace{0.9cm}
\abstracts{
Quantum algebras are a mathematical tool which provides us with 
a class of symmetries wider than that of Lie algebras, which are contained 
in the former as a special case. After a self-contained introduction to 
the necessary mathematical tools ($q$-numbers, $q$-analysis, $q$-oscillators,
$q$-algebras), the su$_q$(2) rotator model and its extensions, 
the construction of deformed exactly soluble models 
(u(3)$\supset$so(3) model, Interacting Boson Model,
Moszkowski model), the 3-dimensional $q$-deformed harmonic oscillator 
and its relation to the nuclear shell model, the use of deformed bosons in 
the description of pairing correlations, and
the symmetries of the anisotropic quantum harmonic oscillator with rational
ratios of frequencies, which underly the structure of superdeformed and 
hyperdeformed nuclei, are discussed in some detail. A brief description
of similar applications to the structure of molecules and of atomic 
clusters, as well as an outlook are also given. 
}

\vfil

\rm\baselineskip=10pt
\centerline{\bf Table of contents}

\begin{enumerate}

\item Introduction

\item $q$-numbers

\item $q$-deformed elementary functions

\item $q$-derivatives

\item $q$-integration

\item  $Q$-numbers

\item $Q$-deformed elementary functions

\item  $Q$-derivative

\item   $Q$-integration

\item The $q$-deformed harmonic oscillator

\item The $Q$-deformed harmonic oscillator

\item The generalized deformed oscillator

\item The physical content of deformed harmonic oscillators

 13.1 Classical potentials equivalent to the $q$-oscillator

 13.2 Classical potentials equivalent to the $Q$-deformed oscillator

 13.3 WKB-EPs for the $q$-deformed oscillator

 13.4 WKB-EPs for the $Q$-deformed oscillator

\item  The quantum algebra su$_q$(2)

\item  Realization of su$_q$(2) in terms of $q$-deformed bosons

\item The quantum algebra su$_q$(1,1)

\item Generalized deformed su(2) and su(1,1) algebras

\item Generalized deformed parafermionic oscillators

\item The su$_q$(2) rotator model

\item Comparison of the su$_q$(2) model to other models

20.1 The Variable Moment of Inertia (VMI) model

20.2 Comparison of the su$_q$(2) model to the VMI and related models

20.3 The hybrid model

20.4  Other models

\item Electromagnetic transitions in the su$_q$(2) model

21.1 The collective model of Bohr and Mottelson

21.2 The Interacting Boson Model (IBM)

21.3 The su$_q$(2) model

21.4 Comparison to experiment

\item  Superdeformed bands

\item The physical content of the su$_q$(2) model

\item The u$_{p,q}$(2) rotator model

\item  Generalized deformed su(2) models

\item Quantum algebraic description of vibrational and transitional 
nuclear spectra

26.1 The Interacting Boson Model

26.2 Generalized VMI

26.3 Modification of the su$_q$(2) model

\item  A toy Interacting Boson Model with su$_q$(3) symmetry

27.1 The su$_q$(3) algebra

27.2 The su$_q$(2) limit

27.3  The so$_q$(3) limit

\item The 3-dimensional $q$-deformed harmonic oscillator and the nuclear 
shell model

28.1 Simplified so$_q$(3) subalgebra and so$_q$(3) basis states 

28.2 Definitions of so$_q$(3) irreducible tensor operators, tensor and 
scalar products 

28.3 Construction of so$_q$(3) vector operators and spherical vector operators

28.4 A choice for the physical angular momentum 

28.5 A choice for the Hamiltonian 

28.6 The Modified Harmonic Oscillator of Nilsson 

28.7 Connection between the 3-dimensional $q$-deformed harmonic oscillator 
and the Modified 

{}$\qquad$ Harmonic Oscillator of Nilsson 

28.8 3-dimensional $q$-deformed harmonic oscillator with $q$-deformed 
spin--orbit term 

\item Further development of u$_q$(3)$\supset$so$_q$(3) models 

29.1 Matrix elements of a $q$-deformed quadrupole operator in the 
u$_q$(3)$\supset$so$_q$(3) basis 

29.2 Matrix elements of the usual quadrupole operator in the 
u$_q$(3)$\supset$so$_q$(3) basis 

29.3 Transformation between so$_q$(3) and so(3) basis states  

\item  The question of complete breaking of symmetries and some 
applications

\item  $q$-deformation of the Interacting Boson Model (IBM)

\item Deformed versions of other collective models

\item Fermion pairs as deformed bosons: approximate mapping

33.1 The single-j shell model

33.2 Fermion pairs of zero angular momentum

33.3 Mapping using the $q$-deformed oscillator

33.4 Mapping using the $Q$-deformed oscillator

\item Fermion pairs as deformed bosons: exact mapping

34.1 An appropriate generalized deformed oscillator

34.2 Related potentials

\item  The seniority scheme

35.1 Uncovering a dynamical symmetry

35.2 Comparison to experiment

35.3 Other approaches

\item Anisotropic quantum harmonic oscillators with rational ratios of 
frequencies

36.1 A deformed u(2) algebra

36.2 The representations

36.3 The ``angular momentum'' quantum number

36.4 Multisections of the isotropic oscillator

36.5 Connection to W$_3^{(2)}$ 

36.6 Discussion

\item The use of quantum algebras in molecular structure

\item The use of quantum algebras in the structure of atomic clusters 

\item Outlook

Acknowledgements 

References

\end{enumerate}

\vfil
\rm\baselineskip=14pt
\section{Introduction}

Quantum algebras \cite{Dri798,Jim247,Jim10,KR101,Skl262} (also called quantum 
groups) are deformed versions of the
usual Lie algebras, to which they reduce when the deformation parameter 
$q$ is set equal to unity. From the mathematical point of view they are
Hopf algebras \cite{Abe1977}. Details about their mathematical properties 
can be found in \cite{CP94,BL95,GAS96,KS97,CD96,KAS95,JOS95}. 
Their use in physics became popular with the 
introduction of the $q$-deformed harmonic oscillator (sec. 10) as a tool for 
providing a boson realization of the quantum algebra su$_q$(2) (sec. 14), 
although 
similar mathematical structures had already been known (sec. 11). 
Initially used for solving the quantum Yang--Baxter equation \cite{Jim10}, 
quantum algebras
have subsequently found applications in several branches of physics, as, for 
example, in the description of spin chains\cite{BMNR90,GPPR94},
anyons\cite{LS93,Roy96,Ubr97}, 
quantum optics\cite{Buz92,CEK90,Kua94,Maz97,Zhe93},
 rotational  and vibrational nuclear and molecular spectra, and in conformal 
field theories. By now several kinds of generalized deformed oscillators 
(sec. 12)
and generalized deformed algebras (sec. 17)  have been introduced.  

It is clear that quantum algebras provide us with a class of symmetries
which is richer than the class of Lie symmetries, which are contained 
in the former as a special case. It is therefore conceivable that quantum 
algebras can turn out to be appropriate for describing symmetries of 
physical systems which are outside the realm of Lie algebras. 

Here we shall confine ourselves to applications of quantum algebras in nuclear 
structure physics. The structure of this review is as follows:
 In order to make this review self-contained, we are going
first to give a brief account of the necessary tools: $q$-numbers and 
$q$-analysis (secs 2--9), $q$-deformed oscillators (secs 10--13), $q$-deformed 
 algebras (secs 14--18). 
The remainder will be devoted to specific applications in nuclear 
structure problems, starting with phenomenology and advancing towards more
microscopic subjects.  The su$_q$(2) rotator model (secs 19--23) and its 
extensions (secs 24--26),  as well as  the formulation of deformed exactly 
soluble models (u(3)$\supset$so(3) model (secs 28, 29), Interacting Boson 
Model (secs 27, 30, 31), Moszkowski model (sec. 32))
will be covered in some detail. In addition the 3-dimensional $q$-deformed 
harmonic oscillator and its relation to the nuclear shell model 
(sec. 28) will be extensively discussed. Subsequently, the use of quantum 
algebraic techniques for the description of pairing correlations in nuclei 
(secs 33--35), 
as well as the symmetries of the anisotropic quantum harmonic oscillators
with rational ratios of frequencies (sec. 36) will also be considered in some 
detail. The latter are of current interest in connection with the symmetries 
underlying superdeformed and hyperdeformed nuclear bands (sec. 36).  
Finally, a brief account of applications of the same techniques to the 
structure of molecules (sec. 37) and of atomic clusters (sec. 38), as well as 
an outlook (sec. 39) will be given. 
An earlier review of the subject has been given in \cite{RJP109}. 

\section{$q$-numbers}

The 
{\sl $q$-number} corresponding to the ordinary number $x$  is defined as 
$$ [x]= {q^x-q^{-x}\over q-q^{-1}},\eqno(2.1)$$
where $q$ is a parameter. The same definition holds if $x$ is an operator. 
We remark that $q$-numbers remain invariant under the substitution 
$q\rightarrow q^{-1}$. 

If $q$ is real, $q$-numbers can easily be put in the form
$$ [x] = {\sinh(\tau x)\over \sinh(\tau)}, \eqno(2.2)$$
where $q=e^{\tau}$ and $\tau$ is real. 

If $q$ is a phase factor, $q$-numbers can be written as 
$$ [x]= {\sin(\tau x)\over \sin(\tau)},\eqno(2.3)$$
where $q=e^{i\tau}$ and $\tau$ is real. 

In both cases it is clear that in the limit $q\to 1$ (or, equivalently, 
$\tau\to 0$) $q$-numbers (or operators) tend to the ordinary numbers 
(or operators): 
$$ \lim_{q\to 1} [x] = x .\eqno(2.4)$$

A few examples of $q$-numbers are given here:
$$ [0]=0,\qquad [1] =1, \qquad [2]= q+q^{-1}, \qquad [3] = q^2 +1+q^{-2}.
\eqno(2.5)$$

Identities between $q$-numbers exist. They are, however, different from the 
familiar identities between usual numbers. As an exercise one can show 
(using the definition of $q$-numbers) that
$$ [a] [b+1]-[b] [a+1] = [a-b].\eqno(2.6)$$

The {\sl $q$-factorial} of an integer $n$ is defined as
$$ [n]!=[n] [n-1] \ldots [2] [1].\eqno(2.7)$$

The {\sl $q$-binomial coefficients} are defined as
$${m \brack n} ={[m]!\over [m-n]! [n]!},\eqno(2.8)$$
while the {\sl $q$-binomial expansion} is given by
$$ [a \pm b]^m = \sum_{k=0}^m {m\brack k} a^{m-k} (\pm b)^k.\eqno(2.9)$$

In the limit $q\to 1$ we obviously have 
$$ [n]!\to n! \qquad {\rm and} \qquad {m\brack n} \to {m\choose n},
\eqno(2.10)$$
where $n!$ and ${m\choose n}$ are the standard factorial and binomial 
coefficients respectively. 

It should be noticed that two-parameter deformed numbers have also been 
introduced
$$ [x]_{p,q}= {q^x-p^{-x}\over q-p^{-1} }.\eqno(2.11)$$
In the special case $p=q$ they reduce to the usual $q$-numbers. 

\section{$q$-deformed elementary functions}

In addition to $q$-deformed numbers and operators, $q$-deformed elementary 
functions can be introduced. The {\sl $q$-exponential function} is defined as
$$ e_q (a x) = \sum_{n=0}^{\infty}  {a^n\over [n]!} x^n ,\eqno(3.1)$$
while the {\sl $q$-trigonometric functions} are defined as
$$ \sin_q (x)= \sum_{n=0}^{\infty} (-1)^n {x^{2n+1}\over [2n+1]!},\qquad
 \cos_q (x)= \sum_{n=0}^{\infty} (-1)^n {x^{2n}\over [2n]!}.\eqno(3.2)$$
It should also be noticed that $q$-deformed polynomials, such as 
$q$-deformed Hermite polynomials \cite{FV45,VdJ267,CGY1517} 
and $q$-deformed Laguerre polynomials \cite{FV45}
also exist (see also subsec. 36.3). 

The definitions of the $q$-exponential function and the $q$-trigonometric 
functions given above are not unique; for a different set of definitions, 
based on the Tsallis statistics, see \cite{Borges}. 

\section{$q$-derivatives}

Proceeding along this path one can build a new differential calculus, based on 
$q$-deformed quantities (see \cite{GN945,BAZG1379} for concise expositions). 
For this purpose the {\sl $q$-derivative}
is defined as
$$ D^q_x f(x) = {f(qx)-f(q^{-1} x)\over (q-q^{-1})x}.\eqno(4.1)$$
The similarity between the present definition and the one of $q$-numbers 
(eq.(2.1)) is clear. 

Using the definition of the $q$-derivative one can easily see that
$$ D^q_x (a x^n) = a [n] x^{n-1},\eqno(4.2)$$
$$ D^q_x e_q(ax) = a e_q(ax).\eqno(4.3)$$
>From the definition of the $q$-derivative one can further derive 
the sum rule
$$ D_x^q (f(x)+g(x)) = D_x^q f(x)+ D_x^q g(x),\eqno(4.4)$$ 
as well as the rule
$$D^q_{x_2} [ a x_1 \pm b x_2]^m =\pm [m] b [a x_1 \pm b x_2]^{m-1},
\eqno(4.5)$$
where $a$ and $b$ are costants and $[a x_1 \pm b x_2]^m$ is given 
by the $q$-binomial expansion (eq. (2.9)). One can also prove the 
{\sl $q$-integration
by parts} formula 
$$ D_x^q (f(x)g(x)) = { f(q x) g(q x) - f(q^{-1}x) g(q^{-1}x)\over 
(q-q^{-1})x }.\eqno(4.6) $$
From this, the following two forms of the Leibnitz rule can be derived
$$ D_x^q (f(x)g(x))= (D_x^q f(x)) g(q^{-1}x)+ f(qx) (D_x^q g(x)),\eqno(4.7)$$
$$ D_x^q (f(x)g(x))= (D_x^q g(x)) f(q^{-1}x)+ g(qx) (D_x^q f(x)).\eqno(4.8)$$ 
In addition one can show the property
$$D^q_x f(qx) = q D^q_x f(x)\vert_{x=qx},\eqno(4.9)$$
and the chain rules 
$$D^q_{ax} f(x) = {1\over a} D_x^q f(x)\qquad,\eqno(4.10)$$
$$ D^q_x f(x^n) = [n] x^{n-1} D^{q^n}_{x^n} f(x^n),\eqno(4.11)$$
where $a$ is a constant. 
Another useful result is 
$$ D^{q^n}_x f(x)= {1\over [n]} \sum_{k=0}^{n-1} D^q_x f(q^{2k-(n-1)} x).
\eqno(4.12)$$

\section{ $q$-integration}

The  $q$-integration (see \cite{GN945,BAZG1379} for concise expositions)
in the interval $[0,a]$ is defined by
$$\int_0^a f(x) d_qx = a(q^{-1}-q)\sum_{n=0}^{\infty} q^{2n+1} f(q^{2n+1}a),
\eqno(5.1)$$
while for the interval $[0,\infty)$ one has 
$$\int_0^{\infty} f(x) d_qx = (q^{-1}-q)\sum_{n=-\infty}^{\infty} 
q^{2n+1} f(q^{2n+1}).\eqno(5.2)$$
The {\sl indefinite $q$-integral} is defined as 
$$ \int f(x) d_qx= (q^{-1}-q)\sum_{n=0}^{\infty}  q^{2n+1}  x f(q^{2n+1} x)
+ \rm{constant},\eqno(5.3)$$
where $0 < q <1$. For entire functions  $f(x)$ one can easily see that 
this $q$-integral approaches the Riemann integral as $q\to 1$, and also 
that the operators of $q$-dif\-fer\-ent\-iat\-ion and 
$q$-integration are inverse 
to each other
$$ D_x^q \int f(x) d_qx = f(x) = \int D_x^q f(x) d_qx .\eqno(5.4)$$
One can also easily see that 
$$\int a x^{n-1}  d_qx = {1\over [n]} a x^n + \rm{constant}, \eqno(5.5)$$
$$ \int e_q (ax) d_qx= {1\over a} e_q(ax) +\rm{constant}.\eqno(5.6)$$
From (4.6) one can also prove the following formulae of integration 
by parts 
$$\int_0^a f(qx) (D_x^q g(x)) d_qx= f(x)g(x) \vert_{x=0}^{x=a} -
\int_0^a (D_x^q f(x)) g(q^{-1}x) d_q x,\eqno(5.7)$$
$$\int_0^a f(q^{-1}x) (D_x^q g(x)) d_q x = f(x) g(x) \vert_{x=0}^{x=a}
-\int_0^a (D_x^q f(x)) g(qx) d_qx.\eqno(5.8)$$

The following formulae can also be proved
$$\int f(x) d_{aq}x = a\int f(x) d_qx , \eqno(5.9)$$
$$ \int f(x^n)  d_{q^n} x^n = [n] \int x^{n-1} f(x^n) d_qx, \eqno(5.10)$$
$$\int f(x) d_qx= {1\over [n]} \sum_{k=0}^{n-1} q^{2k-(n-1)} \int
f(q^{2k-(n-1)}x) d_{q^n}x.\eqno(5.11)$$

The {\sl $q$-analogue for Euler's formula} for the function $\Gamma (x)$ is
$$\int_0^{\zeta} e_q(-x) x^n d_qx= [n] [n-1] [n-2] \ldots [1] =[n]!.
\eqno(5.12)$$
A proof of this formula can be found in \cite{GN945}.

\section{$Q$-numbers}

The definition of $q$-numbers given in sec. 2 is not the only possible one. 
We have already seen the two-parameter deformed numbers, defined in eq. 
(2.11).  Furthermore, a different definition of quantum numbers 
(the $Q$-numbers) has been used in mathematics since the early ninetenth 
century, with      
rich literature existing  on this subject \cite{Exton,Andrews}. 
{\sl $Q$-numbers} are defined as 
$$ [x]_Q = {Q^x - 1 \over Q - 1}, \eqno(6.1)$$ 
where $x$ can be a number or an operator and $Q$ is a deformation parameter. 
$Q$ is a real number ($Q\ne 0,1$). The notation $Q=e^T$, where $T$ a real
number, will be often used.  
The subscript $Q$ will be used in this review in order to distinguish 
deformed numbers defined as in eq. (6.1) from these defined by eq. (2.1). 
It is clear that in the limit $Q\to 1$ (or, equivalently, $T\to 0$) 
$Q$-numbers become ordinary numbers, i.e. $[x]_Q \to x$.

A few examples of $Q$-numbers are given here:
$$ [0]_Q=0, \qquad [1]_Q=1,\qquad [2]_Q= Q+1, \qquad [3]_Q= Q^2+Q+1.
\eqno(6.2)$$
 
$Q$-numbers clearly do not remain invariant under the substitution 
$Q\rightarrow Q^{-1}$. One can easily prove that
$$ \left[x\right]_Q = Q^{x-1} \left[x\right]_{1/Q}.\eqno(6.3)$$

$Q$-numbers are connected to $q$-numbers through the relation \cite{GN945}
$$ [x]= q^{1-x}[x]_Q, \qquad {\rm with} \qquad Q=q^2.\eqno(6.4)$$

The definitions of $Q$-factorials and $Q$-binomial coefficients still look
like the ones given in eqs. (2.7)--(2.8):
$$ [n]_Q! = [n]_Q [n-1]_Q \ldots [1]_Q,\eqno(6.5)$$
$$ {m \brack n}_Q = {[m]_Q! \over [m-n]_Q! [n]_Q!}.\eqno(6.6)$$
As it can be easily seen from eq. (6.1) under the substitution $Q\rightarrow
Q^{-1}$ one obtains
$$ \left[n\right]_Q!= Q^{n(n-1)/2}\left[n\right]_{1/Q},\eqno(6.7)$$ 
and
$$\left[ \begin{array}{c} n\\ k\end{array}\right]_Q =Q^{k(n-k)} 
\left[\begin{array}{c} n \\ k \end{array}\right]_{1/Q}.\eqno(6.8) $$
$Q$-factorials are connected to $q$-factorials by
$$[n]!= q^{-n(n-1)/2}[n]_Q!,  \qquad {\rm with} \qquad Q=q^2.\eqno(6.9)$$

\section{$Q$-deformed elementary functions}

The definitions of $Q$-deformed elementary functions \cite{Exton}
 look similar 
to these given in sec. 3. The {\sl $Q$-deformed exponential function}
is defined as
$$ e_Q(ax)=\sum _{n=0}^{\infty} {a^n\over [n]_Q!} x^n,\eqno(7.1)$$
and satisfies the property
$$ e_Q(x) e_{1/Q}(-x) =1.\eqno(7.2)$$
(Notice that $e_Q(x) e_Q(-x) \neq 1.$)

The {\sl $Q$-deformed trigonometric functions} are defined as
$$ \sin_Q (x)= \sum_{n=0}^{\infty} (-1)^n {x^{2n+1}\over [2n+1]_Q!},
\eqno(7.3)$$
$$ \cos_Q (x)= \sum_{n=0}^{\infty} (-1)^n {x^{2n}\over [2n]_Q!}.\eqno(7.4)$$
One can easily show that 
$$ \sin_Q(x)= {1\over 2i} \left( e_Q(ix)-e_Q(-ix) \right),\eqno(7.5)$$
$$ \cos_Q(x)= {1\over 2}  \left( e_Q(ix)+e_Q(-ix) \right).\eqno(7.6)$$
Instead of the familiar identity $\sin^2(x)+\cos^2(x)=1$ one has 
$$ \sin_Q(x) \sin_{1/Q}(x)+\cos_Q(x)\cos_{1/Q}(x)=1.\eqno(7.7)$$
The above defined $Q$-deformed functions are examples of 
$Q$-deformed hypergeometric functions\cite{Exton,Slater,MV93}. In addition
$Q$-deformed polynomials, which are counterparts of the ordinary non deformed 
polynomials, can be defined,  such as $Q$-deformed Hermite polynomials and 
$Q$-de\-fo\-rmed Laguerre polynomials \cite{Exton}.

\section{ $Q$-derivative}

Given the function $f(x)$ one defines its $Q$-derivative $D^Q_x$ 
\cite{Exton} by
the expression
$$ D^Q_x f(x) = { f(Qx) -f(x) \over (Q-1)x}. \eqno(8.1)$$
The similarity between this definition and the one of $Q$-numbers (eq. (6.1))
is clear. 

One can easily prove that
$$ D^Q_x x^n \equiv {Q^n x^n -x^n \over (Q -1)x} = [n]_Q x^{n-1}, \eqno(8.2) $$
which looks exactly like eq. (4.2). In addition one has 
$$ D^Q_x e_Q(ax)= a e_Q(ax),\eqno(8.3)$$
$$ D^Q_x e_{1/Q} (ax)= a e_{1/Q} (aQx),\eqno(8.4)$$
$$ D^Q_x \sin_Q(ax)= a\cos_Q(ax), \eqno(8.5)$$
$$ D^Q_x \cos_Q(ax)=-a\sin_Q(ax).\eqno(8.6)$$
One can also easily see that $\sin_Q(ax)$ and $\cos_Q(ax)$ are the linearly 
independent solutions of the $Q$-differential equation
$$ (D^Q_x)^2 u(x) +a^2 u(x) =0,\eqno(8.7)$$
while the functions $\sin_{1/Q} (ax)$ and $\cos_{1/Q}(ax)$ satisfy the 
equation
$$ (D^Q_x)^2 u(x) + a^2 u(Q^2 x) =0.\eqno(8.8)$$

The following {\sl Leibnitz rules} can also be shown:
$$	D^Q_x\left(f_1(x) f_2(x)\right)= \left(D^Q_x f_1(x)\right)f_2(Qx) +
	f_1(x)\left(D^Q_x f_2(x)\right), \eqno(8.9)$$
$$	D^Q_x\left(f_1(x) f_2(x)\right)= \left(D^Q_x f_1(x)\right)f_2(x) +
	f_1(Qx)\left(D^Q_x f_2(x)\right).\eqno(8.10)$$
One can further obtain
$$ D^Q_x {f_1(x) \over f_2(x)} = {\left(D^Q_x f_1(x)\right) f_2(x) -f_1(x)
	\left(D^Q_x f_2(x) \right) \over f_2(Qx) f_2(x) }.\eqno(8.11) $$
	For the second derivative of $f(x)$ one has
$$ (D^Q_x)^2 f (x)= (Q-1)^{-2} Q^{-1} x^{-2}\left\{f(Q^2 x) -(Q+1) f(Q x) 
+Q f(x) \right\}, \eqno(8.12)$$
	and by mathematical induction we obtain the general formula
$$ (D^Q_x)^n f(x) =(Q-1)^{-n} Q^{-n(n-1)/2} x^{-n} \sum_{k=0}^n\left[ 
\begin{array}{c}
n \\ k \end{array}\right]_Q (-1)^k Q^{k(k-1)/2} f(Q^{n-k}x). \eqno(8.13)$$

\section{  $Q$-integration}

In a way analogous to that of sec. 5
 the definite $Q$-integral of the function $f(x)$
in the interval $[0, 1]$ is defined \cite{Exton} as follows
$$ \int_0^1 f(x) d_Q x = (1 -Q) \sum_{s=0}^{\infty} f(Q^s) Q^s, \eqno(9.1)$$
	assuming that $Q$ is real and $|Q| < 1$, while
for the definite integral of $f(x)$ in the interval
	$[0, \infty]$, we have
$$ \int_0^{\infty} f(x) d_Q x = (1-Q) \sum_{s=-\infty}^{\infty} Q^s f(Q^s).
\eqno(9.2)$$
For the indefinite $Q$-integral of $f(x)$ one has
$$ \int f(x) d_Q x = (1-Q) x \sum_{s=-\infty}^{\infty} Q^s f(Q^sx) + 
{\rm constant}. \eqno(9.3)$$
One can easily check that $Q$-differentiation and $Q$-integration are 
operations inverse to each other
$$ D^Q \int f(x)d_Q x = f(x).\eqno(9.4)$$
The formula for $Q$-integration by parts reads
$$ \int\left(D^Q f_1(x)\right)f_2(x) d_Q x= f_1(x) f_2(x) -
	\int f_1(Qx) \left(D^Q f_2(x) \right) d_Q x. \eqno(9.5)$$

\vfill\eject 
\section{The $q$-deformed harmonic oscillator}

The interest for possible applications of quantum algebras in physics has been
triggered in 1989 by the introduction of the $q$-deformed harmonic oscillator
\cite{Bie873,Mac4581,SF983},
of which earlier equivalent versions existed \cite{AC524,Kur111}. 

 The $q$-deformed harmonic oscillator 
\cite{Bie873,Mac4581,SF983,Song821,Ng1023,Yan459}  is defined 
in terms
of the creation and annihilation operators $a^\dagger$ and $a$ and the number
operator $N$, which satisfy the commutation relations
$$ [N, a^\dagger] = a^\dagger, \quad [N, a] =-a, \eqno(10.1)$$
$$ a a^\dagger - q^{\mp 1} a^\dagger a = q^{\pm N}  .\eqno(10.2)$$
In addition the following conditions of hermitian conjugation hold
(in the case of $q$ being a real number or $q$ being a root of unity)
$$ (a^\dagger)^\dagger =a, \qquad N^\dagger =N.\eqno(10.3)$$
Eq. (10.1) is the same as in ordinary quantum mechanics, while eq. (10.2) is 
modified by the presence of the deformation parameter $q$. 
For $q\to 1$ it is clear that eq. (10.2) goes to the usual
boson commutation relation $[a, a^\dagger]=1$. 
 An immediate consequence of (10.2) is that
$$ a^\dagger a = [N], \quad a a^\dagger = [N+1]. \eqno(10.4)$$
Thus the number operator $N$ is {\it not} equal to $a^\dagger a$, as in the 
ordinary case. The operators $a^\dagger$ and $a$ are referred to as 
{\sl  $q$-deformed boson creation and annihilation operators} respectively. 

The basis of the Fock space is defined by repeated action of the creation 
operator $a^\dagger$ on the vacuum state, which is annihilated by $a$: 
$$ a\vert 0\rangle=0,\qquad    |n> = {(a^\dagger)^n\over\sqrt{ [n]!}} |0>. 
\eqno(10.5)$$
The action of the operators on the
basis is given by
$$N |n> = n |n>, \eqno(10.6)$$
$$a^\dagger |n> = \sqrt{[n+1]} |n+1>, \eqno(10.7)$$
$$a |n> = \sqrt{[n]} |n-1>. \eqno(10.8)$$
We remark that these equations look very similar to the ones of the 
ordinary case, the only difference being that $q$-numbers appear under 
the square roots instead of usual numbers. 

The Hamiltonian of the $q$-deformed harmonic oscillator is
$$H={\hbar \omega \over 2} (a a^\dagger + a^\dagger a), \eqno(10.9)$$
and its eigenvalues in the basis given above are
$$ E(n)= {\hbar \omega \over 2} ([n]+[n+1]). \eqno(10.10)$$
One can easily see that for $q$ real the energy eigenvalues increase more 
rapidly than the ordinary case, in which the spectrum is equidistant, i.e.
the spectrum gets ``expanded''. 
In contrast, for $q$ being a phase factor ($q=e^{i\tau}$ with $\tau$ real) 
the eigenvalues of the energy increase less rapidly than the ordinary
(equidistant) case, i.e. the spectrum is ``compressed''. 
In particular, for $q$ real ($q=e^\tau$) the eigenvalues can be written as 
$$ E(n) ={\hbar \omega \over 2} {\sinh \left(\tau\left(n+{1\over 2}\right)
\right) \over \sinh{\tau \over 2}} ,\eqno(10.11)$$
while for $q$ being a phase factor ($q=e^{i\tau}$) one has 
$$ E(n) = {\hbar\omega\over 2} {\sin\left(\tau\left( n+{1\over 2}\right)
\right) \over \sin{\tau \over 2}}.\eqno(10.12)$$
In both cases in the limit $q\to 1$ ($\tau \to 0$) the ordinary expression
$$ E(n)= \hbar \omega \left( n+{1\over 2}\right) \eqno(10.13)$$
is recovered.  

In addition, the following commutation relation holds
$$ [a, a^\dagger] = [N+1] -[N]. \eqno(10.14)$$
For $q$ being a phase factor, this commutation relation takes the form
$$[a, a^\dagger] = {\cos{(2N+1)\tau\over 2}\over\cos{\tau\over 2}}.
\eqno(10.15)$$

It is useful to notice that the $q$-deformed boson operators $a^\dagger$ and
$a$ can be expressed in terms of usual boson operators $\alpha^\dagger$ and
$\alpha$ (satisfying $[\alpha, \alpha^\dagger]=1$ and $N=\alpha^\dagger 
\alpha$) through the relations \cite{Song821,KD415}
$$a=\sqrt{[N+1]\over N+1} \alpha =\alpha\sqrt{[N]\over N}, \qquad 
a^\dagger=\alpha^\dagger \sqrt{[N+1]\over N+1}=\sqrt{[N]\over N} 
\alpha^\dagger.\eqno(10.16)$$
The square root factors in the last equation have been called {\sl
$q$-deforming functionals}. 

For $q$ being a primitive root of unity, i.e. $q=e^{2\pi i/k}$ 
($k=2$, 3,~\dots), it is clear the the representation of eqs (10.5)--(10.8)
becomes 
finite-dimensional and has dimension $k$, since only the vectors $|0>$, 
$|1>$, \dots, $|k-1>$ can be present. This case has been related to the 
system of two anyons \cite{FT163}. In what follows we are going to assume 
that $q$ is {\sl not} a primitive root of unity. 

A discussion on the position and momentum operators of the $q$-deformed 
oscillator can be found in \cite{CK917}. 

\section{The $Q$-deformed harmonic oscillator}

A different version of the deformed harmonic oscillator can be
obtained by defining \cite{KD415,BF171,Jan91} the operators $b$, $b^+$ through
the equations
$$ a= q^{1/2} b q^{-N/2}, \quad a^\dagger=q^{1/2}  q^{-N/2} b^\dagger .
\eqno(11.1)$$
Eqs. (10.1) and  (10.2) then give
$$ [N, b^\dagger ] = b^\dagger, \quad [N, b]=-b , \eqno(11.2)$$
$$ b b^\dagger -  q^2 b^\dagger b =1 . \eqno(11.3)$$
This oscillator has been first introduced by Arik and Coon \cite{AC524}
and later considered also by Kuryshkin \cite{Kur111}.  
One then easily finds that
$$b^\dagger b= [N]_Q, \quad b b^\dagger = [N+1]_Q, \eqno(11.4)$$
where $Q=q^2$ and $Q$-numbers are defined in (6.1). 
The basis is defined by
$$ b\vert 0\rangle=0, \qquad   |n>= {(b^\dagger)^n\over \sqrt{[n]_Q!}}|0>, 
\eqno(11.5)$$
while the action of the operators on the basis is given by  
$$ N |n> = n |n>, \eqno(11.6)$$
$$ b^\dagger |n>= \sqrt {[n+1]_Q} |n+1>, \eqno(11.7)$$
$$ b |n> = \sqrt{[n]_Q} |n-1> .\eqno(11.8)$$
The Hamiltonian of the corresponding deformed harmonic oscillator
has the form
$$H = {\hbar \omega\over 2} (b b^\dagger + b^\dagger b), \eqno(11.9)$$
the eigenvalues of which are
$$E(n)={\hbar \omega \over 2} ([n]_Q+[n+1]_Q). \eqno(11.10)$$
One can easily see that for $Q=e^T$, where $T>0$ and real, the spectrum 
increases more rapidly than the ordinary (equidistant) spectrum, while 
for $Q=e^T$, with $T<0$ and real, the spectrum is increasing less rapidly 
than the ordinary (equidistant) case. 

From the above relations, it is clear that the following
commutation relation holds
$$[b, b^\dagger] = Q^N.\eqno(11.11)$$

\section{The generalized deformed oscillator}

In addition to the oscillators described in the last two sections,
many kinds of deformed oscillators have been introduced in the literature
(see \cite{BD100} for a list). 
All of them can be accommodated within the common mathematical framework 
of the {\it generalized deformed oscillator} \cite{Das789,DY4157}, 
which is defined 
as the algebra generated by the operators $\{1, a, a^\dagger, N\}$ and the 
{\sl structure function} $\Phi(x)$, satisfying the relations 
$$ [a, N]=a, \qquad [a^\dagger, N]=-a^\dagger, \eqno(12.1)$$
$$ a^\dagger a =\Phi(N) =[N], \qquad aa^\dagger = \Phi(N+1) =[N+1],
\eqno(12.2)$$
where $\Phi(x)$ is a positive analytic function with $\Phi(0)=0$ and $N$ is 
the number operator. 
>From eq. (12.2) we conclude that 
$$N=\Phi^{-1} (a^\dagger a),\eqno(12.3)$$
and that the following commutation and anticommutation relations are 
obviously satisfied:
$$ [a, a^\dagger]=[N+1]-[N], \qquad \{a,a^\dagger\}=[N+1]+[N].\eqno(12.4)$$
The {\sl structure function} $\Phi(x)$ is characteristic to the deformation
scheme. In Table 1 the structure functions corresponding to different 
deformed oscillators are given. They will be further discussed at the end of
this section. 

It  can be proved that
the generalized deformed algebras possess a Fock space of 
 eigenvectors \hfill\break
$|0>,|1>,\ldots,|n>,\ldots$
of the number operator $N$
$$N|n>=n|n>,\quad <n|m>=\delta_{nm}, \eqno(12.5) $$
if the {\it vacuum state} $|0>$ satisfies the following relation:
$$ a|0>=0. \eqno(12.6)$$
 These eigenvectors are generated by the formula:
 $$ \vert n >= {1 \over \sqrt{ [n]!}} {\left( a^\dagger \right)}^n \vert 0 >,
\eqno(12.7) $$
where
 $$[n]!=\prod_{k=1}^n [k]= \prod_{k=1}^n \Phi(k). \eqno(12.8) $$
The generators  $a^\dagger$ and $a$ are the creation and
annihilation operators of this deformed oscillator algebra:
$$a\vert n> = \sqrt{[n]} a\vert n-1>,\qquad
 a^\dagger \vert n> = \sqrt{[n+1]} a\vert n+1>. \eqno(12.9) $$

These eigenvectors are also eigenvectors of the energy operator 
$$H={\hbar \omega \over 2} (aa^\dagger +a^\dagger a), \eqno(12.10)$$
corresponding to the eigenvalues 
$$ E(n)= {\hbar\omega \over 2} (\Phi(n)+\Phi(n+1)) =
{\hbar \omega \over 2} ([n]+[n+1]).\eqno(12.11)$$

For $$\Phi(n)=n  \eqno(12.12)$$ 
one obtains the results for the ordinary harmonic oscillator.
For $$\Phi(n)= {q^n -q^{-n}\over q-q^{-1}}=[n]  \eqno(12.13)$$
one has the results for the $q$-deformed harmonic oscillator, while the 
choice
$$ \Phi(n) ={Q^n-1\over Q-1}= [n]_Q  \eqno(12.14)$$
leads to the results of the $Q$-deformed harmonic oscillator. Many more 
cases are shown in Table 1, on which the following comments apply:


\begin{table}[bt]
\begin{center}
\caption{ Structure functions of special deformation schemes}
\bigskip
\begin{tabular}{|c c p{2.0 in}|}
\hline
\ & $\Phi(x)$ & Reference \\
\hline\hline
\romannumeral 1 & $x$ & 
harmonic  oscillator, bosonic algebra\\[0.05in]
\romannumeral 2&  ${ {q^x- q^{-x} }  \over {q- q^{-1} } } $ &
$q$-deformed harmonic oscillator \cite{Bie873,Mac4581}
\\[0.05in]
\romannumeral 3& ${ {q^x- 1 } \over {q- 1 } } $ & Arik--Coon,
Kuryshkin, or $Q$-deformed oscillator \cite{AC524,Kur111} \\[0.05in]
\romannumeral 4&  ${ {q^x- p^{-x} }  \over {q- p^{-1} } } $ &
2-parameter deformed oscillator \cite{BJM820,JBM775,CJ711}
\\[0.05in]
\romannumeral 5& $x(p+1-x)$ & parafermionic oscillator
 \cite{OK82} \\[0.05in]
\romannumeral 6& $ { \sinh (\tau x) \sinh (\tau (p+1-x) )}\over
{ \sinh^2(\tau) } $ & $q$-deformed parafermionic oscillator
\cite{FV1019,OKK591} \\[0.05in]
\romannumeral 7& $x\cos^2(\pi x/2) + (x+p-1)\sin^2(\pi x /2)  $&
parabosonic oscillator \cite{OK82} \\[0.05in]
\romannumeral 8&
$\begin{array}{c}
\frac{\sinh(\tau x)}{\sinh(\tau)} 
\frac{\cosh(\tau (x+2N_0-1))}{\cosh(\tau)} \cos^2 (\pi x/2) +\\
+ \frac{\sinh(\tau (x+2N_0-1))}{\sinh(\tau)} 
\frac{\cosh(\tau x)}{\cosh(\tau)} \sin^2 (\pi x/2)
\end{array}$
 & $q$-deformed parabosonic oscillator
 \cite{FV1019,OKK591} \\[0.05in]
\romannumeral 9 &
$\sin^2 {\pi x/2}$ & fermionic algebra \cite{JBSZ123} \\[0.05in]
\romannumeral 10 & $ q^{x-1} \sin^2 {\pi x/2}$ &
$q$-deformed fermionic algebra \cite{Hay129,CK72,FSS179,Gang819,Sci219,PV613}
 \\[0.05in]
\romannumeral 11& 
$\frac{1-(-q)^x}{1+q}$ & generalized $q$-deformed fermionic algebra
 \cite{VPJ335} \\[0.05in]
\romannumeral 12& $x^n$ & \cite{Das789} \\[0.05in]
\romannumeral 13& ${ {sn(\tau x)} \over {sn(\tau )} }$ & \cite{Das789}
 \\[0.05in]
\hline
\end{tabular}
\end{center}
\end{table}

i) Two-parameter deformed oscillators have been introduced 
\cite{BJM820,JBM775,CJ711}, in analogy to the one-parameter deformed 
oscillators. 

ii) Parafermionic oscillators \cite{OK82} of order $p$ represent particles
of which the maximum number which can occupy the same state is $p$. 
Parabosonic oscillators \cite{OK82} can also be introduced. The deformed 
oscillator realization of the parafermionic and parabosonic oscillator has 
been studied in  \cite{BD100}, while in \cite{Jing98} an equivalent realization
for the parabosonic oscillator has been introduced. 

iii) $q$-deformed versions of the parafermionic and parabosonic oscillators 
have also been introduced \cite{FV1019,OKK591}. 

iv) $q$-deformed versions of the fermionic algebra \cite{JBSZ123} have also
been introduced \cite{Hay129,CK72,FSS179,Gang819,Sci219,PV613}, as well as 
$q$-deformed versions 
of generalized $q$-deformed fermionic algebras \cite{VPJ335}. It has been
proved, however, that $q$-deformed fermions are fully equivalent to the 
ordinary fermions \cite{BD1589,JX891}.       

\section{The physical content of deformed harmonic oscillators}

In order to get a feeling about the physical content of the various 
deformed harmonic oscillators 
it is instructive to construct potentials giving spectra similar to these of
the oscillators. 

\subsection{Classical potentials equivalent to the $q$-oscillator}

Let us consider the $q$-deformed harmonic oscillator first. For small values 
of $\tau$ one can 
 take Taylor expansions
of the functions appearing there and thus find an expansion of the $q$-number 
$[n]$ of eq. (2.1) in terms of powers of $\tau^2$. The final result is
$$[n]= n\pm {\tau^2  \over 6} (n-n^3) +{\tau^4 \over 360} (7n -
10 n^3 + 3 n^5) \pm {\tau^6\over 15120} (31 n -49 n^3 +21 n^5 -3 n^7)
+\ldots, \eqno(13.1)$$
where the upper (lower) sign corresponds to $q$ being a phase (real).
Using this expansion the energy of the $q$-deformed harmonic oscillator
of eq. (10.10) can be rewritten as
$$E(n)/(\hbar \omega)=(n+{1\over 2}) (1\pm {\tau^2 \over 24}) \mp {\tau^2 \over 6}
(n+{1\over 2})^3 +\ldots \eqno(13.2)$$
On the other hand, one can consider the potential
$$V(x)=V_0+k x^2 +\lambda x^4 + \mu x^6 +\xi x^8 +\ldots.\eqno(13.3)$$
If $\lambda$, $\mu$, $\xi$ are much smaller than $k$, one can
consider this potential as a harmonic oscillator potential plus some
perturbations and calculate the corresponding spectrum through
the use of perturbation theory \cite{BDK6153} (see also subsec. 34.2). 
 In order to keep the subsequent formulae simple, we measure
$x$ in units of $(\hbar/(2m\omega))^{1/2}$.
 Using standard first order perturbation theory one finds that the
corresponding spectrum up to the order considered is
$$E(n)= E_0+(2\kappa +25 \mu) (n+{1\over 2}) + (6\lambda+245 \xi)
(n+{1\over 2})^2 +20 \mu (n+{1\over 2})^3 + 70 \xi (n+{1\over 2})^4.
\eqno(13.4)$$
By equating coefficients of the same powers of $(n+{1\over 2})$
in eqs. (13.2) and (13.4), we can determine the coefficients
appearing in the expansion of the potential given in eq.
(13.3). The final result for the potential, up to the order
considered, is
$$V(x)= ({1\over 2} \pm {\tau^2\over 8}) x^2 \mp
{\tau^2 \over 120} x^6 .\eqno(13.5)$$
We see therefore that to lowest order one can think of the $q$-oscillator 
with a small value of the parameter $\tau$ as a harmonic oscillator 
perturbed by a $x^6$ term. 
It is clear that if we go to higher order, the next term to appear
will be proportional to $ \tau^4 x^{10}$.

The results of this subsection are corroborated by an independent study
of the relation between the $q$-deformed harmonic oscillator and the 
ordinary anharmonic oscillator with $x^6$ anharmonicities \cite{AZB2555}.  

A different interpretation of the $q$-deformed oscillator as the nonlinear 
oscillator with the frequency dependent on the amplitude has been given 
in \cite{MMZ95}. 

\subsection{Classical potentials equivalent to the $Q$-deformed oscillator}

Similar considerations can be made for the $Q$-oscillator. 
Defining $Q=e^T$ it is instructive to construct the expansion of
the $Q$-number of eq. (6.1) in powers of $T$. Assuming that $T$ is
small and taking Taylor expansions in eq. (6.1) one finally has
$$[n]_Q= n+{T\over 2} (n^2 -n) +{T^2\over 12} (2n^3-3n^2+1) +
{T^3\over 24} (n^4-2n^3+n^2) +\ldots\eqno(13.6)$$
Then the corresponding expansion of the energy levels of the
oscillator of eq. (11.10) is
$$E(n)/(\hbar \omega)= E'_0 + (1-{T\over 2} +{T^2\over 8}
-{T^3\over 16}+\ldots) (n+{1\over 2}) + ({T\over 2} -{T^2\over 4}
+{5 T^3\over 48} -\ldots) (n+{1\over 2})^2   $$
$$+({T^2\over 6}-{T^3\over 12}+\ldots)  (n+{1\over 2})^3
+({T^3\over 24} -\ldots) (n+{1\over 2})^4  .\eqno(13.7)$$
Comparing this expansion to eq. (13.4) and equating
equal powers of $(n+{1\over 2})$ we arrive at the following
expression for the potential
$$V(x)={T^2\over 12} + ({1\over 2} -{T\over 4}-{T^2\over 24}+{T^3
\over 48}) x^2 + ({T\over 12}-{T^2\over 24}-{T^3\over 144})x^4 
+ ({T^2\over 120} -{T^3\over 240}) x^6 + {T^3\over 1680} x^8 .
\eqno(13.8)$$
Thus to lowest order one can think of the $Q$-oscillator with small
values of the parameter $T$ as a harmonic oscillator perturbed by a 
$x^4$ term. A similar expression is found, for example, by Taylor expanding 
the modified P\"oschl--Teller potential \cite{PT1933},
 which, among
other applications, has been recently used in the description of
hypernuclei \cite{GLM283,LGM2098,LGM303,GLMP391}.
The modified P\"oschl--Teller potential (see also subsec. 34.2) has the
form
$$ V(x)_{PT}= -{A\over \cosh(ax)^2}. \eqno(13.9)$$
Its Taylor expansion is
$$ V(x)_{PT}= A(-1+ a^2 x^2 -{2\over 3}a^4 x^4 +{17\over 45} a^6 x^6 -\ldots).
\eqno(13.10)$$
We remark that this expansion contains the same powers of $x$
as the expansion (13.8).  Furthermore, the signs of the coefficients
of the same powers of $x$ in the two expansions are the same for
$T<0$.

\subsection{WKB-EPs for the $q$-deformed oscillator}

The potentials obtained above are only rough lowest order estimates. More 
accurate methods exist for constructing WKB equivalent potentials (WKB-EPs)
giving (within the limits of WKB approximation) the same spectrum as the 
above mentioned oscillators. A method by which this can be achieved has been
given by Wheeler \cite{Whe1976} and is described by Chadan and Chabatier
\cite{CS1977}. Applying this method to the $q$-deformed oscillator (sec. 10) 
with $q$ being a phase factor one finds the potential \cite{BDK795,BDKJMP}
$$V(x)=\left( {\tau\over 2\sin(\tau/2)}\right)^2 {m \omega^2 \over 2} x^2
\left[ 1-{8\over 15} \left({x\over 2 R_e}\right)^4 +{4448\over 1575}
\left({x\over 2R_e}\right)^8 -{345344\over 675675} \left({x\over 2R_e}
\right)^{12} +\dots\right],\eqno(13.11)$$
where 
$$R_e= {1\over \tau} \left( {\hbar^2\over 2 m}\right)^{1/2}
\left( {2\sin(\tau/2)\over \hbar \omega}\right)^{1/2},\eqno(13.12)$$
while for the $q$-deformed oscillator with $q$ real one has 
$$V(x)=  \left({\tau\over 2\sinh(\tau/2)}\right)^2 {m \omega^2\over 2} x^2
\left[ 1+{8\over 15}\left({x\over 2 R_h}\right)^4+{4448\over 1575} \left(
{x\over 2 R_h}\right)^8 +{345344\over 675675}\left( {x\over 2 R_h}\right)^{12}
+\dots\right],\eqno(13.13)$$
where
$$ R_h ={1\over \tau} \left({\hbar^2\over 2 m}\right)^{1/2} 
\left( 2\sinh (\tau/2) \over \hbar \omega\right)^{1/2}.\eqno(13.14)$$
The results of this subsection are corroborated by an independent study of 
the relation between the $q$-deformed harmonic oscillator and the ordinary
anharmonic oscillator with $x^6$ anharmonicities\cite{AZB2555}.  

\subsection{WKB-EPs for the $Q$-deformed oscillator}

Using the same technique one finds that the WKB 
equivalent potential 
for the $Q$-deformed harmonic oscillator (sec. 11) takes the form 
\cite{Ioann93}
$$ V(x) = V_{min} + {(\ln Q)^2 \over Q} \left( {Q+1\over Q-1}\right) ^2 
{1\over 2} m \omega^2 x^2 $$
$$\left[ 1 -{2\over 3} \left( {x\over R'}\right)^2
+ {23\over 45} \left( {x\over R'}\right)^4 -{134\over 315} \left( {x\over R'}
\right)^6 + {5297\over 14172} \left( {x\over R'}\right)^8 -\ldots \right],
\eqno(13.15)$$
where 
$$V_{min} = {\hbar \omega (\sqrt{Q}-1) \over 2 \sqrt{Q} (\sqrt{Q}+1)},
\eqno(13.16)$$
and
$$ R'= \left( {\hbar \sqrt{Q} (Q-1) \over \omega m (Q+1) }\right)^{1/2} 
{(\ln Q)^{-1} \over \sqrt{2}}.\eqno(13.17) $$

We remark that this WKB-EP contains all even powers of $x$, in contrast to 
the WKB-EPs for the $q$-oscillator (eqs (13.11), (13.13)), which contain 
only the powers
$x^2$, $x^6$, $x^{10}$, \dots. This is in agreement with the lowest order 
results obtained in subsecs 13.1 and 13.2.

In this section approximate expressions for some potentials having similar 
energy spectrum with some  deformed oscillators have been given.
Another point of view is the search for phase space coordinates having exact
symmetries, like the  operators involved in $q$-deformed algebras in the case 
of real $q$ \cite{Raj93}. In the case of $q$ being a root of unity the 
corresponding phase space is discrete \cite{BDEF150}

\section{The quantum algebra su$_q$(2)}

Quantum algebras are generalized versions of the usual Lie algebras, to
which they reduce when the deformation parameter $q$ is set equal to unity. 
A simple example of a quantum algebra is provided by su$_q$(2)
\cite{Bie873,Mac4581}, which
is generated by the operators
$J_+$, $J_0$, $J_-$, satisfying  the commutation relations
$$ [ J_0, J_{\pm} ] = \pm J_{\pm} , \eqno (14.1)$$
$$ [ J_+ , J_- ] = [ 2 J_0 ] , \eqno (14.2)$$
with $J_0^{\dagger} = J_0$, $(J_+)^{\dagger} = J_-$. 
We remark that eq. (14.1) is the same as in the case of the ordinary su(2) 
algebra, while eq. (14.2) is different, since in the usual su(2) case it 
reads $$[J_+, J_-]=2J_0.\eqno(14.3)$$
In the rhs of eq. (14.3) one has the first power of the $J_0$ operator, while
in the rhs of eq. (14.2) one has the $q$-operator $[2J_0]$, defined in sec. 2.
Because of eqs (2.2) and (2.3) it is clear that if one writes the rhs of eq. 
(14.2) in expanded form all odd powers of $J_0$ will appear: 
$$ [J_+,J_-]= {1\over \sinh(\tau)} \left( {2\tau J_0\over 1!}+
{(2\tau J_0)^3\over 3!} +{(2\tau J_0)^5\over 5!}+\dots\right) \qquad {\rm for}
\qquad q=e^\tau,\eqno(14.4)$$
$$ [J_+,J_-]= {1\over \sin(\tau)} \left( {2\tau J_0\over 1!} -
{(2\tau J_0)^3\over 3!} +{(2\tau J_0)^5\over 5!} -\dots\right) \qquad
{\rm for} \qquad q=e^{i\tau}.\eqno(14.5)$$
Thus su$_q$(2)
can be loosely described as a nonlinear generalization of the usual su(2):
While in usual Lie algebras the commutator of two generators is always 
producing a linear combination of generators, in the case of quantum 
algebras the commutator of two generators can contain higher powers 
of the generators as well.  

The irreducible representations (irreps) $D^J$ of su$_q$(2)  
(which have dimensionality $2J+1$) are
determined by highest weight states with $J=0, {1\over 2}, 1,
\ldots$. The basic states $|J, M>$ (with $-J\leq M \leq J$) are
connected with highest weight states $|J, J>$ as follows
$$|J, M> = \sqrt{[J+M]!\over [2J]! [J-M]!} (J_-)^{J-M} |J,J>,
\eqno (14.6)$$
with $J_+|J,J> =0$ and $<J,J|J,J>=1$.
The action of the generators of the algebra on these basic vectors is 
given by
$$ J_0 |J,M>= M |J,M>,\eqno(14.7)$$
$$ J_{\pm} |J,M>=\sqrt{ [J\mp M] [J\pm M+1]} |J,M\pm 1>.\eqno(14.8)$$
These expressions look similar to the ones of the usual su(2) algebra, the 
only difference being that $q$-numbers appear under the square root instead
of ordinary numbers. 

The second order Casimir operator of su$_q$(2) is determined from the condition
that it should commute with all of the generators of the algebra. The resulting
operator is
$$C^q_2 = J_- J_+ +[J_0] [J_0+1] = J_+J_- +[J_0][J_0-1]. \eqno (14.9)$$
Its eigenvalues in the above mentioned basis are given by 
$$C^q_2 |J, M> = [J] [J+1] |J, M> , \eqno(14.10)$$
while for the usual su(2) the eigenvalues of the Casimir operator are $J(J+1)$.
One can easily check that for real $q$ ($q=e^\tau$ with $\tau$ real)
the eigenvalues $[J][J+1]$ produce 
a spectrum increasing more rapidly than $J(J+1)$ (an expanded spectrum),
while for $q$ being a phase factor ($q=e^{i\tau}$ with $\tau$ real) the 
eigenvalues $[J][J+1]$ correspond to a spectrum increasing less rapidly 
than $J(J+1)$ (a compressed spectrum). 

It should be noticed that the generators $J_+$, $J_0$, $J_-$ of su$_q$(2)
are connected to the generators $j_+$, $j_0$, $j_-$ of the usual su(2),
which satisfy the commutation relations 
$$[j_0, j_{\pm}]=\pm j_{\pm}, \qquad [j_+, j_-]=2j_0, \eqno(14.11)$$ 
through the relations \cite{CZ237,CGZ676}
$$ J_0=j_0, \qquad J_+=\sqrt{[j_0+j] [j_0-1-j]\over (j_0+j) (j_0-1-j)}j_+,
         \qquad J_-=j_-\sqrt{[j_0+j][j_0-1-j]\over [j_0+j][j_0-1-j]},
\eqno(14.12)$$
where $j$ is determined by the relation for the second order Casimir operator
of su(2)
$$ C= j_-j_++j_0(j_0+1) = j_+ j_-+j_0(j_0-1) = j(j+1).\eqno(14.13)$$

\section{Realization of su$_q$(2) in terms of $q$-deformed bosons}

Realizations of Lie algebras in terms of (ordinary) bosons are useful not only
as a convenient mathematical tool, but also because of their applications in 
physics \cite{KM375}. In the case of quantum algebras it turns out
that boson realizations are possible in terms of the $q$-deformed boson 
operators already introduced in sec. 10. 

In the case of su$_q$(2) the generators can be mapped onto $q$-deformed 
bosons as follows \cite{Bie873,Mac4581}
$$ J_+= a_1^\dagger a_2, \qquad J_-=a_2^\dagger a_1, \qquad J_0={1\over 2}
(N_1-N_2), \eqno(15.1)$$
where $a_i^\dagger$, $a_i$ and $N_i$ are $q$-deformed boson creation, 
annihilation and number operators as these introduced in sec. 10. 
One can easily prove that the boson images satisfy the commutation 
relations (14.1) and (14.2). For example, one has
$$[J_+,J_-]= a_1^\dagger a_2 a_2^\dagger a_1-a_2^\dagger a_1 a_1^\dagger a_2=
[N_1] [N_2+1]-[N_1+1][N_2] = [N_1-N_2]=[2J_0],\eqno(15.2)$$
where use of the identity (2.6) has been made. 

In the $q$-boson picture the normalized highest weight vector is
$$|JJ> ={(a_1^\dagger)^{2J}\over \sqrt{[2J]!}} |0>,\eqno(15.3)$$
while the general vector $|JM>$ is given by
$$ |JM> = {(a_1^\dagger)^{J+M}\over \sqrt{[J+M]!}} {(a_2^\dagger)^{J-M}
\over \sqrt{[J-M]!}} |0>.\eqno(15.4)$$  

It should be  noticed that it was the search for a boson realization of 
the su$_q$(2) algebra that led to the introduction of the $q$-deformed 
harmonic oscillator in 1989 \cite{Bie873,Mac4581}. 

Starting from su$_q$(2) one can formulate a $q$-deformed version of angular
momentum theory. Some references are listed here:

i) Clebsch-Gordan coefficients for su$_q$(2) can be  found in 
\cite{KK443,GKK2769,Rue1085,Nom1954,KK4009,STK959,AM1139,AS981,AS1435}.

ii) 6-j symbols for su$_q$(2) can be found in 
\cite{KK2717,STK1746,RR3761,Mae2598}.

iii) 9-j symbols for su$_q$(2) can be found in \cite{STK2863}. 

iv) 3n-j symbols (with n=1, 2, 3, 4, 5) for su$_q$(2) can be found in 
\cite{Nom3851}. 

v) The $q$-deformed version of the Wigner--Eckart theorem can be found 
in \cite{Kli2919,Nom2345,STK690}. 

vi) Irreducible tensor operators for su$_q$(2) can be found in \cite{STK690}. 

In addition, it should be noticed that a two-parameter deformation of su(2),
labelled as su$_{p,q}$(2) has been introduced 
\cite{BJM820,CJ711,BH165,BH629,Kib9234,Jing543,JVdJ2831}. Clebsch-Gordan 
coefficients for su$_{p,q}$(2) have been discussed in
\cite{SW5563,DMM43,MM5177,Que19,KAS9439}. 

The way in which the algebra su$_q$(2) can be realized in terms of the 
$Q$-deformed bosons of sec. 11 is given in subsec. 35.1 (see eqs 
(35.13)--(35.16)). 

\section{The quantum algebra su$_q$(1,1)}

In this section we shall give a brief account of the algebra su$_q$(1,1)
\cite{KD415,UA237,Aiz1115}.

In the classical case the so(2,1) generators satisfy the
commutation relations \cite{LSK87}
$$ [K_1, K_2] =-i K_3, \quad\quad [K_2, K_3]= i K_1, \quad\quad
[K_3, K_1] =i K_2, \eqno(16.1)$$
which differ from the classical so(3) commutation relations in the sign
of the r.h.s. of the first commutator. 
Defining
$$ K_+=K_1 +i K_2, \quad\quad K_-=K_1-i K_2, \quad\quad
K_3=K_z, \eqno(16.2)$$
one obtains the su(1,1) commutation relations
$$[K_z, K_{\pm}] = \pm K_{\pm}, \quad\quad [K_+, K_-]=-2 K_z,
\eqno(16.3)$$
which differ from the familiar su(2) commutation relations in 
the sign of the r.h.s. of the last commutator. 
The generators of su(1,1) accept the following boson representation
\cite{AIG27}
$$K_+= a_1^\dagger a_2^\dagger, \quad\quad K_-=a_1 a_2, \quad\quad
K_z = {1\over 2} (a_1^\dagger a_1 + a_2^\dagger a_2 +1) , \eqno(16.4)$$
where $a_1^\dagger$, $a_1$, $a_2^\dagger$, $a_2$ satisfy usual boson 
commutation relations.

The second order Casimir operator of so(2,1) is \cite{LSK87}
$$C_2[{\rm so}(2,1)]= -(K_1^2+K_2^2-K_3^2),\eqno(16.5)$$
while for su(1,1) one has 
$$C_2 [{\rm su}(1,1)]= [K_0] [K_0-1] -K_+ K_- = [K_0] [K_0+1] -K_- K_+.
\eqno(16.6)$$

In the quantum case, the generators of su$_q$(1,1) satisfy the
commutation relations \cite{KD415,UA237,Aiz1115}
$$[K_0, K_{\pm}]=\pm K_{\pm}, \quad\quad [K_+, K_-]=-[2 K_0]. 
\eqno(16.7)$$

The generators of su$_q$(1,1) accept the following boson
representation 
$$K_+= a_1^\dagger a_2^\dagger, \quad\quad K_-=a_1 a_2, \quad \quad
K_0= {1\over 2} (N_1 +N_2 +1), \eqno(16.8)$$
where the bosons $a_i^\dagger$, $a_i$ ($i=1,2$) satisfy the usual $q$-boson 
commutation relations. 

The second order Casimir operator of su$_q$(1,1) is 
$$C_2[{\rm su}_q(1,1)]=[K_0][K_0-1]-K_+K_-=[K_0][K_0+1]-K_-K_+. \eqno(16.9)$$
Its eigenvalues are \cite{UA237,Aiz1115}
$$C_2[{\rm su}_q(1,1)] |\kappa \mu> = [\kappa] [\kappa-1] |\kappa \mu>,
\eqno(16.10)$$
where
$$\kappa ={1+|n_1 -n_2|\over 2}, \quad
\mu ={1+n_1+n_2\over 2}, \eqno(16.11)$$
since the basis has the form
$| \kappa \mu> = | n_1 > | n_2 >$, with 
$$|n_i> = {1 \over \sqrt{[n_i]!}} (a^\dagger_i)^{n_i} |0>. \eqno(16.12)$$
In this basis the possible values of $\mu$ are given by
$$\mu=\kappa, \kappa+1, \kappa+2, \ldots,\eqno(16.13)$$
up to infinity, while $\kappa$ may be any positive real number.
The action of the generators on this basis is given by 
$$ K_0 \vert \kappa \mu \rangle = \mu \vert \kappa \mu \rangle ,\eqno(16.14) $$
$$ K_{\pm} \vert \kappa \mu \rangle = \sqrt{ [\mu\pm \kappa]
[\mu\mp \kappa\pm 1]} \vert \kappa \mu\pm 1\rangle.\eqno(16.15)$$

Clebsch-Gordan and Racah coefficients for su$_q$(1,1) can be found in 
\cite{STK959,AS1435,STK1746,Mae2598,Aiz1937,Alv92}. 
Furthemore, a two-parameter deformed version of 
su$_q$(1,1), labelled as su$_{p,q}$(1,1), has been introduced 
\cite{CJ711,Kib9234,BBCH341,Jing495,Kli9339}. 

\section{Generalized deformed su(2) and su(1,1) algebras}

In the same way that in addition to the $q$-deformed oscillators one can 
have generalized deformed oscillators, it turns out that 
generalized deformed su(2) algebras, containing su$_q$(2) as a special case
and  having representation theory similar 
to that of the usual su(2),  can be constructed \cite{BDK871,Pan5065}. It has 
been proved that it is possible to construct an algebra 
$$ [J_0, J_{\pm}]=\pm J_{\pm}, \qquad [J_+,J_-]=\Phi(J_0(J_0+1))-
\Phi(J_0(J_0-1)),\eqno(17.1)$$
where $J_0$, $J_+$, $J_-$ are the generators of the algebra and 
$\Phi(x)$ is any increasing entire function defined for $x\geq -1/4$. 
Since this algebra is characterized by the function $\Phi$, we use for it
the symbol su$_{\Phi}$(2). The appropriate basis $|l,m>$ has the 
properties
$$ J_0|L,M> = M |L,M>, \eqno(17.2)$$
$$ J_+|L,M> = \sqrt{\Phi(L(L+1))-\Phi(M(M+1))} |L,M+1>, \eqno(17.3)$$
$$ J_-|L,M> = \sqrt{\Phi(L(L+1))-\Phi(M(M-1))} |L, M-1>,\eqno(17.4) $$
where 
$$ L=0, {1\over 2}, 1, {3\over 2}, 2, {5\over 2}, 3, \ldots,\eqno(17.5)$$
  and 
$$ M= -L, -L+1, -L+2, \ldots, L-2, L-1, L.\eqno(17.6)$$
The Casimir operator is
$$ C= J_-J_+ +\Phi(J_0(J_0+1))=J_+J_-+\Phi(J_0(J_0-1)), 
\eqno(17.7)$$
its eigenvalues indicated by
$$C |L,M> = \Phi(L(L+1)) |L,M>.\eqno(17.8)$$
The usual su(2) algebra is recovered for 
$$ \Phi(x(x+1))= x(x+1),\eqno(17.9)$$
while the quantum algebra su$_q$(2) 
$$ [J_0, J_{\pm}]=\pm J_{\pm}, \qquad [J_+, J_-]=[2 J_0]_q,\eqno(17.10) $$
occurs for
$$ \Phi(x(x+1))= [x]   [x+1]  .\eqno(17.11)$$

The su$_{\Phi}$(2) algebra occurs in several cases, in which the rhs of the 
last equation in (17.1) is an odd function of $J_0$ \cite{BDK2197}. It can be 
seen that other
algebraic structures, like the quadratic Hahn algebra QH(3) \cite{GLZ217}
and the finite W algebra  $\bar{\rm W}_0$ \cite{Bow945}
 can be brought into the su$_\Phi$(2) form, the 
advantage being that the representation theory of su$_\Phi$(2) is already 
known. It can also be proved that several physical systems, like 
the isotropic oscillator in a 2-dim curved space with constant curvature
\cite{Hig309,Lee489}, the Kepler system in a 2-dim curved space with constant 
curvature \cite{Hig309,Lee489}, and the system of two identical particles in 
two dimensions \cite{LM3649} can also be put into
an su$_\Phi$(2) form. More details can be found in \cite{BDK2197}. 

A unified description of several deformed su(2), su(1,1) and osp(1,2) 
algebras has been given in \cite{BDKL311}, using Verma modules techniques. 
Nonlinear deformed su(2) algebras involving {\sl two} deforming functions 
have been constructed in \cite{BKDLQ1189} and have been endowed with a 
double Hopf algebraic structure, referred to as a two-colour quasitriangular
Hopf algebra,  in \cite{BDKLQ369}.   

Nonlinear sl(2) algebras subtending generalized angular momentum theories 
have been introduced in \cite{ABCD3075} and their representation theory 
has been developed. In particular the quadratic sl(2) and cubic sl(2) 
algebras have been studied in detail \cite{ABCD3075}.  

A general method for developing nonlinear deformations of the su(2) and 
su(1,1) algebras has been developed in \cite{DQ127,DQ227,DQ961}. Using this 
method nonlinear deformations of the su(1,1) algebra related to the 
P\"oschl--Teller potential have been obtained in \cite{Ni313,CLG6473}. 

\section{Generalized deformed parafermionic oscillators}

It turns out that the generalized deformed su$_\Phi$(2) algebras mentioned 
in the last section are related to 
 generalized deformed parafermionic oscillators, which we will therefore 
describe here.

 It has been proved \cite{Que245} that any generalized 
deformed parafermionic algebra of order $p$ can be written as a generalized 
oscillator (sec. 12) with structure function 
$$ F(x)= x (p+1-x) (\lambda +\mu x+\nu x^2 +\rho x^3 +\sigma x^4 +\ldots),
\eqno(18.1)$$
where $\lambda$, $\mu$, $\nu$, $\rho$, $\sigma$, \dots are real constants 
satisfying the conditions
$$ \lambda + \mu x + \nu x^2 + \rho x^3 + \sigma x^4 +\ldots > 0, \qquad 
x \in \{ 1,2,\ldots, p\}.\eqno(18.2)$$

Considering an su$_{\Phi}$(2) algebra \cite{BDK871} with structure function 
$$ \Phi(J_0(J_0+1))= A J_0(J_0+1) + B (J_0(J_0+1))^2 + C (J_0(J_0+1))^3,
\eqno(18.3) $$
and making the correspondence 
$$ J_+ \to A^{\dag}, \qquad J_-\to A, \qquad J_0\to N,\eqno(18.4)$$
one finds that the su$_{\Phi}$(2) algebra is equivalent to a generalized 
deformed parafermionic  oscillator of the form
$$F(N)= N (p+1-N)$$ $$ [ -(p^2(p+1)C +p B)+ (p^3 C +(p-1)B) N+
((p^2-p+1)C +B) N^2+ (p-2) C N^3 + C N^4], \eqno(18.5)$$
if the condition
$$ A+ p(p+1) B + p^2 (p+1)^2 C =0 \eqno(18.6)$$
holds. The condition of eq. (18.2) is always satisfied for $B>0$ and $C>0$. 

In the special case of $C=0$ one finds that the su$_{\Phi}$(2) algebra 
with structure function
$$ \Phi(J_0(J_0+1))= A J_0(J_0+1) + B (J_0(J_0+1))^2\eqno(18.7)$$
is equivalent to a generalized deformed parafermionic oscillator 
characterized by
$$ F(N)= B N (p+1-N) (-p+(p-1)N+ N^2),\eqno(18.8)$$
if the condition
$$ A+ p(p+1) B=0\eqno(18.9)$$ 
is satisfied. The condition of eq. (18.2) is satisfied for $B>0$. 

Including higher powers of $J_0(J_0+1)$ in eq. (18.3) results in higher powers
of $N$ in eq. (18.5) and higher powers of $p(p+1)$ in eq. (18.6). If, however, 
one sets $B=0$ in eq. (18.7), then eq. (18.8) vanishes, indicating that no 
parafermionic oscillator equivalent to the usual su(2) rotator can be 
constructed.  

It turns out that several other mathematical structures, like the finite W 
algebras $\bar {\rm W}_0$ \cite{Bow945}
and W$^{(2)}_3$ (see subsec. 36.5) can be put into the generalized 
deformed parafermionic oscillator form. The same is true for several 
physical systems, such as the isotropic oscillator and the Kepler problem 
in a 2-dim curved space with constant curvature \cite{Hig309,Lee489}, and the 
Fokas--Lagerstrom \cite{FL325}, Smorodinsky--Winternitz \cite{SW444}, 
and Holt \cite{Holt1037} potentials. Further details can be found in 
\cite{BDK2197}. 

A detailed discussion of the representation theory of several deformed 
oscillator algebras can be found in \cite{QV141,QV115}. 

\section{The su$_q$(2) rotator model}

It has been suggested by Raychev, Roussev and Smirnov \cite{RRS137}
 and independently by Iwao \cite{Iwao363,Iwao368} that 
rotational spectra of deformed nuclei can be described by the $q$-deformed 
rotator, which corresponds to the 2nd order 
Casimir operator of the quantum algebra su$_q$(2), already studied in 
sec. 14. We shall show here that this assumption works and discuss the 
reasons behind this success, as well as the relation \cite{PLB251}
 between the 
su$_q$(2) model and the Variable Moment of Inertia (VMI) model (see sec. 20).

The $q$-deformed rotator corresponds to the  Hamiltonian
$$H = {1\over 2 I} C_2({\rm su}_q(2)) + E_0, \eqno (19.1)$$
where $I$ is the moment of inertia and $E_0$ is the bandhead energy
(for ground state bands $E_0=0$).
For $q$ real, i.e. with $q=e^{\tau}$ with $\tau$ real, the
energy levels of the $q$-rotator are 
$$E(J) = {1\over 2I} [J] [J+1] +E_0 = {1\over 2I}
{\sinh(\tau J) \sinh(\tau (J+1))\over  \sinh^2(\tau) } +E_0.\eqno(19.2)$$
For $q$ being a phase, i.e. $q=e^{i\tau}$ with $\tau$ real, 
 one obtains
$$E(J) ={1\over 2I} [J] [J+1] +E_0= {1\over 2I}
{\sin (\tau J) \sin(\tau(J+1))\over \sin^2 (\tau)} +E_0. \eqno (19.3)$$

 Raychev {\it et al.} \cite{RRS137} have  found that good fits of rotational 
spectra of
even--even rare earths and actinides are obtained with eq. (19.3).
It is easy to check that eq. (19.2) fails in describing such spectra.
In order to understand  this difference, it is useful to make
Taylor expansions of the  quantities in the numerator of eq. (19.2)
(eq. (19.3)) and collect together the terms containing the same
powers of $J(J+1)$ (all other terms cancel out), finally summing up
the coefficients of each power. In the first case the final
result is
$$E(J) = E_0 + {1\over 2I}{1\over (\sqrt{\pi \over 2\tau} I_{1/2}(\tau)
)^2} (\sqrt{\pi\over 2\tau} I_{1/2}(\tau) J(J+1) +\tau
\sqrt{\pi\over 2\tau} I_{3/2}(\tau) (J(J+1))^2 $$
$$+{2\tau^2\over 3} \sqrt{\pi \over 2\tau} I_{5/2}(\tau) (J(J+1))^3 
+{\tau^3\over 3} \sqrt{\pi\over 2\tau} I_{7/2} (\tau) (J(J+1))^4
+\ldots \eqno(19.4)$$
where $\sqrt{\pi\over 2\tau} I_{n+{1\over 2}} (\tau)$ are the
modified spherical Bessel functions of the first kind \cite{AS72}. 

In the second case (eq. (19.3)) following the same procedure one
obtains
$$E(J) =E_0+ {1\over 2I} {1\over (j_0(\tau))^2} (j_0(\tau) J(J+1)
 -\tau j_1(\tau) (J(J+1))^2 $$ $$+{2\over 3} \tau^2 j_2(\tau) (J(J+1))^3
-{1\over 3} \tau^3  j_3(\tau) (J(J+1))^4+{2\over 15} \tau^4
j_4(\tau) (j(j+1))^5-\ldots), \eqno(19.5)$$
where $j_n(\tau)$ are the spherical Bessel functions of the first
kind \cite{AS72}.

Both  results are of the form 
$$ E(J)= AJ(J+1)+B(J(J+1))^2+C (J(J+1))^3+D(J(J+1))^4+\dots.\eqno(19.6)$$
Empirically it is known that nuclear rotational spectra do show such 
a behaviour, the coefficients $A$, $B$, $C$, $D$, \dots having alternating 
signs (starting with $A$ positive) and magnitudes dropping by about 
3 orders of magnitude each time one moves to the next higher power of $J(J+1)$
\cite{deVoigt949,XWZ2337}.

It is interesting to check if the empirical characteristics of
the coefficients $A$, $B$, $C$, $D$ are present in the case of the
expansions of eqs. (19.2), (19.3), especially for small values of
$\tau$. (Since we deal with rotational spectra, which
are in first order approximation described by the usual algebra
su(2), we expect $\tau$  to be relatively small, i.e.
the deviation of su$_q$(2) from su(2) to be small. This is in
agreement to the findings of \cite{RRS137}, where $\tau$ is
 found to be around 0.03.)

One can easily see that in eq. (19.2) it is impossible to get alternating
signs, while in eq. (19.3) the condition of alternating signs is readily 
fulfilled. This fact has as a result that the energy levels given by eq. (19.2)
increase more rapidly than the levels given by the $J(J+1)$ rule,
while the levels given by eq. (19.3) increase less rapidly than $J(J+1)$. 
In order to check the order of magnitude of the coefficients
for small values of $\tau$, it is useful to expand the spherical
Bessel functions appearing in eq. (19.3) and keep only the lowest order
term in each expansion. The result is
$$E(J) = E_0+{1\over 2I} (J(J+1)-{\tau^2\over 3} (J(J+1))^2
+{2\tau^4\over 45} (J(J+1))^3 $$
$$-{\tau^6\over 315} (J(J+1))^4 +{2\tau^8\over 14175}
(J(J+1))^5 -\ldots ).
\eqno(19.7)$$
We remark that each term contains a factor $\tau^2$ more
than the previous one. For $\tau$ in the area of 0.03, $\tau^2$
is of the order of $10^{-3}$, as it should. We conclude therefore that
eq. (19.6) is suitable for fitting rotational spectra, since its
coefficients have the same characteristics as the empirical
coefficients of eq. (19.6).  

Extended comparisons of the su$_q$(2) predictions to experimental data for 
ground state bands of rare earth and actinide nuclei can be found in 
\cite{RRS137,PLB251,CGST180,HVW1337}.
More recently, the su$_q$(2) formalism has been used for the 
description of $\beta$- and $\gamma$-bands of deformed rare earths and 
actinides, with satisfactory results \cite{MDRRB95}.

It is necessary for $E(J)$ to be an increasing function of $J$. In order
to guarantee this in eq. (19.3) one must have 
$$ \tau (J+1) \leq {\pi \over 2}.\eqno(19.8)$$
In the case of $\tau=0.036$ (as in $^{232}$U in \cite{PLB251}),
one finds $J\leq 42$, this limiting value being larger than the highest 
observed $J$ in ground state bands in the actinide region \cite{Sakai84}. 
Similarly, for $\tau=0.046$ (as in $^{178}$Hf in \cite{PLB251}),
one finds $J\leq 32$, this limiting value being again higher than the highest 
observed $J$ in ground state bands in the rare earth region \cite{Sakai84}. 

It should be mentioned  that a method of fixing the deformation parameter 
by demanding that the relevant Hilbert space be large enough to contain 
all the irreps of the system has been suggested \cite{Johal051}
in the framework of Barnett--Pegg theory, with good results. 

\section{ Comparison of the su$_q$(2) model to other models}

\subsection{The Variable Moment of Inertia (VMI) model}

In lowest order approximation rotational nuclear spectra can be described by
the formula 
$$E(J)={J(J+1)\over 2\Theta} ,\eqno(20.1)$$
where $\Theta$ is the moment of inertia of the nucleus, which is assumed to
be constant. However, in order to get closer
agreement to experimental data, one finds that he has to include higher
order terms in this formula, as shown in eq. (19.6). 

Another way to improve agreement with experiment is to let the moment of 
inertia $\Theta$ to vary as a function of the angular momentum $J$. 
One thus obtains the {\sl Variable Moment of Inertia (VMI) model} 
\cite{MGB1864}. In this model the 
levels of the ground state band are given by
$$ E(J) = {J(J+1)\over 2\Theta(J)} + {1\over 2} C (\Theta(J)-\Theta_0)^2,
\eqno(20.2)$$
where $\Theta(J)$ is the moment of inertia of the nucleus at the state with
angular momentum $J$, while $C$ and $\Theta_0$ are the two free parameters 
of the model, fitted to the data. The parameter $\Theta_0$ corresponds to 
the ground state moment of inertia, while instead of the parameter $C$ 
it has been found meaningful to use the parameter combination 
$$ \sigma= {1\over 2 C \Theta_0^3},\eqno(20.3)$$
which is related to the softness of the nucleus. The moment of inertia 
at given $J$ is determined through the variational condition 
$$ {\partial E(J)\over \partial \Theta(J)} \vert _J = 0,\eqno(20.4)$$
which is equivalent to the cubic equation 
$$ \Theta(J)^3 -\Theta(J)^2 \Theta_0 -{J(J+1) \over 2C} =0. \eqno(20.5) $$
This equation has only one real root, which can be written as 
$$ \Theta(J) = \root 3 \of { {J(J+1)\over 4C} + {\Theta_0^3\over 27}
+ \sqrt{ {(J(J+1))^2\over 16 C^2} + { \Theta_0^3 J(J+1)\over 54C} } } $$
$$ + \root 3 \of { {J(J+1)\over 4C} + {\Theta_0^3\over 27} -
\sqrt{ {(J(J+1))^2 \over 16C^2} + {\Theta_0^3 J(J+1)\over 54C} } } +
{\Theta_0 \over 3}.\eqno(20.6)$$
Expanding the roots in this expression one obtains 
$$ \Theta(J) = \Theta_0 ( 1+ \sigma J(J+1) -2 \sigma^2 (J(J+1))^2 
+7 \sigma^3 (J(J+1))^3 -30 \sigma^4 (J(J+1))^4 +\ldots ). \eqno(20.7)$$
Using eq. (20.7) in eq. (20.2) one obtains the following expansion for the 
energy
$$E(J) = {1\over 2\Theta_0} ( J(J+1) -{1\over 2} \sigma (J(J+1))^2  
+\sigma^2 (J(J+1))^3 -3 \sigma^3 (J(J+1))^4 +\dots ). \eqno(20.8)$$
Empirically it is known \cite{MGB1864} that for rotational 
nuclei the softness parameter $\sigma$ is of the order of $10^{-3}$. 
Therefore the expansion of eq. (20.8) has the same characteristics as the 
expansion of eq. (19.6) (alternating signs, successive coefficients
falling by about 3 orders of magnitude). 

\subsection{Comparison of the su$_q$(2) model to the VMI and related models}

We now turn to the comparison of the expansion of eq. (19.5) to
the Variable Moment of Inertia (VMI) model, discussed in the previous 
subsection. 
 Comparing eqs (19.5) and (20.8) we see that both expansions have
the same form. The moment of inertia parameter $I$ of (19.5)
corresponds to the ground state moment of inertia $\Theta_0$
of (20.8). The small parameter of the expansion is $\tau^2$
in the first case, while it is the softness parameter
$1/(2C\Theta_0^3)$ in the second. However, the numerical
coefficients in front of each power of $J(J+1)$ are not the same.

In \cite{PLB251}  a comparison is made between the
 parameters obtained by fitting the same spectra by the su$_q$(2) 
and VMI formulae. 
The agreement between $1/(2I)$ and $1/(2\Theta_0)$ is very good,
as it is the agreement between $\tau^2$ and $\sigma$ as well.
Therefore the known \cite{MGB1864}
 smooth variation of $\Theta_0$ and
$\sigma$ with the ratio $R_4=E(4)/E(2)$ is expected to hold
for the parameters $I$ and $\tau^2$ as well. This is indeed seen 
in \cite{PLB251}. 

 The difference between the expansions of eqs (19.5) (or (19.7)) and (20.8) is
also demonstrated by forming the dimensionless ratios
$AC/(4 B^2)$ and $A^2D/(24 B^3)$ in eq. (19.6).  In the case of eq.
(20.8) both quantities are equal to 1, as expected, since it is known
that  the VMI is equivalent \cite{DDK235,KDD333}
to the Harris expansion \cite{Harris65}, in which
both quantities are known to be equal to 1. In the case of
eq. (19.7) the corresponding values are $AC/(4B^2)=1/10$ and
$A^2 D/ (24 B^3)=1/280$. According to the empirical values
of these ratios given in \cite{XWZ2337,MWZ2545}, the ratios given from eq.
(19.7) are better than the ratios given by eq. (20.8), especially
the second one.

\subsection{ The hybrid model}

The hybrid model of nuclear collective motion 
\cite{Mosh156,Mosh257,CMV605,CFHM1367}
has been introduced in order to provide a link
between the two successful ways of describing low-lying nuclear 
excitations: the extended form of the Bohr-Mottelson model (BMM)
\cite{GG449,HSMG147}
and the Interacting Boson Model (IBM) \cite{IA1987,IVI91,Talmi93}
(see secs 27, 31 for more details). 
The hybrid model combines the 
advantages of both models, i.e. the geometrical significance of the 
collective coordinates inherent in the extended BMM, and the use of 
group theoretical concepts characterizing IBM. In the framework of 
the rotational limit of the hybrid model, associated with the 
chain u(6)$\supset$su(3),  Partensky and Quesne \cite{PQ340,PQ2837}, 
starting from the fact that in the geometrical description the square 
of the deformation is proportional to the moment of inertia of the 
ground state band, proved that  the energy levels of the 
ground state band are given by 
$$ E(J)= {A\over J(J+1)+B} J(J+1),\eqno(20.9)$$
with $A$ being a free parameter and $B$ given by 
$$ B= 8 N^2 +22 N-15, \eqno(20.10)$$
where $N$ is the sum of the number of valence proton pairs (or proton-hole 
pairs, when more than half of the proton valence shell is filled) $N_{\pi}$
and the number of the valence neutron pairs (or neutron-hole pairs, when more
than half of the neutron valence shell is filled) $N_{\nu}$. We remark 
that in the framework of this model the moment of inertia 
$$ \Theta(J,N) = {J(J+1)+B \over 2A}\eqno(20.11)$$
depends on both the angular momentum $J$ and the valence pair number $N$. 

It is instructive to expand $E(J)$ in powers of J(J+1) \cite{MRR67}
$$ E(J) = A \sum_{k=0}^{\infty} (-1)^k \left( {J(J+1)\over B} \right)^{k+1}$$ 
$$ = 
{A\over B} ( J(J+1)- {1\over B} (J(J+1))^2 + {1\over B^2}
(J(J+1))^3 -{1\over B^3} (J(J+1))^4 +\ldots). \eqno(20.12)$$
Comparing the present expansion to the one of eq. (19.7) for the su$_q$(2)
model one has 
$$ \tau = \sqrt{3\over B} = \sqrt{3\over 8 N^2 + 22 N-15},\eqno(20.13)$$
obtaining thus a connection between the $\tau$ parameter of the su$_q$(2)
model and a microscopic quantity, the valence nucleon pair number. 
>From this equation it is clear that $\tau$ decreases for increasing $N$.
Thus the minimum values of $\tau$ are expected near the midshell regions, 
where the best rotators are known to be located. This is a reasonable 
result, since $\tau$ describes the deviations from the su(2) (rigid 
rotator) limit. It is clear that the minimum deviations should occur in the
case of the best rigid rotators. 

From eq. (20.13) one can obtain for each nucleus the value of $\tau$ from 
the number of valence nucleon pairs present in it. These predictions for 
$\tau$ have been compared in \cite{MRR67} to the values of
$\tau$ found empirically by fitting the corresponding spectra, with good 
results in both the rare earth and the actinide regions.  In addition eq.
 (20.13) indicates that in a given shell nuclei characterized by the same
valence nucleon pair number $N$ will correspond to the same value of 
$\tau$. Such multiplets have been studied in the framework of the hybrid 
model by \cite{Bon351}.   

It is worth noticing that fits of $\gamma_1$-bands in the rare earth region
(Er, Yb isotopes) \cite{MDRRB95}
give $\tau$ parameters very similar to the ones coming 
from fitting the corresponding ground state bands, in addition exhibiting
the same as the one mentioned above behaviour of $\tau$ as a function of $N$.

Taking further advantage of the above connection between the su$_q$(2) 
model and the hybrid model, the parameter $\tau$ has been connected 
\cite{MRR557} to
the nuclear deformation parameter $\beta$, as well as to the 
electromagnetic transition probabilities B(E2:$2_1^+\rightarrow 0_1^+$). 
Since both $\beta$ and B(E2:$2_1^+\rightarrow 0_1^+$) are known to increase 
with increasing collectivity, $\tau$ is expected to decrease with increasing 
$\beta$ or increasing B(E2:$2_1^+\rightarrow 0_1^+$). This expectation is 
corroborated by the results reported in \cite{MRR557}. 

\subsection{ Other models}

It should be noticed that an empirical formula very similar to that of 
eq. (19.3) has been recently proposed by Amal'ski\u{\i} \cite{Amal70}
 on completely different physical grounds. The formula reads 
$$ E(J) = A \sin^2\left( {\pi J \over B}\right) ,\eqno(20.14)$$
where $A$ and $B$ are free parameters. This formula should be compared 
to eq. (19.3), with which it is almost identical. 

A different formula, also giving very good fits of 
rotational spectra, has been introduced by Celeghini, Giachetti, Sorace 
and Tarlini \cite{CGST180}, based on the $q$-Poincar\'e rotator. 

\section{Electromagnetic transitions in the su$_q$(2) model} 

We have already seen that the su$_q$(2) formalism provides an alternative 
to the VMI model, the deformation parameter $q$ being connected to the 
softness parameter of the VMI model. 

The stretching effect present in rotational energy levels,
which can equally well be described in terms of the VMI model
and the su$_q$(2) symmetry, should also manifest itself in the
B(E2) transition probabilities among these levels.
If deviations from the su(2) symmetry are observed in the
energy levels of a band, relevant deviations should also
appear in the B(E2) transitions connecting them.
In the case of the VMI model no way has been found for
making predictions for the B(E2) transition probabilities
connecting the levels of a collective band. The su$_q$(2)
symmetry naturally provides such a link. Before studying the su$_q$(2) case,
though, it is useful to recall the predictions of other models on this matter.

\subsection{The collective model of Bohr and Mottelson}

In rotational
bands one has \cite{BM75}
$$B(E2:J+2\rightarrow J) = {5\over 16\pi} Q_0^2
|C^{J+2,2,J}_{K,0,-K}|^2, \eqno(21.1)$$
i.e. the B(E2) transition probability depends on the relevant
Clebsch-Gordan coefficient of su(2), while  $Q_0^2$ is
the intrinsic electric quadrupole moment
and $K$ is the projection of the angular momentum $J$
on the symmetry axis of the nucleus in the body-fixed frame.
For $K=0$ bands, as the ground state bands,  one then has 
\cite{BM75}
$$  B(E2:J+2\rightarrow J)= {5\over 16\pi} Q_0^2
{3\over 2} {(J+1)(J+2)\over (2J+3)(2J+5)}.\eqno(21.2)$$
It is clear that the B(E2) values should saturate 
with increasing $J$.

\subsection{The Interacting Boson Model (IBM)}

It is also instructive to mention what happens in the case of the 
Interacting Boson Model (IBM) \cite{IA1987}, the successful 
algebraic model of nuclear structure with which we are going to deal
in more detail later (see sections 27, 31). 
In the case of the su(3) limit of the IBM,
which is the limit applicable to deformed nuclei, the
corresponding expression is \cite{AI201} 
$$B(E2:J+2\rightarrow J)= {5\over 16\pi} Q_0^2 {3\over 2}
{(J+1)(J+2)\over (2J+3)(2J+5)} {(2N-J)(2N+J+3)\over (2N+3/2)^2},
\eqno(21.3)$$
where $N$ is the total number of bosons. Instead of saturation
one then gets a decrease of the B(E2) values at high $J$,
which finally reach zero at $J=2N$. This is a well known
disadvantage of the simplest version of the model (IBM-1)
 due to the small number of collective bosons ($s$ ($J=0$) and
$d$ ($J=2$)) taken into account. It can be corrected by the
inclusion of higher bosons ($g$ ($J=4$), $i$ ($J=6$), etc),
which approximately restore saturation (see \cite{IA1987,Bon88} for full list
of references).

Another way to avoid the problem of decreasing B(E2)s in the
su(3) limit of IBM at high $J$ is the recently proposed \cite{Muk229}
transition from the compact su(3) algebra to the noncompact
sl(3,R) algebra. The angular momentum at which this transition
takes place is fitted to experiment. In this way an increase of the
B(E2) values at high $J$ is predicted, which agrees well with
the experimental data for $^{236}$U.

\subsection{The su$_q$(2) model} 

In order to derive a formula similar to (21.2) in the su$_q$(2)
case, one needs to develop an su$_q$(2) angular
momentum theory. As mentioned in sec. 15, this has already been achieved. 
It turns out that an equation similar to (21.1)
 holds in the q-deformed case,
 the only difference being that the Clebsch-Gordan
coefficient of the su$_q$(2) algebra must be used instead.
These coefficients have the form \cite{STK959} 
$$ _q C^{J+2,2,J}_{K,0,-K} = q^{2K} \sqrt{
[3] [4] [J-K+2] [J-K+1] [J+K+1] [J+K+2]\over [2] [2J+2] [2J+3]
[2J+4] [2J+5]}. \eqno(21.4)$$
 For $K=0$ bands one then has
$$B_q(E2;J+2\rightarrow J)={5\over 16\pi} Q_0^2
{[3] [4] [J+1]^2 [J+2]^2 \over [2] [2J+2] [2J+3] [2J+4] [2J+5]}.
\eqno(21.5)$$
For $q=e^{i\tau}$ this equation takes the form
$$B_q(E2; J+2\rightarrow J) = {5\over 16\pi} Q_0^2 
{\sin(3 \tau) \sin(4\tau) \over \sin(2\tau) \sin(\tau)}$$
$$ {(\sin(\tau(J+1)))^2 (\sin(\tau(J+2)))^2\over
\sin(\tau(2J+2)) \sin(\tau(2J+3))
\sin(\tau(2J+4))\sin(\tau(2J+5))}.\eqno(21.6)$$
  It is
useful to get an idea of the behaviour of this expression as a
function of $J$, especially for the small values of $\tau$
found appropriate for the description of ground state spectra.
Expanding all functions and keeping corrections of the leading
order in $\tau$ only, one has
$$B_q(E2, J+2\rightarrow J) = {5\over 16\pi} Q_0^2 {3\over 2}
{(J+1)(J+2)\over (2J+3)(2J+5)}
(1+{\tau^2\over 3}(6 J^2 +22 J+12)).\eqno(21.7)$$
We see that the extra factor, which depends on $\tau^2$,
contributes an extra increase with $J$, while the usual  su(2)
expression reaches saturation at high $J$ and IBM even predicts
a decrease.

\subsection{Comparison to experiment}

Is there any experimental evidence for such an increase? In order
to answer this question one should discover cases in which
the data will be consistent with the su$_q$(2) expression but
inconsistent with the classical su(2) expression. (The opposite
cannot happen, since the classical expression is obtained from
the quantum expression for the special parameter value $\tau=0$.)
Since error bars of B(E2) values are usually large, in most cases
both symmetries are consistent with the data. One should expect
the differences to show up more clearly in two cases:

i) In rare earth nuclei not very much deformed (i.e. with an
$R_4=E(4)/E(2)$ ratio around 3.0). These should be deformed enough
so that the su$_q$(2) symmetry will be able to describe them,
having, however, at the same time values of $\tau$ not very small.
Since in several of these nuclei backbending (or upbending) occurs
at $J=14$ or 16, one can expect only 5 or 6 experimental points
to compare the theoretical predictions with.

ii) In the actinide region no backbending occurs up to around
$J=30$, so that this is a better test ground for the two
symmetries.
 However, most nuclei in this region are well deformed, so
that small values of $\tau$ should be expected, making the
distinction between the two theoretical predictions difficult.

A few characteristic examples (4 rare earths and an actinide)
are given in \cite{BFRRS3275}.
In all cases it is clear that the su$_q$(2) curve follows the
experimental points, while the su(2) curve has a different shape
which cannot be forced to go through all the error bars.
Several comments are now in place:

i) For a given nucleus the value of the parameter $\tau$ obtained from
fitting the B(E2) values among the levels of the ground state band
should be equal to the value obtained from fitting the energy levels
of the ground state band. In \cite{BFRRS3275}  it is clear that both
 values
are similar, although in most cases the value obtained from the
B(E2)s is smaller than the value obtained from the spectra. It
should be taken into account, however, that in most cases the
number $n'$ of levels fitted is different (larger) than the number
$n$ of the B(E2) values fitted. In the single case ($^{184}$W)
in which $n=n'$, the two $\tau$ values are almost identical, as
they should.

ii) One can certainly try different fitting procedures. Using
the value of $\tau$ obtained from the B(E2) values for fitting
the spectrum one gets a reasonably good description of it,
although the squeezing of the spectrum is not as much as it should
have been (with the exception of $^{184}$W). Using the value of
$\tau$ obtained from the spectrum for fitting the B(E2) values
one obtains an increase more rapid than the one shown by the data
(again with the exception of $^{184}$W). One can also try to make an
overall fit of spectra and B(E2)s using a common value of $\tau$.
Then both the squeezing of the spectrum and the rise of the B(E2)s
can be accounted for reasonably well although not exactly.
 One should notice, however,
that the experimental uncertainties of the B(E2)s are much higher
than the uncertainties of the energy levels.

iii) Concerning energy levels, the rigid rotator model and the
su(3) limit of the IBM
predict a $J(J+1)$ increase, while the su$_q$(2) model and the VMI
model predict squeezing, which is seen experimentally.

iv) Concerning the B(E2) values, the VMI makes no prediction,
 the rigid rotator predicts saturation
at high $J$, the su(3) limit of the IBM predicts decrease,
while the su$_q$(2) model predicts an increase. The evidence
presented in \cite{BFRRS3275}
 supports the su$_q$(2) prediction, but
clearly much more work, both experimental and analytical,
is needed before final conclusions can be drawn. The modified
su(3) limit of IBM described in \cite{Muk229}, as well as the su(3) 
limit of the $sdg$-IBM \cite{LJ1686} also support the
increase of the B(E2) values at high $J$.
Increasing BE(2) values are also predicted in the framework of the 
Fermion Dynamical Symmetry Model \cite{PCWF2224}, as well as by a binary 
cluster model \cite{BMP2095}. 
There is also empirical evidence for increasing B(E2) values in the recent 
systematics by Zamfir and Casten \cite{ZC1280}. 

v) It is clear that much further work is needed as far as comparisons of the 
su$_q$(2) predictions to experimental BE(2) values are concerned
for safe conclusions to be reached. 

vi) Since the quadrupole operator is not a member of the symmetry algebra 
su$_q$(2) under consideration, it is clear that the B(E2) values studied 
here do not contain any dynamical deformation effects, but only the kinematical
ones (through the use of the $q$-deformed Clebsch-Gordan coefficients). 
A more complete approach to the problem will be the construction of a 
larger algebra, of which the quadrupole operator will be a member and it will
also be an irreducible tensor under su$_q$(2) or so$_q$(3). This task 
will be undertaken in sec. 29. 

\section{ Superdeformed bands}

One of the most impressive experimental discoveries in nuclear phy\-sics 
during the last decade was that of superdeformation \cite{Twin811}
(see \cite{Hod365,NT533,JK321}
 for relevant reviews). The energy levels of superdeformed bands follow
the $J(J+1)$ rule much more closely than the usual rotational bands. 
Levels with $J$ larger than 60 have been observed. A compilation of 
superdeformed bands has been given in \cite{HW43}.
 The best examples have been found in the 
A$\approx$150 mass region, while additional examples have been found in the
A$\approx$80, 
 A$\approx$130 and A$\approx$190 regions. It is understood that the 
superdeformed bands in the A$\approx$150 region correspond to elongated 
ellipsoidal shapes with an axis ratio close to 2:1, while in the A$\approx$130
and A$\approx$190 regions the  ratios 3:2 and 1.65:1 respectively 
appear closer to reality. 

Since the su$_q$(2) model has been found suitable for describing normal
deformed bands, it is plausible that it will also be successful in describing 
superdeformed bands as well. A test has been performed in \cite{BDRRS67}.
The su$_q$(2) model has
been found to give good results in all mass regions, the deformation 
parameter $\tau$ being smaller than in the case of normal deformed bands,
thus indicating smaller deviations from the su$_q$(2) symmetry. 
In particular, $\tau$ has been found to obtain values about 0.01 in the 
A$\approx$130 and A$\approx$190 regions, while it obtains even smaller 
values, around 0.004, in the A$\approx$150 region, which contains the best 
examples of superdeformed bands observed so far. These results should be 
compared to the values of $\tau$ around 0.03 obtained in the case of 
normal deformations. 

Concerning the corresponding B(E2) values, the experimental information 
is still quite poor for allowing a meaningful comparison of the su$_q$(2)
predictions to experiment. 

A new effect, called {\sl $\Delta I=2$ staggering}, has been recently 
observed in a few superdeformed bands 
\cite{Fli4299,Fli373,Ced3150,Sem3671,Kru2109}. If $\Delta I=2$ staggering 
is present, then, for example, the levels with $I=2$, 6, 10, 14, \dots 
are displaced relatively to the levels with $I=0$, 4, 8, 12, \dots, i.e. 
the level with angular momentum $I$ is displaced relatively to its neighbours 
with angular momentum $I\pm 2$. In most cases, though, the effect is smaller 
than the experimental error. $\Delta I=2$ staggering has also been seen in 
some normal deformed bands \cite{RJR19,TW2006,WT1821}. 
Although many theoretical attempts 
for the explanation of the effect have been made (see 
\cite{BDDKMMRR} for a relevant 
list), it seems plausible that what is seen in most cases is simply a series 
of backbendings \cite{RJR19,HS637,HL1789}. 
The same effect has been observed in rotational 
bands of diatomic molecules \cite{BDDLMRR}. In this case the effect is clearly 
larger than the experimental error and it is also clear that it is due 
to interband interactions \cite{BDDKMMRR}. Furthermore in diatomic molecules a
$\Delta I=1$ staggering effect has been observed \cite{RMD2759,MDRTB}, i.e. a 
relative 
displacement of the levels with even $I$ with respect to the levels with 
odd $I$. This effect is clearly larger than the experimental error. 
A description of the $\Delta I=1$ staggering effect in terms of 
two $q$-deformed rotators with slightly different parameter values has 
been attempted \cite{RMD2759}. 

\section{ The physical content of the su$_q$(2) model}

From the above it is clear that the su$_q$(2) model offers a way of describing 
nuclear stretching, i.e. the departure of deformed nuclei from the su(2)
symmetry of the rigid rotator, similar to the one of the VMI model. 
 The parameter $\tau$ describes this departure quantitatively, vanishing 
in the rigid rotator limit. Therefore the deformation parameter $\tau$ 
should not be confused with nuclear deformation; it is in fact related 
to nuclear softness, as already discussed in sec. 20. 

On the other hand, the increase of the moment of inertia with increasing $J$
in the framework of the VMI model means that collectivity gets increased
\cite{BM75}.
The su$_q$(2) model is an alternative way for describing this increase 
in collectivity. But increased collectivity implies increased B(E2)
transitions. Therefore it is not surprising that the su$_q$(2) model 
predicts B(E2) values increasing with $J$. 

Given the su$_q$(2) generators $J_+$, $J_-$, $J_0$, it is instructive to
define as usual the operators $J_x$, $J_y$, $J_z$ by
$$    J_+= J_x +i J_y, \qquad J_-=J_x-iJ_y, \qquad J_0=J_z.\eqno(23.1)$$
The su$_q$(2) commutation relations can then be rewritten in the form 
$$ [J_x, J_y ] = {i\over 2} [2 J_z], \qquad [J_y, J_z]=i J_x, \qquad
[J_z, J_x] = i J_y,\eqno(23.2)$$
which is a generalization of the so(3) commutation relations, obtained 
in the limit $q \rightarrow 1$. We remark that while in the classical 
so(3) case the three commutation relations have exactly the same form, 
in the quantum case the first commutation relation differs (in the right 
hand side) from the other two, thus indicating that in the framework of 
the problem under study the z-direction is not any more equivalent 
to the x- and y- directions. This is of course a phenomenological way 
to describe the softness of deformed nuclei by adding the appropriate 
perturbations to the pure su(2) Hamiltonian and has nothing to do with
the isotropy of space, as implied in \cite{BNW375}. 

It is worth remarking at this point that the Casimir operator of su$_q$(2)
is also invariant under the usual su(2) \cite{CCZB163}. 
Therefore the quantum number $J$ characterizing the irreps of su$_q$(2), 
and as a result the nuclear levels, is exactly the same as the quantum 
number $J$ used in the case of the usual su(2). Therefore there is no
reason for a ``total reformulation of quantum mechanics'', as implied 
in \cite{BNW375}, the su$_q$(2) generators being connected
to their su(2) counterparts by the $q$-deforming functionals 
of sec. 14.   
In other words, one continues to believe in usual angular momentum theory 
and usual quantum mechanics. All what is done in the framework of the 
su$_q$(2) model is to add to the usual su(2) Hamiltonian  
several perturbations which have a special form making them 
suitable to be summed up, including the original su(2) term, 
into the form of the su$_q$(2) Hamiltonian.  

\section{The u$_{p,q}$(2) rotator model}

An extension of the su$_q$(2) model is the u$_{p,q}$(2) model
\cite{Kib9358,BMK13,BMK385,Kib9521}, which is based
on a two-parameter deformed algebra (see sec. 15 for a list of references). 
For $p=q$ (using the definition of $(p,q)$-numbers of eq. (2.11)) 
this model reduces to the su$_q$(2) one. 
This model has been successfully applied to superdeformed nuclear bands 
\cite{BMK13,BMK385}. When Taylor expanded, it becomes clear that the 
eigenvalues of the  Hamiltonian of this model (which is the second order 
Casimir operator of u$_{p,q}$(2)) contain 
terms of the form $J(J(J+1))^n$, in addition to the $(J(J+1))^n$ ones. 
It is therefore closer to the modification of su$_q$(2) which will be 
discussed in subsec. 26.3.  

\section{ Generalized deformed su(2) models}

Another formula giving very good results for rotational spectra has been 
introduced by Holmberg and Lipas \cite{HL552}, and rediscovered in
\cite{HWZ1617}. In this case the energy levels are given by 
$$ E(J)= a \left[\sqrt{1+b J(J+1)}-1\right] .\eqno(25.1)$$
Taylor expansion of the square root immediately shows that the present 
formula is a special case of eq. (19.6). This formula can be derived from the
collective model of Bohr and Mottelson \cite{BM75}.
 
It has been argued in \cite{MWZ2545}  that the Hamiltonian of eq. (25.1)
gives better agreement to rotational nuclear spectra than the one 
coming from the su$_q$(2) symmetry. Using the techniques described in 
detail in sec. 17 one can construct a generalized deformed algebra
 su$_{\Phi}$(2), characterized by a function $\Phi(J(J+1)$,  giving the 
spectrum 
of eq. (25.1) exactly. In this particular case the algebra 
 is characterized by the structure function 
$$\Phi(J(J+1)) = a \left[\sqrt{1+b J(J+1)}-1\right].\eqno(25.2)$$
It is of interest to check if this choice of structure function also
improves the agreement between theory and experiment in the case of 
the electromagnetic transition probabilities connecting these energy 
levels. In order to study this problem, one has to construct the relevant
generalized Clebsch-Gordan coefficients. This problem is still open. 

\section{Quantum algebraic description of vibrational and transitional 
nuclear spectra}

We have already seen  that the su$_q$(2) model describes successfully 
deformed and superdeformed bands. 
It is not surprising that the applicability of the su$_q$(2) formalism
is limited to the rotational region (where the ratio $R_4= E(4)/E(2)$
obtains values between 3.0 and 3.33), since it is based on a deformation
of the rotation algebra. For describing nuclear spectra in the vibrational
($2.0 \leq R_4 \leq 2.4$) and transitional ($2.4 \leq R_4 \leq 3.0$)
regions it is clear that an extension of the model is needed. In order to
be guided towards such an extension, we briefly review the existing 
experience of other successful models.

\subsection{The Interacting Boson Model}

 In the rotational (su(3)) limit \cite{AI201} of the
 Interacting Boson Model (IBM) (see secs 27, 31 for more details) 
the spectrum is described by a $J(J+1)$ expression, while 
in the vibrational (u(5)) \cite{AI253} and transitional (o(6))
 \cite{AI468} limits 
expressions of the form $J(J+c)$  with $c>1$ appear. In the u(5) limit,
in particular, the energy levels are given by
$$E(N, n_d, v, n_{\Delta}, J, M_J )=  E_0 + \epsilon n_d +\alpha n_d (n_d+4)
+\beta 2v (v+3) +\gamma 2J(J+1), \eqno(26.1)$$
where $N$ is the total number of bosons, $n_d$ is the number of d-bosons, 
$v$ is the seniority, $n_{\Delta}$ is the ``missing'' quantum number
in the reduction from o(5) to o(3), $M_J$ is the third component of the 
angular momentum $J$, while $E_0$, $\epsilon$, $\alpha$, $\beta$, 
$\gamma$ are free parameters. The ground state band, in particular, 
is characterized by quantum numbers $n_d=0$, 1, 2, \dots, $v=n_d$, 
$n_{\Delta}=0$, $J=2 n_d$, so that the energy expression for it reads
$$E(J)= E_0+ {\epsilon\over 2} J + {\alpha\over 4} J(J+8) +
{\beta \over 2} J(J+6) + 2 \gamma  J(J+1). \eqno(26.2)$$
In the o(6) limit the energy is given by
$$E(N,\sigma,\tau,\nu_{\Delta},J,M_J)= E_0+\beta 2\tau(\tau+3) +
\gamma 2J(J+1) + \eta 2\sigma(\sigma+4), \eqno(26.3)$$
where $\sigma$ is the quantum number characterizing the irreducible
representations (irreps) of o(6), $\tau$ is the quantum number 
characterizing the irreps of o(5), $\nu_{\Delta}$ is the missing
quantum number in the reduction from o(5) to o(3), while $E_0$, $\beta$,
$\gamma$, $\eta$ are free parameters. The ground state band is 
characterized by the quantum numbers $\sigma=N$, $\tau=0$, 1, 2, \dots, 
$\nu_{\Delta}=0$, $J=2\tau$, so that the relevant energy expression 
takes the form
$$E(J)= E_0 + {\beta\over 2} J(J+6) +\gamma 2J(J+1) +\eta 2N(N+4).
\eqno(26.4)$$ 

The message from eqs (26.2) and (26.4) is that nuclear anharmonicities 
are described by expressions in which $J$ and $J^2$ appear with 
different coefficients, and not with the same coefficient as in $J(J+1)$. 
The earliest introduction of this idea is in fact the Ejiri formula 
\cite{Eji1189}
$$ E(J) = k J(J+1) + a J , \eqno(26.5)$$
which has been subsequently justified microscopically  in \cite{DDK632}.

\subsection{Generalized VMI}

 The two-parameter VMI model is known to continue giving good fits in the 
transitional and even in the vibrational region. In these regions, 
however, the accuracy of the model is substantially improved by adding
a third parameter, which essentially allows for treating $J$ and $J^2$
with a different coefficient \cite{BK27,BK1879,BS1014}.

The usual VMI model has been briefly reviewed in subsec. 20.1. 
One of the (essentially equivalent) three-parameter extensions of the 
model, which give improved fits of vibrational and transitional 
spectra, is the generalized VMI (GVMI) model \cite{BK27,BK1879}
in which the energy  levels are described by
$$ E(J) = {J+x J(J-2)\over \Phi(J)} +{1\over 2} k (\Phi(J)-\Phi_0)^2, 
\eqno(26.6)$$
which can be easily rewritten in the form
$$E(J) = {J(J+x')\over 2\Phi'(J)} +{1\over 2} k' (\Phi'(J)-\Phi'_0)^2,
\eqno(26.7)$$
where $x'=x^{-1} -2$. It is clear that for $x=1/3$ the GVMI reduces 
to the usual VMI, while for transitional and vibrational nuclei $x$
obtains lower values, so that $x'$ becomes greater than 1. 
 The variational condition determining the moment of 
inertia still has the form of eq. (20.4), while the expansion of the 
energy turns out to be 
$$ E(J)= {1\over 2\Phi'_0} (J(J+x') -{\sigma'\over 2} (J(J+x'))^2
+ (\sigma')^2 (J(J+x'))^3 -3 (\sigma')^3 (J(J+x'))^4 +\dots), 
\eqno(26.8)$$
where 
$$\sigma' = {1\over 2 k' (\Phi'_0)^3}.\eqno(26.9)$$
We remark that an expansion in terms of $J(J+x')$ is obtained, as compared
to an expansion in terms of $J(J+1)$ in the case of the usual VMI. 
The physical content of the parameters is clear: the centrifugal 
stretching effect is accounted for by the softness parameter $\sigma'$, 
as in the case of the usual VMI, while anharmonicities, important in the 
vibrational region, are introduced by $x' >1$. Since centrifugal 
stretching and anharmonicities are two effects of different origins, it is 
reasonable to describe them by two different parameters. 

\subsection{Modification of the su$_q$(2) model}

The evidence coming from the IBM and the generalized VMI model described
above, suggests a model in which the spectrum is given by 
$$ E(J)= {1\over 2I} [J]_q [J+c]_q , \eqno(26.10)$$
which contains 3 parameters: the moment of inertia $I$, the deformation 
parameter $q$ and the new parameter $c$, which is expected to be 1 in 
the rotational limit and larger than 1 in the vibrational and 
transitional  regions. This energy expression can be expanded as 
$$E(J) = {1\over 2I} {1\over (j_0(\tau))^2} ( j_0(\tau) J(J+c)
-\tau j_1(\tau) (J(J+c))^2 $$ $$+ {2\over 3} \tau^2 j_2(\tau) (J(J+c))^3
 -{1\over 3} \tau^3 j_3(\tau) (J(J+c))^4 + {2\over 15} \tau^4
j_4(\tau) (J(J+c))^5 -\dots ), \eqno(26.11)$$
which is  similar to eq. (19.5) with $J(J+1)$ replaced by $J(J+c)$.  

It is expected that the deformation parameter $\tau$, which plays the 
role of the small parameter in the expansion, as the softness parameter
does in the case of the VMI, will describe the centrifugal stretching 
effect, while the parameter $c$ will correspond to the anharmonicity 
effects. These expectations are corroborated from  fits of 
the experimental data reported in \cite{BDFRR497}.
The following comments can be made:  

i) The anharmonicity parameter $c$
is clearly decreasing with increasing $R_4$, i.e. with increasing 
collectivity. It obtains high values (8-18) in the vibrational region, while 
in the rotational region it stays close to 1. (It should be noted that 
by fixing $c=1$ in the rotational region the fits are only very slightly
changed, as expected.) In the transitional region its values are close to
3. 
 
ii) The deformation parameter $\tau$, which corresponds to the centrifugal
stretching, is known from the su$_q$(2) model to obtain values close 
to 0.3-0.4 in the rotational region, a fact also seen here. The same 
range of values appears in the vibrational region as well, while in the 
transitional region $\tau$ reaches values as high as 0.6. It is not 
unreasonable for this parameter, which is connected to the softness of 
the nucleus, to obtain its highest values in the region of $\gamma$-soft
nuclei. 

iii) It is worth remarking that eq. (26.10) coincides for 
$q=1$ and $c$=integer=N  with 
the eigenvalues of the Casimir operator of the algebra
o(N+2) in completely symmetric states \cite{DGI873,DGI123}.
 In the rotational region
the fits gave N=1, which corresponds to o(3), as expected, 
while in the transitional region the fits gave approximately N=3, which
corresponds to o(5), which is a
subalgebra contained in both the u(5) and o(6) limits of the IBM. 
A unified description of 927 low lying states of 271 nuclei from five 
major shells along these lines has been given in \cite{DGGRR1853}.  

iv) It is also worth remarking that a special case of the expression of eq. 
(26.10) occurs in the q-deformed version of the o(6) limit of the Interacting 
Boson Model, which will be reported below (see eq. (31.7)).  

v) The su$_q$(2) symmetry is known to make specific predictions for 
the deviation of the behaviour of the B(E2) values from the rigid 
rotator model (subsec. 21.3). It will be interesting to connect the spectrum 
of eq. (26.10) to some deformed symmetry, at least for special values of $c$,
and examine the implications of such a symmetry for the B(E2) values.
Such a study in the framework of the q-deformed version of the o(6) 
limit of IBM, mentioned in iv), is also of interest. 

vi) It is worth noticing that an expansion in terms of $J(J+c)$ can also be
obtained from a generalized oscillator (sec. 12)
 with a structure function
$$ F(J) = [ J(J+c)]_Q, \eqno(26.12)$$
where $[x]_Q$ stands for the $Q$-numbers introduced in sec. 6
and $Q=e^T$, with $T$ real. This is similar to an oscillator 
successfully used
for the description of vibrational spectra of diatomic molecules
(see sec. 37). 
It can also be considered as a deformation of the oscillator corresponding
to the Morse potential (see sec. 37). 

We have therefore introduced an extension of the su$_q$(2) model
of rotational nuclear spectra, which is applicable in the vibrational
and transitional regions as well. This extension is in agreement with
the Interacting Boson Model and the Generalized Variable Moment of 
Inertia model.  
In addition to the overall scale
parameter, the model contains two parameters, one related to the 
centrifugal stretching and another related to nuclear anharmonicities. 
In the rotational region the model coincides with the usual su$_q$(2)
model, while in the transitional region an approximate o(5) 
symmetry is seen. These results give additional motivation in pursueing
the construction of a deformed version of the Interacting Boson Model.
This problem will be discussed in the next section as well as in sec. 31. 

\section{ A toy Interacting Boson Model with su$_q$(3) symmetry}

The Interacting Boson Model (IBM) \cite{IA1987} is a very popular algebraic 
model of nuclear structure. 
In the simplest version of IBM  low lying
collective nuclear spectra are described in terms of $s$ ($J=0$)
and $d$ ($J=2$) bosons, which are supposed to be correlated
fermion pairs.
 The symmetry of the simplest version
of the model is u(6), which contains u(5) (vibrational), su(3)
(rotational) and o(6) ($\gamma$-unstable) chains of subalgebras (see also 
sec. 31).
A simplified version of the model, having the su(3) symmetry
with su(2) and so(3) chains of subalgebras also exists \cite{BSR719}.
It can be considered as a toy model for two-dimensional nuclei, but
it is very useful in demonstrating the basic techniques used in the full IBM.

In the present section we will construct the $q$-deformed version of 
this toy model. Since this project requires the construction of a realization 
of su$_q$(3) in terms of $q$-deformed bosons, we will also use this 
opportunity to study su$_q$(3) in some detail.  

\subsection{The su$_q$(3) algebra}

In the classical version of the toy IBM \cite{BSR719}
 one introduces
bosons with angular momentum $m = 0, \pm 2$, represented by
the creation (annihilation) operators $a^\dagger_0$, $a^\dagger_+$, 
$a^\dagger_-$
($a_0$, $a_+$, $a_-$). They satisfy usual boson commutation
relations
$$[a_i, a_j^\dagger]=\delta_{ij}, \quad [a_i, a_j]=[a_i^\dagger, 
a_j^\dagger]=0.\eqno(27.1)$$
The 9 bilinear operators
$$\Lambda_{ij} = a^\dagger_i a_j \eqno(27.2)$$
satisfy then the commutation relations
$$[\Lambda_{ij}, \Lambda_{kl}] = \delta_{jk}\Lambda_{il}
-\delta_{il} \Lambda_{kj}, \eqno(27.3)$$
which are the standard u(3) commutation relations.
The total number of bosons
$$N=\Sigma_i \Lambda_{ii} = {a^\dagger_0} {a_0} + {a^\dagger_+} {a_+} 
+ {a^\dagger_-} {a_-}
\eqno(27.4)$$
is kept constant. Since we are dealing with a system of bosons, only the 
totally symmetric irreps $\{N,0,0\}$ of u(3) occur. 

In the quantum case one has the u$_q$(3) commutation relations given in Table 
2 \cite{STK437}, where $A_{ij}$ are the generators of u$_q$(3)
and the $q$-commutator is defined as
$$[A, B]_q = AB -qBA.\eqno(27.5)$$

\begin{table}[bth]
\begin{center}
\caption{ u$_q$(3) commutation relations $^{210}$, 
given in the form 
$[A,B]_a =C$. $A$ is given in the first column, $B$ in the 
first row. 
$C$ is given at the intersection of the row containing $A$ with the 
column containing $B$. $a$, when different from 1, follows $C$, enclosed 
in parentheses.}
\bigskip

\begin{tabular}{|c| c c c c c c |}
\hline
  & $A_{11}$ & $A_{22}$ & $A_{33}$ & $A_{12}$ & $A_{23}$ &
    $A_{13}$  \\
\hline
$A_{11}$ &0& 0 & 0 & $A_{12}$ & 0 & $A_{13}$ \\
$A_{22}$ &0&0&0& $-A_{12}$ & $A_{23}$ & 0\\
$A_{33}$ &0&0&0&0& $-A_{23}$ & $-A_{13}$  \\
$A_{12}$ & $-A_{12}$ & $A_{12}$ &0&0& $A_{13}$ ($q$)& 0 ($q^{-1}$)
 \\
$A_{23}$ & 0 & $-A_{23}$ & $A_{23}$ & $-q^{-1} A_{13} (q^{-1})$ &
0 & 0 ($q$)  \\
$A_{13}$ & $-A_{13}$ & 0& $A_{13}$ & 0($q$) & 0 ($q^{-1}$) & 0
 \\
$A_{21}$ & $A_{21}$ & $-A_{21}$ & 0& $-[A_{11}-A_{22}]$ & 0 &
$A_{23}q^{A_{11}-A_{22}}$  \\
$A_{32}$ & 0 & $A_{32}$ & $-A_{32}$ & 0 & $-[A_{22}-A_{33}]$ &
$-q^{-A_{22}+A_{33}} A_{12}$ \\
$A_{31}$ & $A_{31}$ & 0& $-A_{31}$ & $-q^{-A_{11}+A_{22}}A_{32}$ &
$-A_{21} q^{A_{22}-A_{33}}$ & $-[A_{11}-A_{33}]$
\\
\hline
\end{tabular}

\vskip 0.2truein

\begin{tabular}{| c | c c c | }
\hline
  &
    $A_{21}$ & $A_{32}$ & $A_{31}$  \\
\hline
$A_{11}$ & $-A_{21}$ &0& $-A_{31}$\\
$A_{22}$ & $A_{21}$ & $-A_{32}$&0\\
$A_{33}$ & 0& $A_{32}$ & $A_{31}$ \\
$A_{12}$ &
$[A_{11}-A_{22}]$ & 0 & $-q^{-A_{11}+A_{22}} A_{32}$ \\
$A_{23}$ &
 0 & $[A_{22}-A_{33}]$ & $A_{21} q^{A_{22}-A_{33}}$ \\
$A_{13}$ &
$-A_{23} q^{A_{11}-A_{22}}$ & $q^{-A_{22}+A_{33}} A_{12}$ &
$[A_{11} -A_{33}]$ \\
$A_{21}$ &
 0 & $-q A_{31} (q)$ & 0 ($q^{-1}$) \\
$A_{32}$ &
 $A_{31} (q^{-1})$ & 0 & 0 ($q$) \\
$A_{31}$ &
 0 ($q$) &
0 ($q^{-1}$) & 0\\
\hline
\end{tabular}
\end{center}
\end{table}

In order to obtain a realization of u$_q$(3) in terms of
the $q$-bosons described in sec. 10,
one starts with  \cite{Kul195}
$$A_{12}= a_1^\dagger a_2, \quad A_{21}= a_2^\dagger a_1,\quad
A_{23}= a_2^\dagger a_3 , \quad A_{32}= a_3^\dagger a_2 .\eqno(27.6)$$
One can easily verify that the u$_q$(3) commutation relations
involving these generators are satisfied. For example, one
has
$$[A_{12}, A_{21}]=  [N_1-N_2], \qquad
  [A_{23}, A_{32}]= [N_2 -N_3], \eqno(27.7) $$
using the identity of eq. (2.6) and the identifications
$$N_1 = A_{11}, \quad N_2= A_{22}, \quad N_3= A_{33}.\eqno(27.8) $$
One can now determine the boson realizations of $A_{13}$
and $A_{31}$ from other commutation relations, as follows
$$ A_{13}= [A_{12}, A_{23}]_q = a_1^\dagger a_3 q^{-N_2},\eqno(27.9)$$
$$ A_{31}= [A_{32}, A_{21}]_{q^{-1}}= a_3^\dagger a_1 q^{N_2}.\eqno(27.10) $$
Using eq. (2.6) once more one can verify that
the relation
$$[A_{13}, A_{31}]= [N_1 -N_3] \eqno(27.11)$$
is fulfilled by the boson images of (27.9), (27.10). It is by now
a straightforward task to verify that all commutation relations
of Table 2 are fulfilled by the boson images obtained above.

Before turning to the study of the two limits of the model, we give for 
completeness some additional information on su$_q$(3):

i) The irreps of su$_q$(3) have been studied in 
\cite{STK437,ST375,MSK595,MSK1031,PC4017,Que5977,Cap5942,Yu399,Yu5881,YY3025}.

ii) Clebsch-Gordan coefficients for su$_q$(3) have been given 
in \cite{STK437,Yu5881,SK263,SK690}.

iii) The Casimir operators of su$_q$(3) and their eigenvalues have been 
given explicitly in \cite{RP2020}. Using the Elliott quantum 
numbers \cite{Ell128,Ell562,Ell557}
$$\lambda =f_1-f_2, \qquad \mu=f_2,\eqno(27.12)$$
where $f_i$ represents the number of boxes in the $i$-th line of the 
corresponding Young diagram, the irreps of su$_q$(3) are labelled as 
$(\lambda, \mu)$. The eigenvalues of the second order Casimir operator
then read
$$ C_2= \left[{\lambda\over 3}-{\mu\over 3}\right]^2+\left[{2\lambda\over 3}
+{\mu\over 3}+1\right]^2+\left[{\lambda\over 3}+{2\mu\over 3}+1\right]^2-2,
\eqno(27.13)$$
while the eigenvalues of the third order Casimir operator are
$$C_3=2\left[{\lambda\over 3}-{\mu \over 3}\right] \left[{2\lambda \over 3}
+{\mu\over 3}+1\right] \left[ {\lambda\over 3}+{2\mu\over 3}+1\right].
\eqno(27.14)$$
In the limit $q\to 1$ these reproduce the ordinary results for su(3):
$$C_2={2\over 3} (\lambda^2+\mu^2+\lambda \mu + 3\lambda+3\mu),\eqno(27.15)$$
$$C_3={2\over 27} (\lambda-\mu) (2\lambda+\mu+3)(\lambda+2\mu+3).\eqno(27.16)$$

iv) Irrducible tensor operators under u$_q$(3) have been considered in
\cite{SK360}. 

v) A different deformation of sl(3) has been studied in \cite{BDF74}.

vi) Coupling (Wigner--Clebsch--Gordan) coefficients for the 
algebras u$_q$(n) have been considered in \cite{AS5925}. 

\subsection{The su$_q$(2) limit}

We shall study the su$_q$(2) limit of the model first, since it is 
technically less demanding. 

So far we have managed to write a boson realization of u$_q$(3)
in terms of 3 $q$-bosons, namely $a_1$, $a_2$, $a_3$. Omitting
the generators involving one of the bosons,
one is left with an su$_q$(2) subalgebra. Omitting the generators
involving $a_3$, for example, one is left with $A_{12}$,
$A_{21}$, $N_1$, $N_2$, which satisfy usual su$_q$(2)
commutation relations if the identifications
$$ J_+= A_{12}, \quad J_-=A_{21}, \quad J_0={1\over 2} (N_1-N_2)
\eqno(27.17)$$
are made. $J_0$ alone forms an so$_q$(2) subalgebra. Therefore
the relevant chain of subalgebras is
$$ {\rm su}_q(3)\supset {\rm su}_q(2) \supset {\rm so}_q(2) .\eqno(27.18)$$

The second order Casimir operator of su$_q$(2) has been given in eq. (14.9).
Substituting the above expressions for the generators one finds
$$C_2({\rm su}_q(2))= \left[{N_1+N_2\over 2}\right] \left[{N_1+N_2\over 2}
+1\right].\eqno(27.19)$$

All of the above equations go to their classical counterparts by
allowing $q\rightarrow 1$, for which $[x]\rightarrow x$, i.e.
$q$-numbers become usual numbers. In the classical case \cite{BSR719} out 
of the three bosons ($a_0$, $a_+$, $a_-$) forming su(3), one chooses
to leave out $a_0$, the boson with zero angular momentum,
 in order to be left with the su(2) subalgebra
formed by $a_+$ and $a_-$, the two bosons of angular momentum two.
The choice of the su$_q$(2) subalgebra made above  is then consistent
with the following correspondence between classical bosons and
$q$-bosons
$$ a_+\rightarrow a_1, \quad a_-\rightarrow a_2, \quad
a_0\rightarrow a_3.\eqno(27.20)$$
(We have opted in using different indices for usual bosons and
$q$-bosons  in order to avoid confusion.)

In the classical case  the states of the system are characterized by
the quantum numbers characterizing the irreducible representations
(irreps) of the algebras appearing in the classical
counterpart of the chain of eq. (27.18). For su(3) the total number
of bosons $N$ is used. For su(2) and so(2) one can use the
eigenvalues of $J^2$ and $J_0$, or, equivalently, the eigenvalues
of $a^\dagger_+ a_+ + a_-^\dagger a_-$ and $L_3= 4 J_0$, for which we use the
symbols $n_d$ (the number of bosons with angular momentum 2)
and $M$. Then the basis in the classical case can be written as \cite{BSR719}
$$|N,n_d,M> = {(a_0^\dagger)^{N-n_d}\over (N-n_d)!}
{(a^\dagger_+)^{n_d/2+M/4}\over (n_d/2+M/4)!}
{(a_-^\dagger)^{n_d/2-M/4}\over (n_d/2-M/4)!} |0>.\eqno(27.21) $$
In the quantum case for each oscillator one defines the basis as
in sec. 10. Then  the full basis in the $q$-deformed case is
$$ |N, n_d, M>_q = {(a_3^\dagger)^{N-n_d} \over [N-n_d]!}
{(a_1^\dagger)^{n_d/2+M/4}\over [n_d/2+M/4]!}
{(a_2^\dagger)^{n_d/2-M/4}\over [n_d/2-M/4]!} |0>,\eqno(27.22) $$
where $N=N_1 +N_2 +N_3$ is the total number of bosons,
$n_d = N_1 + N_2$ is the number of bosons with angular momentum 2,
and $M$ is the eigenvalue of $L=4J_0$.
$n_d$ takes values from 0 up to $N$, while for a given value
of $n_d$, $M$ takes the values $\pm 2n_d$,
$\pm 2(n_d-2)$, \dots, $\pm 2$ or 0, depending on whether $n_d$
is odd or even.
In this basis the eigenvalues of the second order Casimir
operator of su$_q$(2) are then
$$ C_2({\rm su}_q(2)) |N, n_d, M>_q = \left[{n_d\over 2} \right] 
\left[{n_d\over 2}+1\right]
|N, n_d, M>_q .\eqno (27.23)$$
In the case of $N=5$ one can easily see that the spectrum will
be composed by the ground state band, consisting of states
with $M=0$, 2, 4, 6, 8, 10 and $n_d=M/2$, the first excited
band with states characterized by $M=0$, 2, 4, 6 and
$n_d=M/2+2$, and the second excited band, containing states
with $M=0$, 2 and $n_d=M/2+4$.

In the case that the Hamiltonian has the su$_q$(2)
dynamical symmetry, it can be written in terms of the Casimir
operators of the chain (27.18). Then one has
$$ H = E_0 + A C_2({\rm su}_q(2)) + B C_2({\rm so}_q(2)),\eqno(27.24)$$
where $E_0$, $A$, $B$ are constants.
Its eigenvalues are
$$ E= E_0 + A \left[{n_d \over 2}\right] \left[{n_d\over 2}+1\right] + 
B M^2.\eqno(27.25)$$

Realistic nuclear spectra are characterized by strong
electric qua\-dru\-po\-le transitions among the levels of the
same band, as well as by interband transitions. In the
framework of the present toy model one can define, by analogy
to the classical case \cite{BSR719},
 quadrupole transition operators
$$Q_+= a_1^\dagger a_3 + a_3^\dagger a_2, \qquad
  Q_-= a_2^\dagger a_3 + a_3^\dagger a_1.\eqno(27.26)$$
In order to calculate  transition matrix elements of these
operators one only needs eqs. (10.7), (10.8), i.e. the action of the
$q$-boson operators on the $q$-deformed basis. The selection
rules, as in the classical case, are $\Delta M=\pm 2$,
$\Delta n_d=\pm 1$, while the corresponding matrix elements
are
$$_q<N, n_d+1, M\pm 2| Q_{\pm}| N,n_d,M>_q =
\sqrt{[N-n_d] [{n_d\over 2}\pm{M\over 4} +1]}, \eqno(27.27)$$
$$_q<N, n_d-1, M\pm 2| Q_{\pm} | N, n_d,M>_q =
\sqrt{[N-n_d+1][{n_d\over 2}\mp {M\over 4}]}.\eqno(27.28)$$
From these equations it is clear that both intraband and interband
transitions are possible.

In order to get a feeling of the qualitative changes in the
spectrum and the transition matrix elements resulting from the
$q$-deformation of the model, a simple calculation for
a system of 20 bosons ($N=20$) has been performed in \cite{BDFRR267}.
Two cases are distinguished:
i) $q$ real  ($q=e^{\tau}$, with $\tau$ real),
ii) $q$ a phase factor ($q=e^{i\tau}$, with $\tau$ real). The main conclusions 
are: 

i) When $q$ is real the spectrum is
increasing more rapidly than in the classical case, while when
$q$ is a phase the spectrum increases more slowly than in the
classical case, in agreement with the findings of the
$q$-rotator model.

ii) The transition matrix elements in the case that
$q$ is real increase more rapidly than in the classical case,
while they increase less rapidly than in the classical case when
$q$ is a phase.

iii) Transition matrix elements
are much more sensitive to $q$-de\-fo\-rma\-tion than energy spectra.
This is an interesting feature, showing that $q$-deformed
algebraic models can be much more flexible in the description of
transition probabilities than their classical counterparts.

\subsection { The so$_q$(3) limit}

The classical su(3) toy model has, in addition to the above mentioned su(2)
chain of subalgebras, an so(3) chain. However, the problem of constructing
 the su$_q$(3)$\supset$so$_q$(3) decomposition is a very di\-f\-fi\-cult one. 
Since this decomposition is needed in constructing the $q$-deformed versions
of several collective models, including the Elliott model 
\cite{Ell128,Ell562,Ell557,Har67},  the su(3) limit of the IBM 
\cite{IA1987,AI201,Bon88}, the Fermion Dynamical Symmetry Model (FDSM)
\cite{WFCCG313,WFG227}, the Interacting Vector Boson Model (IVBM) 
\cite{GRR1377,GRR521,GRR147,Ray471}, 
the nuclear vibron model for clustering \cite{DI73},
as well as the su(3) limit of the vibron model \cite{IL3046} for 
diatomic molecules,  
we report here the state of the art in this problem:

i) As far as the completely symmetric irreps of su$_q$(3) are concerned,
the problem has been solved by Van der Jeugt 
\cite{VdJ213,VdJ131,VdJ1799,VdJ947,VdJ948}. This suffices for our
needs in the framework of the toy IBM, since only completely symmetric 
su$_q$(3) irreps occur in it. 

ii) Sciarrino \cite{Sci92} started from so$_q$(3) and obtained a deformed gl(3)
containing so$_q$(3) as a subalgebra. However, it was not clear how to 
impose the Hopf structure on this larger algebra. Trying the other way 
around, he found that by starting from a gl$_q$(3) algebra, which already 
possesses the Hopf structure, one loses the Hopf structure of the 
principal 3-dim subalgebra, which should have been so$_q$(3). 

iii) Pan \cite{Pan257} and Del Sol Mesa {\it et al.} \cite{Del1147} attacked 
the problem through the use of $q$-deforming functionals of secs 10, 14. 

iv) Quesne \cite{Que81} started with $q$-bosonic operators transforming as 
vectors under so$_q$(3) and constructed a $q$-deformed u(3) by 
tensor coupling. 

v) $q$-deformed subalgebras of several $q$-deformed algebras have recently
been studied by Sciarrino \cite{Sci94}. 

vi) A simplified version of the so$_q$(3) subalgebra of u$_q$(3) has been 
constructed in \cite{BRRT95}. Furthermore, explicit
expressions for the irreducible vector and quadrupole tensor operators under
so$_q$(3) have been given and the matrix elements of the latter have been 
calculated. Further details on these developments will be given in sections 
28 and 29.   

In what follows, it suffices to use the solution given in \cite{VdJ213},
since in the model under study only completely symmetric irreps
of u$_q$(3) enter. Using the notation
$$ a_+\rightarrow a_1, \quad a_-\rightarrow a_2, \quad
a_0 \rightarrow a_3, \eqno(27.29)$$
the  basis states are of the form 
$$ | n_+ n_0 n_- > =
{ (a_+^\dagger)^{n_+} (a_0^\dagger)^{n_0} (a_-^\dagger)^{n_-} \over
\sqrt{[n_+]! [n_0]! [n_-]!}} |0>, \eqno(27.30)$$
with $a_i |0>=0$ and
$N_i|n_+ n_0 n_-> = n_i |n_+n_0n_->$, where $i=+,0,-$.
The principal subalgebra so$_q$(3) is  then generated by
$$L_0= N_+-N_-, \eqno(27.31)$$
$$L_+= q^{N_--{1\over 2} N_0}  \sqrt{q^{N_+} + q^{-N_+} } \quad
a_+^\dagger a_0 \quad+\quad a_0^\dagger a_- \quad q^{N_+-{1\over 2}N_0}
\sqrt{q^{N_-}+q^{-N_-}}, \eqno(27.32)$$
$$L_-= a_0^\dagger a_+ q^{N_--{1\over 2} N_0}\sqrt{q^{N_+}+q^{-N_+}}
+ q^{N_+-{1\over 2} N_0} \sqrt{q^{N_-}+q^{-N_-}} a_-^\dagger a_0,
\eqno(27.33)$$
satisfying the commutation relations
$$ [L_0, L_{\pm}] =\pm L_{\pm}, \quad  [L_+, L_-]= [2L_0].
\eqno(27.34)$$
$L_0$ alone generates then the so$_q$(2) subalgebra.
Therefore the relevant chain of subalgebras is
$$ {\rm su}_q(3)\supset {\rm so}_q(3) \supset {\rm so}_q(2) .\eqno(27.35)$$
The so$_q$(3) basis vectors can be written in terms of the vectors
of eq. (27.30) as 
$$ |v(N,L,M)>= q^{-[(L+M)(L+M-1)]/4}
 \sqrt{{[N+L]!![2L+1] [L+M]! [L-M]!\over [N-L]!! [N+L+1]!}} $$
$$\sum_x q^{(2L-1)x/2} \quad s^{(N-L)/2} \quad
{|x, L+M-2x, x-M>\over
\sqrt{[2x]!![L+M-2x]![2x-2M]!!}}, \eqno(27.36)$$
where
$$s= (a_0^\dagger)^2 q^{N_++N_-+1} -\sqrt{{[2N_+][2N_-]\over [N_+]
[N_-]}} a_+^\dagger a_-^\dagger q^{-N_0-{1\over 2}}, \eqno(27.37)$$
$x$ takes values from max(0,$M$) to $[(L+M)/2]$ in steps of 1,
$L = N,$ $N-2,$ $\ldots,$ 1 or 0, $M=-L,$ $-L+1,$ $\ldots,$ $ +L$, and
$[2x]!!=[2x] [2x-2] \ldots [2]$. The action of the generators of
so$_q$(3) on these states  is given by
$$L_0 |v(N,L,M)> = M \quad |v(N,L,M)>, \eqno(27.38)$$
$$L_{\pm}\quad |v(N,L,M)>=
\sqrt{[L\mp M]  [L\pm M+1]} \quad |v(N,L,M\pm 1)>.\eqno(27.39)$$
The second order Casimir operator of so$_q$(3)
has the form
$$C_2({\rm so}_q(3))=  L^2 = L_- L_+ + [L_0] [L_0+1].\eqno(27.40)$$
Its eigenvalues in the above basis are given by
$$C_2({\rm so}_q(3))\quad |v(N,L,M)> = [L] [L+1]\quad |v(N,L,M)>.\eqno(27.41)$$

All of the above equations go to their classical counterparts by
allowing $q\rightarrow 1$, for which $[x]\rightarrow x$, i.e.
$q$-numbers become usual numbers. In the classical case \cite{BSR719}
 the states of the system are characterized by
the quantum numbers characterizing the irreducible representations
(irreps) of the algebras appearing in the classical
counterpart of the chain of eq. (27.35). For su(3) the total number
of bosons $N$ is used. For so(3) and so(2) one can use $L$ and
$M$, respectively. In \cite{BSR719}, however, the eigenvalue of
 $L_0' =2 L_0$ is used, which is $M' =2M$.

Since the rules for the decomposition of the totally symmetric
u$_q$(3) irreps into so$_q$(3) irreps are the same as in the classical
case, it is easy to verify that for a system with
 $N=6$  the spectrum will
be composed by the ground state band, consisting of states
with $M'=0$, 2, 4, 6, 8, 10, 12 and $L=N$, the first excited
band with states characterized by $M'=0$, 2, 4, 6, 8 and
$L=N-2$, the second excited band, containing states
with $M=0$, 2, 4 and $L=N-4$, and the third  excited band,
containing a state with $M'=0$ and $L=N-6$.

In the case that the Hamiltonian has the so$_q$(3)
dynamical symmetry, it can be written in terms of the Casimir
operators of the chain (27.35). Then one has
$$ H = E_0 + A C_2({\rm so}_q(3)) + B C_2({\rm so}_q(2)),\eqno(27.42)$$
where $E_0$, $A$, $B$ are constants.
Its eigenvalues are
$$ E= E_0 + A [L] [L+1] + B M^2.\eqno(27.43)$$
It is then clear that in this simple model the internal structure
of the rotational bands is not influenced by $q$-deformation.
What is changed is the position of the bandheads.

We turn now to the study of electromagnetic transitions. In the
present limit one can define, by analogy
to the classical case,  quadrupole transition operators $Q_{\pm}$
proportional to the so$_q$(3) generators $L_{\pm}$
$$Q_{\pm} = L_{\pm}.\eqno(27.44)$$
In order to calculate  transition matrix elements of these
operators one only needs eq. (27.39).
 The selection
rules, as in the classical case, are $\Delta M'=\pm 2$,
$\Delta L =0$, i.e. only intraband transitions are allowed.
The relevant matrix elements are
$$ <v(N,L,M'+2)| Q_+ | v(N,L,M')> =
\sqrt{\left[L-{M'\over 2}\right] \left[L+{M'\over 2}+1\right]}, \eqno(27.45)$$
$$ <v(N,L,M'-2)| Q_- | v(N,L, M')> =
\sqrt{\left[L+{M'\over 2}\right] \left[L-{M'\over 2}+1\right]}. \eqno(27.46)$$

In order to get a feeling of the qualitative changes in the
spectrum and the transition matrix elements resulting from the
$q$-deformation of this limit of the model,  a simple calculation
has been done in \cite{BonNii}. Again the cases of $q$ being real 
or $q$ being a phase factor have been considered. The main conclusions are:

i) $q$-deformation influences only the
position of bandheads, while it leaves the internal structure of
the bands intact.

ii)  When $q$ is real the bandheads are
increasing more rapidly than in the classical case, while when
$q$ is a phase the bandheads increase more slowly than in the
classical case. This result is in qualitative agreement with the
findings of the su$_q$(2) model.

iii) Transition matrix elements in the case that
$q$ is real have values higher than in the classical case,
while they have values lower than in the classical case when
$q$ is a phase.

\section{The $3$-dimensional $q$-deformed harmonic oscillator
and the nuclear shell model}

Having developed in the previous section several techniques related 
to the u$_q$(3) algebra and its so$_q$(3) subalgebra, it is instructive 
to see how a $q$-deformed version of the 3-dimensional harmonic oscillator 
possessing this symmetry can be built and how this oscillator can be used 
for reproducing the predictions of the Modified Harmonic Oscillator,
first suggested by Nilsson \cite{Nilsson29} and studied in detail in 
\cite{GLNN613,BR14,NR95}. The construction of the Hamiltonian of the 
3-dimensional $q$-deformed harmonic oscillator is a non-trivial problem,
since one has to construct first the square of the $q$-deformed angular 
momentum operator. 

\subsection{
Simplified so$_q$(3) subalgebra and so$_q$(3) basis states}

The so$_q$(3) subalgebra of u$_q$(3) has already been mentioned in 
subsec. 27.3. Here we shall derive a simplified form of it. 

We shall use three independent $q$-deformed boson
operators $b_i$ and $b_i^\dagger$ ($i=+,0,-$),
which satisfy the commutation relations
$$ [N_i,b_i^\dagger ] = b_i^\dagger, \qquad
[N_i,b_i] = -b_i, \qquad
b_i b_i^\dagger -  q^{\pm1} b_i^\dagger b_i = q^{\mp N_i}, \eqno(28.1)$$
where $N_i$ are the corresponding number operators.
(These bosons are the same as the bosons $a$ and $a^\dagger$, with
$i=+, 0, -$, used in the previous section.) 

The elements of the so$_q$(3)
algebra (i.e. the angular momentum operators) acting in the Fock
space of the totally symmetric representations [N,0,0] of u$_q$(3)
(which is in fact 
the space of the 3-dimensional $q$-deformed harmonic oscillator)
have been derived by Van der Jeugt 
\cite{VdJ213,VdJ131,VdJ1799,VdJ947,VdJ948}. 
As it has already been mentioned in the previous section,
a simplified form of the subalgebra ${\rm so}_q(3)\subset {\rm u}_q(3)$,
can be obtained by introducing the operators \cite{BRRT95}
$$ B_0 = q^{-\frac{1}{2}N_0}b_0,  \qquad
 B_0^\dagger = b_0^\dagger q^{-\frac{1}{2}N_0}, \qquad
B_i = q^{N_i+\frac{1}{2}} b_i \sqrt{\frac{[2 N_i]}{[N_i]}},  \qquad
 B_i^\dagger = \sqrt{\frac{[2 N_i]}{[N_i]}} b_i^\dagger q^{N_i+\frac{1}{2}},
\qquad i=+,-, \eqno(28.2) $$
where the $q$-numbers $[x]$ are defined as in eq. (2.1). 
We shall consider only real values for the deformation parameter
	$q$ (i.e. $q=e^\tau$ with $\tau$ being real).  
These new operators satisfy the usual commutation relations
$$ [N_i,B^{\dagger}_i] = B_i^{\dagger}, \qquad
[N_i,B_i] = -B_i ,\eqno(28.3) $$
while in  the Fock space, spanned on the normalized
	eigenvectors of the excitation number operators $N_{+}, N_{0}, N_{-}$,
	they satisfy the relations
$$ B^{\dagger}_{0} B_{0} = q^{- N_{0}+1}[N_{0}], \qquad
 B_{0} B^{\dagger}_{0} = q^{-N_{0}}[N_{0}+1], \eqno(28.4)$$
$$B_i^\dagger B_i = q^{2N_i-1}  [2N_i],  \qquad
 B_i B_i^\dagger =q^{2N_i+1} [2N_i+2], \qquad i=+,-, \eqno(28.5)$$
	and hence, the commutation relations
$$ [B_{0},B^{\dagger}_{0}]= q^{-2N_{0}}, \qquad
[B_i,B_i^\dagger] = [2] q^{4 N_i+1}, \qquad i=+,-. \eqno(28.6)$$
	It was shown in \cite{BRRT95} that the angular momentum
	operators defined in \cite{VdJ213}
and reported in eqs (27.31)-(27.33), when expressed
	in terms of the modified operators of eq. (28.2) ,
	take the simplified form
$$ L_{0} = N_{+}-N_{-}, \qquad
L_{+} = q^{-L_{0}+{1\over 2}}B^{\dagger}_{+} B_{0} + q^{L_{0}-{1\over 2}}
B^{\dagger}_{0} B_{-}, \qquad
L_{-} = q^{-L_{0}-{1\over 2}}B^{\dagger}_{0} B_{+} + q^{L_{0}+{1\over 2}}
B^{\dagger}_{-} B_{0}, \eqno(28.7)$$
	and satisfy the standard $so_q(3)$ commutation relations
$$ [L_0,L_{\pm}] = \pm L_{\pm},  \qquad [L_+,L_-] = [2L_0]. \eqno(28.8)$$
As it has been discussed in detail in \cite{VdJ213} the algebra
of eq. (28.7)  is not a subalgebra of u$_q$(3) when these
algebras are considered as
Hopf algebras, but it is a subalgebra when these algebras are 
considered 
as $q$-deformed enveloping algebras. However,
in our context, working in the Fock space (constructed from three
independent $q$-bosons), we restrict ourselves
to the symmetric representation of u$_q$(3), where the embedding
${\rm so}_q(3)\subset {\rm u}_q(3)$ is valid.
It should also be noted that the operators $L_+$, $L_-$, $L_0$ 
given by eq. (28.7) are not invariant with respect to the replacement 
$q\leftrightarrow q^{-1}$, a fact that restricts us to real $q$. 

	The Casimir operator of so$_q$(3) can be written in the form 
\cite{STK959,STK1746,STK2863,STK690}
$$ C_2^{(q)} = \frac{1}{2}\left\{L_{+}L_{-} + L_{-}L_{+} + [2][L_0]^2\right\}
= L_{-}L_{+} + [L_0][L_0 + 1] = L_{+}L_{-} + [L_0][L_0 - 1] . \eqno(28.9)$$
	One can also define $q$-deformed
	states $|n \ell m\rangle_q$ satisfying the eigenvalue
	equations
$$ {}^q{\bf L}^2 |n \ell m \rangle_q = [\ell][\ell+1] |n \ell m\rangle_q, 
\qquad 
L_0 |n \ell m \rangle_q = m |n \ell m\rangle_q, \qquad
N |n \ell m \rangle_q = n |n \ell m\rangle_q. \eqno(28.10)$$ 
	Here $^q{\bf L}^2 \stackrel{def}{=} C^{(q)}_2$ is the ``square'' of the
	$q$-deformed angular momentum and $ N = N_+ + N_0 + N_-$ is the
	total number operator for the $q$-deformed bosons. These
	states have the form (see \cite{BRRT95} for a derivation) 
$$   |n \ell m\rangle_q = q^{\frac{1}{4}(n-\ell)(n+\ell+1)-\frac{1}{2}m^2}
\sqrt{\frac{[n-\ell]!![\ell+m]![\ell-m]![2\ell+1]}{[n+\ell+1]!!}} \times $$
$$ \times
\sum_{t=0}^{(n-\ell)/2} \sum_{p=\max(0,m)}^{\lfloor(\ell+m)/2\rfloor}
\frac{(-1)^t q^{-(n+\ell+1)t}}{[2t]!![n-\ell-2t]!!}
\frac{(B^\dagger_{+})^{p+t}}{[2p]!!}
\frac{(B^\dagger_{0})^{n+m-2p-2t}}{[n+m-2p]!}
\frac{(B^\dagger_{-})^{p+t-m}}{[2p-2m]!!}
|0\rangle, \eqno(28.11)$$
	where $|0\rangle$ is the vacum state, $[n]!=[n][n-1]\ldots [1]$,~
	$[n]!!=[n][n-2]\ldots [2]\;\mbox{or}\;[1]$,
	and, as shown in \cite{VdJ213,VdJ1799}, form a basis for
	the most symmetric representation $[n,0,0]$ of u$_q$(3),
	corrresponding to the ${\rm u}_q(3) \supset {\rm so}_q(3)$ chain.

\subsection{Definitions of so$_q$(3) 
irreducible tensor operators, tensor and scalar products}

We shall recall now some definitions about
$q$-deformed tensor operators within the framework of the algebra
so$_q$(3). 

An {\sl irreducible tensor} operator of rank $j$
with parameter $q$ according to the algebra so$_q$(3) is a set of
$2j+1$ operators ${\cal T}^{(q)}_{j m}$, satisfying the relations
$$ [L_0,{\cal T}^{(q)}_{j m}] = m\,{\cal T}^{(q)}_{j m}, \qquad
 [L_{\pm},{\cal T}^{(q)}_{j m}]_{q^m} q^{L_0} =
\sqrt{[j \mp m][j \pm m + 1]}\,{\cal T}^{(q)}_{j m\pm1}, \eqno(28.12)$$
where, in order to express the adjoint action of the generators
$L_0,L_{\pm}$ of so$_q$(3) on the components of the tensor operator
${\cal T}^{(q)}_{j m}$, we  use the usual notation of the
{\sl $q$-commutator}
$$ [A,B]_{q^\alpha} = A B - q^{\alpha} B A  .\eqno(28.13)$$
	By $\widetilde{\cal T}^{(q)}_{j m}$ we denote
	the {\sl conjugate irreducible $q$-tensor operator}
$$ \widetilde{\cal T}^{(q)}_{j m} = (-1)^{j-m}q^{-m}{\cal T}^{(q)}_{j, -m} \ ,
\eqno(28.14) $$
	which satisfies the relations
$$ [\widetilde{\cal T}^{(q)}_{j m},L_{0}] = m\;\widetilde{\cal T}^{(q)}_{j m},
\qquad 
 q^{L_0}[\widetilde{\cal T}^{(q)}_{j m},L_{\pm}]_{q^m} =
\sqrt{[j\pm m][j\mp m+1]}\;\widetilde{\cal T}^{(q)}_{j, m\mp 1} \ .
\eqno(28.15)$$
	Then the operator
$$ {\cal P}^{(q)}_{j m} = (\widetilde{\cal T}^{(q)}_{j m})^\dagger =
(-1)^{j-m} q^{-m} ({\cal T}^{(q)}_{j, -m})^\dagger, \eqno(28.16)$$
where ${}^\dagger$ denotes hermitian conjugation, transforms
in the same way (given in eq. (28.12)) as the tensor ${\cal T}^{(q)}_{j m}$,
i.e.  ${\cal P}^{(q)}_{j m}$ also is an irreducible $so_q(3)$
tensor operator of rank $j$.

Let ${\cal A}^{(q_1)}_{j_1 m_1}$ and ${\cal B}^{(q_2)}_{j_2 m_2}$
be two irreducible tensor operators.
We shall define the {\sl tensor} and
{\sl scalar product} of these tensor operators following some of the
	prescriptions, summarized, for example, in 
\cite{STK959,STK1746,STK2863,STK690}.  

For the {\sl tensor product} one can
	introduce the following operator
$$ \left[{\cal A}^{(q_1)}_{j_1}
\times{\cal B}^{(q_2)}_{j_2}\right]^{(q_3)}_{j m}
= \sum_{m_1,m_2} {}^{q_3} C_{j_1 m_1, j_2 m_2}^{j m}
{\cal A}^{(q_1)}_{j_1 m_1} {\cal B}^{(q_2)}_{j_2 m_2}, \eqno(28.17)$$
where ${}^{q_3} C_{j_1 m_1, j_2 m_2}^{j m}$ are the Clebsch-Gordan
coefficients corresponding to the deformation parameter $q_3$. In
general the deformation parameters $q_1, q_2$ and $ q_3$ can be
arbitrary. It turns out, however, if one imposes the condition
that the left hand side of eq. (28.17) transforms as an
irreducible $q$-tensor of rank $j$
in a way that the Wigner-Eckart theorem can be
applied to eq. (28.17)  as a whole, not all of the combinations of
$q_1, q_2$ and $q_3$ are allowed.

	If the tensors ${\cal A}^{(q_1)}_{j_1}$ and ${\cal B}^{(q_2)}_{j_2}$
	depend on one and the same variable and act on a single vector,
	which depends on the same variable, the mentioned requirement will
	be satisfied only if $q_1 = q_2 = q$ and $q_3=1/q$, i.e. the operator
$$ \left[{\cal A}^{(q)}_{j_1}\times{\cal B}^{(q}_{j_2} \right]^{(1/q)}_{j m}
= \sum_{m_1,m_2} {}^{1/q}\!C_{j_1 m_1, j_2 m_2}^{j m}
{\cal A}^{(q)}_{j_1 m_1} {\cal B}^{(q)}_{j_2 m_2} \eqno(28.18)$$
	transforms as an irreducible $q$-tensor of rank $j$ according to the
	algebra $so_q(3)$ . Then, the definition of eq. (28.18)  is
	in agreement with the property
$$ \langle\alpha',\ell'\| [{\cal A}^{(q)}_{j_1} \times
{\cal B}^{(q)}_{j_2} ]^{(1/q)}_{j} \|\alpha,\ell\rangle =$$
$$= \sqrt{[2j+1]}\sum_{\alpha'',\ell''} (-1)^{\ell+j+\ell'}
\left\{\begin{array}{ccc}\ell & j & \ell' \\ j_1 & \ell'' & j_2 \end{array}
\right\}_{q}
\langle\alpha',\ell'\|{\cal A}^{(q)}_{j_1}
\|\alpha'',\ell''\rangle
\langle\alpha'',\ell''\|{\cal B}^{(q)}_{j_2}
\|\alpha,\ell\rangle, \eqno(28.19)$$
	which is a $q$-analogue of the well known classical identity
for the matrix elements of the tensor product of two tensor operators 
\cite{Edm57,VMK88}. 

In what follows  we shall consider also the scalar product of
	two tensor operators depending on two different variables $(1)$ and
	$(2)$ and acting on different vectors, depending on these different
	variables.  In this particular case the {\sl scalar product} of
	irreducible $q$ and $q^{-1}$ -- tensor operators
	${\cal A}^{(q)}_{j}(1),~{\cal B}^{(1/q)}_{j}(2)$ with the same rank,
	acting on {\it different vectors} $(1)$ and $(2)$,
is given by means of the following definition
$$({\cal A}^{(q)}_{j}(1)\cdot{\cal B}^{(1/q)}_{j}(2))^q
= (-1)^j\sqrt{[2j+1]}
\left[{\cal A}^{(q)}_{j}(1)\times{\cal B}^{(1/q)}_{j}(2)\right]^{(q)}_{0 0}
\nonumber\\
= \sum_{m} (-q)^m {\cal A}^{(q)}_{j m}(1)~ {\cal B}^{(1/q)}_{j -m}(2) \ .
\eqno(28.20)$$

\subsection{Construction of so$_q$(3) vector operators
and spherical vector operators} 
	
It should be noticed that the 
simlified angular momentum operators of eq. (28.7)
	satisfy the commutation relations of eq. (28.8), but {\it are not}
	tensor operators in the sense of definition of eq. (28.12).
	One can form, however, the angular momentum operators
$${\cal L}^{(q)}_{\pm1} = \mp\frac{1}{\sqrt{[2]}} q^{-L_0} L_{\pm}, $$
$$ {\cal L}^{(q)}_{0} = \frac{1}{[2]}\left\{q[2L_0] +
(q-q^{-1})L_{-}L_{+}\right\}=  \frac{1}{[2]}\left\{q[2L_0] +
(q-q^{-1})\left( C_2^{(q)} - [L_0][L_0+1] \right) \right\}, \eqno(28.21)$$
	which are tensors of first rank, i.e. so$_q$(3) vectors.

One can easily check that the operators $B^\dagger_0$, $B^\dagger_{\pm}$
and $B_0$, $B_{\pm}$ do not form a spherical vector
(see for example \cite{STK959,STK1746,STK2863,STK690}). 
The situation is the same as in the ``standard'' theory of angular momentum.
Indeed, in the ``standard'' theory, the operators $L_0 = L_3, \,
	L_{\pm}=L_1~\pm~iL_2$, do not form a spherical tensor. 
However, in the ``standard'' theory one constructs the operators
$J_{\pm} = \mp \textstyle\frac{1}{\sqrt{2}}(L_1 \pm i L_2), J_0=L_0$,
which are indeed the components of a spherical tensor of first rank according
	the standard so(3) algebra.

Here we proceed in an analogous way. 
	As shown in \cite{BRRT95}, one can define
	a vector operator $T^\dagger_m$ of the form
$$ T^{\dagger}_{+1} = {1\over \sqrt{[2]}} B^{\dagger}_{+}
q^{-2N_{+} + N - {1\over 2}}, \qquad
T^{\dagger}_{0} = B^{\dagger}_{0}  q^{-2N_{+} + N}, $$
$$ T^{\dagger}_{-1} = {1\over \sqrt{[2]}}\left\{ B^{\dagger}_{-}
q^{2 N_{+} - N - {1\over 2}} - (q - q^{-1}) B_{+} {(B^{\dagger}_{0})}^2
q^{-2 N_{+} + N + {3\over 2}}\right\}. \eqno(28.22)$$
	The corresponding expressions for the conjugate operators
	$\widetilde{T}_m = (-1)^m q^{-m} (T^\dagger_{-m})^\dagger$ are
$$ \widetilde{T}_{+1} = -{1\over \sqrt{[2]}}\left\{
q^{2 N_{+} - N - {3\over 2}} B_{-} - (q-q^{-1}) q^{-2N_{+} + N+ {1\over 2}}
B^{\dagger}_{+} {(B_{0})}^2 \right\}, $$
$$ \widetilde{T}_{0} = q^{-2N_{+} + N} B_{0}, \qquad 
\widetilde{T}_{-1} = -{1\over \sqrt{[2]}}q^{-2N_{+}+N+{1\over 2}} B_{+} .
\eqno(28.23)$$ 
	One can easily check, that the vector operators ${\bf T}^{\dagger}$ and
	$\widetilde{\bf T}$ satisfy the commutation relation
$$ [\widetilde{T}_{-1},T^\dagger_{+1}]_{q^{-2}} = -q^{2N+1}, \qquad
[\widetilde{T}_0,T^\dagger_0] = q^{2N} + q^{-1}(q^2-q^{-2})
T^\dagger_{+1}\widetilde{T}_{-1}, $$
$$[\widetilde{T}_{+1},T^\dagger_{-1}]_{q^{-2}} = -q^{2N-1} + q^{-1}(q^2-q^{-2})
\!\left\{T^\dagger_0\widetilde{T}_0 +
(q-q^{-1}) T^\dagger_{+1}\widetilde{T}_{-1}
\right\}, \eqno(28.24)$$
	and
$$ [\widetilde{T}_0,T^\dagger_{+1}] = 0, \qquad
[\widetilde{T}_{+1},T^\dagger_{+1}]_{q^2} = 0, \qquad
[\widetilde{T}_{+1},T^\dagger_0] = (q^2-q^{-2})
T^\dagger_{+1}\widetilde{T}_0, $$
$$ [\widetilde{T}_{-1},T^\dagger_0] = 0, \qquad
[\widetilde{T}_{-1},T^\dagger_{-1}]_{q^2} = 0,  \qquad
[\widetilde{T}_0,T^\dagger_{-1}] = (q^2-q^{-2})
T^\dagger_0\widetilde{T}_{-1}.\eqno(28.25)$$
Unlike the operators of eq. (28.2), the commutators
$[\widetilde T_m, \widetilde T_n]$ and $ [T^{\dagger}_m,T^{\dagger}_n]$
do not vanish, but are equal to
$$ [T^\dagger_{+1},T^\dagger_0]_{q^2} = 0, \qquad
[T^\dagger_0,T^\dagger_{-1}]_{q^2} = 0, \qquad
[T^\dagger_{+1},T^\dagger_{-1}] = (q-q^{-1}) (T^\dagger_0)^2 , $$
$$ [\widetilde{T}_0,\widetilde{T}_{-1}]_{q^2} = 0, \qquad
[\widetilde{T}_{+1},\widetilde{T}_0]_{q^2} = 0, \qquad
[\widetilde{T}_{+1},\widetilde{T}_{-1}] = (q-q^{-1}) (\widetilde{T}_0)^2, 
\eqno(28.26)$$
	which is in agreement with the results obtained in \cite{Que81}.

\subsection{A choice for the physical angular momentum} 

	It should be noted that the angular momentum operator ${\cal
	L}^{(q)}_M$ (considered as a vector according $so_q(3)$) can be
	represented in the form
$${\cal L}^{(q)}_M = - \sqrt{\frac{[4]}{[2]}} \left[ T^{\dagger} \times
\widetilde{T} \right]_{1M}^{(1/q)} = - \sqrt{\frac{[4]}{[2]}} \sum_{m,n}
{}^{1/q} C_{1m,1n}^{1M} \,T^{\dagger}_m \widetilde T_n.\eqno(28.27)$$
	Its ``square'' differs from $C^{(q)}_2$ and equals to 
\cite{STK959,STK1746,STK2863,STK690}
$$ ({\cal L}^{(q)})^2 \equiv {\cal L}^{(q)}\cdot {\cal L}^{(q)} =
\sum (-q)^{-M} {\cal L}^{(q)}_M{\cal L}^{(q)}_{-M}
=\frac{2}{[2]}C^{(q)}_2 + \left(\frac{q - q^{-1}}{[2]}\right)^2
(C^{(q)}_2)^2.\eqno(28.28)$$
	This difference, however, is not essential. Let us consider, for
	instance, the  expectation values of the scalar operator
of eq. (28.28).  It has the form
$${}_q\langle\ell m|{\cal L}^{(q)}\cdot{\cal L}^{(q)}|\ell m\rangle_q =
\frac{[2\ell][2\ell+2]}{[2]^2}=[\ell]_{q^2} [\ell+1]_{q^2}.\eqno(28.29)$$
	In this sense the replacement of ${}^q {\bf L}^2 \equiv C^{(q)}_2$ with
	${\cal L}^{(q)}\cdot {\cal L}^{(q)}$ is equivalent to the replacement
	$q \to q^2$ and leads to the renormalization of some constant.  For
	this reason we shall accept as the square of the {\it physical}
	angular momentum the quantity $^q{\bf L}^2\equiv C^{(q)}_2$, whose
	eigenvalues are $[\ell][\ell+1]$.

\subsection{A choice for the Hamiltonian} 

	Now let us consider the scalar operator, constructed in terms of
	$T^{\dagger}_M$ and $\widetilde T_M$. We have
$$ X_{0}^{(q)}
= -\sqrt{[3]}[T^\dagger\times \widetilde{T}]^{(1/q)}_{00} 
= -q^{-1}T^\dagger_{+1}\widetilde{T}_{-1}
+ T^\dagger_{0}\widetilde{T}_{0}
- q T^\dagger_{-1}\widetilde{T}_{1}.\eqno(28.30)$$
	We define the Hamiltonian of the three dimensional
	$q$-deformed oscillator (3-dim $q$-HO) as
$${}^qH_{3dim} = \hbar \omega_0 X_0 = \hbar \omega_0 \left(
-q^{-1}T^\dagger_{+1}\widetilde{T}_{-1} +
T^\dagger_{0}\widetilde{T}_{0} - q T^\dagger_{-1}\widetilde{T}_{1} \right).
\eqno(28.31)$$
	The motivations for such an ansatz are:

a) The operator so defined is an so$_q$(3) scalar, i.e. it is
	simultaneously
	measurable with the physical $q$-deformed angular momentum square
	$^q {\bf L}^2$
	and the z-projection $L_0$.

b) Only this so$_q$(3) scalar has the property of being the sum
	of terms conserving the number of bosons. (Each term contains
	an equal number of creation and annihilation operators.) 

c) In the limit $q \to 1$ the Hamiltonian of eq. (28.31)  goes into
$$ \lim_{q\to 1} {}^q H_{3dim} =
\hbar \omega_0 \left(a^{\dagger}_{+1} a_{+1} +
a^{\dagger}_{0} a_{0} + a^{\dagger}_{-1} a_{-1} \right), \eqno(28.32)$$ 
	where $[a_{m},a^{\dagger}_{n}] =  \delta_{m n}$, i.e. in this limit
the Hamiltonian of eq. (28.31)
 coincides with the Hamiltonian of the 3-dimensional
	(spherically symmetric) harmonic oscillator up to an additive constant.

What we have done so far can be summarized in the following way. We have shown 
that in the enveloping algebra of so$_q$(3) there exists an so$_q$(3) scalar
operator, built out of irreducible vector operators according to the 
reduction u$_q$(3)$\supset$so$_q$(3), for the case of the most symmetric 
irreducible representations of u$_q$(3), which are the space of the 3-dim 
$q$-HO. In the limit $q\to 1$ this operator coincides with the 
Hamiltonian of the 3-dim harmonic oscillator. It is therefore reasonable 
to consider this operator as the Hamiltonian of the 3-dim $q$-HO.  
(In the next subsections it will be shown that this Hamiltonian is 
related to the Hamiltonian of the Modified Harmonic Oscillator of 
Nilsson \cite{Nilsson29,GLNN613,BR14,NR95}.) 

It should be underlined that the Hamiltonian of eq. (28.31) does not
commute with the square of the ``classical'' (or ``standard'') angular
momentum ${\bf L}^2$. This is due to the fact that the $q$-deformed
oscillator is really space deformed and the ``standard'' quantum
number $l$ is not a good quantum number for this system. On the other
hand ${}^q {\bf L}^2$ commutes with ${}^qH_{3 dim}$ and the
``quantum angular momentum'' $\ell$ can be used for the
classification of the states of the $q$-oscillator. The
projection of the standard angular momentum on the $z$ axis, $l_z$,
coincides however with the quantum projection, $\ell_z$, i.e. it is a good
	quantum number.

	Eq. (28.31) can be cast in a simpler and
	physically more transparent form. Indeed, making use of
	the third component of the $q$-deformed angular
	momentum (considered as an $so_q(3)$ vector)
$$ {\cal L}_{0}^{(q)}
= -\sqrt{[4]/[2]}\left[T^\dagger\times\widetilde{T}\right]_{1 0}^{(1/q)}
= -T^\dagger_{+1}\widetilde{T}_{-1}
+ (q-q^{-1}) T^\dagger_{0}\widetilde{T}_{0}
+ T^\dagger_{-1}\widetilde{T}_{0},  \eqno(28.33)$$
	we obtain
$$ X_{0}^{(q)} + q {\cal L}_{0}^{(q)}
= -[2] T^\dagger_{+1}\widetilde{T}_{-1}
+ q^2 T^\dagger_{0}\widetilde{T}_{0}.\eqno(28.34)$$
	Since
$$ T^\dagger_{+1}\widetilde{T}_{-1} = \frac{[2N_{+}]}{[2]} q^{-2N_{+}+2N+1} ,
\qquad
T^\dagger_{0}\widetilde{T}_{0} = [N_{0}] q^{-4N_{+}-N_{0}+2N-1}, \eqno(28.35)$$
	we get upon summation
$$ X_{0}^{(q)} = -q {\cal L}_{1 0}^{(q)} + [N+L_{0}] q^{N-L_{0}+1}, 
\eqno(28.36)$$
	and, after some calculations,
$$ {}^qH_{3dim}= \hbar \omega_{0} X_{0}^{(q)} =
\hbar \omega_0\left\{ [N] q^{N+1}
- \frac{q(q-q^{-1})}{[2]} C_2^{(q)}\right\}.\eqno(28.37)$$
	The eigenvalues of such a $q$-deformed Hamiltonian are
$$  {}^qE_{3dim} = \hbar\omega_0 \left\{[n]q^{n+1}-\frac{q(q-q^{-1})}{[2]}
[\ell][\ell+1] \right\}, \qquad \ell = n,n-2,\ldots,0~\mbox{or}~1 .
\eqno(28.38)$$
	In the $q\to 1$ limit we have
	$\lim_{q\to1} {}^qE_{3dim}=\hbar \omega_0 n$,
	which coincides with the classical result.

One should note that the expression of eq. (28.38) can also be put in the form
$$ {}^qE_{3dim}= \hbar \omega_{0} \frac{q^{n+1}}{[2]}
\left\{q^{\ell+1}[n-\ell] + q^{-\ell-1} [n +\ell]\right\}, \eqno(28.39)$$ 
which perhaps looks more elegant.
However,  the form of eq. (28.38) has the advantage that the
	vibrational ($[n]$) and the rotational ($[\ell][\ell+1]$) degrees of
	freedom are clearly separated.

	For small values of the deformation parameter $\tau$, where
	$q=e^{\tau}$, one can expand the rhs of eq. (28.38)
	in powers of $\tau$ obtaining
$$ {}^qE_{3dim}=\hbar\omega_0\; n
-\hbar\omega_0 \tau \left(\ell(\ell+1) - n(n+1)\right)
-\hbar\omega_0 \tau^2 \left(\ell(\ell+1) - \frac{1}{3}n(n+1)(2n+1)\right)
+{\cal O}(\tau^3).\eqno(28.40)$$
To leading order in $\tau$ the expression of eq. (28.40)
closely resembles the one giving the energy eigenvalues
of the Modified Harmonic Oscillator, as we shall see in the next 
subsections. 

\subsection{The Modified Harmonic Oscillator of Nilsson} 

Before examining the connection between the Hamiltonian of the last 
subsection and the Modified Harmonic Oscillator (MHO) of Nilsson 
\cite{Nilsson29,GLNN613,BR14,NR95}, a few words about the MHO 
are in place. 

The basis of the shell model in finite nuclei is 
the assumption of an independent particle motion within a mean field.
The average field in which the particles move  is correctly
described by the Woods--Saxon (WS) potential $V_{WS}({\bf r})$.
However, the corresponding eigenvalue equation can be
solved only numerically. It is therefore desirable for many purposes
to have an exactly soluble model producing approximately the same 
spectrum as the Woods--Saxon potential. 
The  Modified Harmonic Oscillator (MHO) Hamiltonian, 
first suggested by Nilsson \cite{Nilsson29} and studied  in detail 
in \cite{GLNN613,BR14,NR95},  indeed  reproduces 
approximately
the energy spectrum of the WS potential and at the same time
has a simple analytical solution.
The Hamiltonian has the form
$$ V_{MHO} = \frac{1}{2}\hbar \omega \rho^2 -
\hbar\omega k \left\{\mu\left({\bf L}^2 -
\left\langle{\bf L}^2\right\rangle_N\right) + 2{\bf L\cdot S}\right\},
\quad \rho = \sqrt{\frac{M \omega}{\hbar}}\, r, \eqno(28.41)$$
	where ${\bf L}^2$ is the square of the angular momentum and
	${\bf L\cdot S}$ is the spin-orbit interaction.
The subtraction of the average value of
	$\langle{\bf L^2}\rangle$, taken over each $N$-shell, 
$$ \left\langle {\bf L}^2 \right\rangle_N = \frac{N(N+3)}{2}, \eqno(28.42)$$ 
results in avoiding the shell compression induced by the ${\bf L}^2$ term,
	leaving the ``center of gravity'' of each $N$-shell unchanged.
	The $V_{MHO}$ (without the spin-orbit term ${\bf L\cdot S}$) 
indeed reproduces effectively the Woods--Saxon radial potential.
	A better agreement with the experimental data can be achieved if the
	parameters of the potential are let to vary smoothly from shell to
	shell.

It should be noted, however, that from a mathematical point of view
the introduction of an ${\bf L^2}$-dependent term (and of the corresponding
correction $\langle{\bf L}^2\rangle_N$) in the MHO
potential is not so innocent, as it may look at first sight,
because this term depends on the state in which the particle occurs,
and it this sense the potential of eq. (28.41) is ``non-local'' and
``deformable'' (with variable deformation).  This effect is amplified
by the fact that the constant $\mu^{\prime}=k\mu$ is allowed to vary
 from shell to shell, i.e. depends on $N$. 

The eigenvalues of the MHO (if one neglects the spin--orbit term) are
$$E_{nl}=\hbar \omega n -\hbar \omega \mu' \left( l(l+1)-{1\over 2}
n(n+3)\right), \eqno(28.43)$$
where $\mu'$ is allowed to vary from shell to shell. 

\subsection{Connection between the 3-dimensional $q$-deformed harmonic
oscillator and the Modified Harmonic Oscillator of Nilsson} 

Comparing eqs (28.40) and (28.43) we see that eq. (28.40) closely resembles, 
to leading order in $\tau$, eq. (28.43), thus establishing a connection 
between the 
3-dimensional $q$-deformed harmonic oscillator and the Modified Harmonic 
Oscillator of Nilsson, with the parameter $\tau$ of the 3-dim $q$-HO playing
the role of the parameter $\mu^\prime =k \mu$ of the MHO. 
(Remember that the spin--orbit term has been omitted in both cases.) 

Quantitatively, one obtains a good fit of
the energy spectrum produced by the MHO potential of eq. (28.41) by
choosing for $\hbar\omega_{0}$ and $\tau$ of the 3-dimensional $q$-deformed  
harmonic oscillator the values
$\hbar \omega_{0} =0.94109$ and $\tau=0.021948$ \cite{RRIT241}.
(It has been assumed that $\hbar\omega=1$ for the MHO and
the spin--orbit term has been neglected.)
The value of $\tau$ is close to the values usually 
adopted for $\mu^{\prime}=k\mu$. Indeed the MHO fit
in the $^{208}_{82}$Pb region yields \cite{BR14,NR95}
$\mu^{\prime}=0$ for $N=2$, $\mu^{\prime}= 0.0263$ for $N=3$,
and $\mu^{\prime} = 0.024$ for $N \geq 4$, while
for $k$, fixed by the condition that the observed order of
sub-shells be reproduced, the relevant values are $k=0.08$ for $N=2$,
$k=0.075$ for $N=3$ and $k=0.07$ for $N \geq 4$.

It appears surprising at first that $\hbar \omega_0$
differs slightly from $1$. Indeed the correction term $n(n+1)$
in the 3-dim $q$-HO spectrum is slightly different from the
corresponding piece $n(n+3)/2$ in the MHO. Some
compression of the 3-dim $q$-HO spectrum is therefore inevitable.
From a physical point of view, one may say that the
mean radius of the deformed oscillator is slightly larger than
the radius of the classical isotropic HO.

When comparing the MHO and 3-dim $q$-HO spectra it
should be remembered that the constant $\mu'$ in the MHO potential takes 
different values for shells with different
values of $N$, while in the 3-dim $q$-HO model the parameters $\tau$ and
$\hbar\omega_0$ have  the same values for all shells.  The
comparison shows that the $q$-deformation of the 3-dimensional harmonic
oscillator effectively reproduces the non-locality and the
``deformations'' induced in the MHO model through the terms
${\bf L}^2$,~$\langle{\bf L}^2\rangle_N$
and the variability of $\mu^{\prime}$ with the shell number $N$.

\vfill\eject
\subsection{3-dimensional $q$-deformed harmonic oscillator
with $q$-deformed spin--orbit term}

For a full comparison with the MHO and a more realistic description
of the single-particle spectrum, one needs to include a $q$-deformed
spin--orbit term ${\cal L}^{(q)}\cdot{\cal S}^{(1/q)}$ in the
$q$-deformed harmonic oscillator Hamiltonian ${}^qH_{3dim}$. To this
purpose one introduces the spin operators $S_{+},S_{0},S_{-}$,
which are elements of another (independent) su$_q$(2) algebra. These
operators satisfy the commutation relations of eq. (28.8), i.e.
$$[S_{0},S_{\pm}] = \pm S_{\pm},  \qquad\qquad
[S_{+},S_{-}] = [2S_{0}], \eqno(28.44)$$
and act in the {\it two-dimensional} representation space of
this algebra. The orthonormalized basis vectors of this space will be
denoted by $|\frac{1}{2} m_s\rangle_q$, in analogy with the usual
(non-deformed) case. 

One can now define the $q$-deformed total angular momentum as 
\cite{STK959,STK1746,STK2863,STK690}
$$ J_{0} = L_{0} + S_{0}, \qquad
J_{\pm} = L_{\pm} q^{S_{0}} + S_{\pm} q^{-L_{0}},  \eqno(28.45)$$
	where the $q$-deformed orbital angular momentum is given by eq. (28.7).
	The operators of eq. (28.45) satisfy the same
	commutation relations as the operators of the $q$-deformed orbital
	angular momentum (eq. (28.8)) and spin (eq. (28.44)), and the
	corresponding expression for the Casimir operator of the algebra
of eq. (28.45)  can be written in the form
$$ C_{2,J}^{(q)} = J_{-}J_{+} + [J_0][J_0+1].\eqno(28.46)$$
	As in the case of the $q$-deformed orbital momentum we shall consider
	$^qJ^2\equiv C_{2,J}^{(q)}$ as the total angular momentum ``square''.

	The common eigenvectors of $C_{2,J}^{(q)}, C_{2,L}^{(q)}$
	and $C_{2,S}^{(q)}$ can be written in the usual form
$$ |n (\ell \case{1}{2}) j m\rangle_q = \sum_{m_{\ell},m_{s}}
{}^qC_{\ell m_\ell, \case{1}{2} m_s}^{j m}
|n \ell m_\ell\rangle_q |\case{1}{2} m_s\rangle_q, \eqno(28.47)$$
	where ${}^qC_{\ell m_\ell, s m_s}^{j m}$ are the
	$q$-deformed Clebsch-Gordan coefficients
\cite{STK959,STK1746,STK2863,STK690}. 

One can define the spin--orbit term as a scalar product, according
	to the definition of eq. (28.20)
$$ {\cal L}^{(q)}\cdot {\cal S}^{(1/q)}
= \sum_{M=0,\pm1} (-q)^{M} {\cal L}_{M}^{(q)}{\cal S}_{-M}^{(1/q)} =
 \frac{1}{[2]}\left\{ C_{2,J}^{(q)} - C_{2,L}^{(q)} - C_{2,S}^{(q)}
- \frac{(q-q^{-1})^2}{[2]}C_{2,L}^{(q)}C_{2,S}^{(q)} \right\}, \eqno(28.48)$$
where ${\cal S}^{(1/q)}$ is a vector operator according to the
algebra of eq. (28.44), constructed from the
$q$-deformed spin operators taking into account the rule
of eq. (28.21), but for deformation parameter $1/q$.
One can then suggest the following Hamiltonian for the 3-dim $q$-deformed
harmonic oscillator with spin--orbit interaction
$$  {}^qH = \hbar\omega_{0}
\left\{ X_0^{(q)} - \kappa [2]{\cal L}^{(q)}\cdot {\cal S}^{(1/q)}\right\} =$$
$$=  \hbar\omega_{0}\left\{
[N] q^{N+1} - \frac{q(q-q^{-1})}{[2]} C_{2,L}^{(q)}
-\kappa\left( C_{2,J}^{(q)} - C_{2,L}^{(q)} - C_{2,S}^{(q)}
- \frac{(q-q^{-1})^2}{[2]}C_{2,L}^{(q)}C_{2,S}^{(q)} \right)\right\}.
\eqno(28.49)$$
In eq. (28.49) the various factors have been chosen in accordance with
the usual convention used in the classical (spherically symmetric) shell model
with spin--orbit coupling.
The eigenvectors of this Hamiltonian are given by eq. (28.47), 
while the corresponding eigenvalues are
$$ {}^qE_{n \ell j} = \hbar\omega_{0}\left\{
[n] q^{n+1} - \frac{q(q-q^{-1})}{[2]} [\ell][\ell+1] \right. $$
$$  \left.   -\kappa\left( [j][j+1] - [\ell][\ell+1] -
{\textstyle [\frac{1}{2}][\frac{3}{2}]}
- \frac{(q-q^{-1})^2}{[2]} [\ell][\ell+1]
{\textstyle [\frac{1}{2}][\frac{3}{2}]} \right)\right\}.\eqno(28.50)$$

In order to compare the expression for the $q$-deformed $L$--$S$ interaction
(the term proportional to $\kappa$) 
with the classical results one needs  the expansion of the
expectation value of the term given in eq. (28.48) in powers of $\tau$
($q=e^\tau$, with $\tau$ real)
$$ {}_q\langle n(\ell \case{1}{2})jm|{\cal L}^{(q)}\cdot {\cal S}^{(1/q)}
|n(\ell \case{1}{2})jm \rangle_q = \cases{ \frac{1}{2} \ell
\left(1+\frac{\tau^2}{6}(4 \ell^2- 7) + \ldots \right) 
& if $j=\ell+1/2$  \cr
 -\frac{1}{2}(\ell+1)\left(1+\frac{\tau^2}{6}(4 \ell^2 + 8\ell -3)
+ \ldots \right) &  if $j=\ell-1/2$  \cr } \eqno(28.51)$$
It can be easily seen that the expression of eq. (28.48) introduces
some corrections to the classical expressions for the spin--orbit
interaction, which are proportional to $\tau^2$. These corrections
are small for light nuclei, where the values of $\ell$ are small,
but they are non-negligible for heavier nuclei,
where shells with higher values of $\ell$ are important.

One can easily see that the level scheme produced by the 3-dim $q$-HO
with spin--orbit interaction reproduces very well the level scheme 
of the MHO with spin--orbit interaction \cite{RRIT241}. 

\section{Further development of u$_q$(3)$\supset$so$_q$(3) models}

The u$_q$(3)$\supset$so$_q$(3) mathematical framework developed in the 
last section for the case of the 3-dimensional $q$-deformed harmonic 
oscillator is of more general interest, since one can use it for studying
$q$-deformed versions of various models having the u(3)$\supset$so(3) 
symmetry, such as the Elliott model \cite{Ell128,Ell562,Ell557,Har67}
and the su(3) limit of the Interacting Boson Model. Since the quadrupole 
degree of freedom plays a central role in nuclear collectivity, one needs 
to be able to calculate matrix elements of the quadrupole operator in 
the u$_q$(3)$\supset$so$_q$(3) basis. For the quadrupole operator itself
two choices appear:

a) The quadrupole operator can be a $q$-deformed second rank tensor operator,
with its matrix elements being calculated in the $q$-deformed basis.
In this case one assumes that observables are described by $q$-deformed 
operators and physical states are described by $q$-deformed wavefunctions.

b) The quadrupole operator can be the usual (non-deformed) operator, with its 
matrix elements being calculated in the $q$-deformed basis. In this (less 
extreme) case one assumes that observables are described by the usual 
physical operators, while the states are described by $q$-deformed 
wavefunctions, corresponding to highly correlated eigenfunctions 
of the undeformed Hamiltonian. 

In what follows both cases will be examined. 

\subsection{Matrix elements of a $q$-deformed quadrupole operator
in the u$_q$(3)$\supset$so$_q$(3) basis}

We are going to compute the reduced matrix elements of an 
so$_q$(3) quadrupole operator (second rank tensor operator under so$_q$(3)~)
in the u$_q$(3)$\supset$so$_q$(3) basis 
(for the
most symmetric u$_q$(3) representation and for $q$ being real). 

   The generalization of the Wigner--Eckart theorem in the case of the
   algebra so$_q$(3) is \cite{Kli2919,Nom2345}
$$ \langle\alpha',L' M'|T^j_m|\alpha,L M\rangle = (-1)^{2j}\;
\frac{_qC_{LM,jm}^{L'M'}}{\sqrt{[2L'+1]}}
\langle\alpha',L'\|T^j\|\alpha,L\rangle, \eqno(29.1)$$
   where $|\alpha,L M\rangle$ are orthonormalized basis vectors of the
   irreducible representation $_qD^L$ of the algebra so$_q$(3) and
   $_qC_{L_1 M_1,L_2 M_2}^{LM}$ are the Clebsch--Gordan coefficients
   \cite{} of the same algebra. 

Our purpose is to calculate the reduced matrix elements
   of the $q$-deformed quadrupole operator $Q^2$, namely the quantities
$ \langle\lambda,L+2\|Q^2\|\lambda,L\rangle$ and
$\langle\lambda,L\|Q^2\|\lambda,L\rangle$, 
   where the operator $Q^2$ is
$$ Q^2_M = \sqrt{\frac{[3][4]}{[2]}} A^2_M, \qquad
A^2_M = {[T^{\dagger} \otimes \widetilde{T}]}^{2}_{M} =
\sum_{m,n}{}_{q^{-1}}C_{1m,1n}^{2M} T^{\dagger}_{m} \widetilde{T}_{n}.
\eqno(29.2)$$
It is clear that this operator is a second rank tensor under so$_q$(3),
according to the definitions given in the previous section, while  
in eq. (29.2)  the overall constant factor
guarantees the agreement between the present expression and the 
usual classical operator in the limit $q\to 1$.  
Other reduced matrix elements do not occur, since in the most symmetric 
representation of u$_q$(3) only states with equal parity of $\lambda$ and $L$ 
occur. 

Using the Wigner-Eckart theorem (eq. (29.1)) one  obtains for the reduced 
matrix elements of the quadru\-po\-le  operator $Q^2$ of  eq. (29.2)
the following results (see \cite{BRRT95} for a proof)
$$   \langle{\lambda,L+2}\|Q^{2}\|{\lambda,L}\rangle
= {q^{\lambda-{1\over 2}} \over [2]} \sqrt{[3][4] \over [2]}
\sqrt{[\lambda-L][\lambda+L+3][2L+4][2L+2] \over 
[2L+3]} , \eqno(29.3)$$
$$  \langle{\lambda,L}\|Q^{2}\|{\lambda,L}\rangle
= -{q^{\lambda-{1\over 2}}\over [2]} \sqrt{[2L][2L+1][2L+2] \over
[2L-1][2L+3]} \left\{ q^{L-{1\over 2}}[\lambda-L] +
q^{-L+{1\over 2}}[\lambda+L+3] \right\}, \eqno(29.4)$$
which are in agreement with the classical case in the limit $q\rightarrow 1$.
Taking into account the Wigner--Eckart theorem and the symmetry properties
of the $q$-deformed Clebsch--Gordan coefficients it can also be shown that
   the reduced matrix element given in eq. (29.3)
has the following symmetry property
$$ \langle\lambda,L+2\|Q^2\|\lambda,L\rangle =
\langle\lambda,L\|Q^2\|\lambda,L+2\rangle .\eqno(29.5)$$
   For small values of the deformation parameter $\tau$
   \ ($q= e^\tau$, with  $\tau$ being real) the following Taylor 
expansions  for the reduced matrix elements of eqs (29.3) and (29.4) 
are obtained
$$  \langle\lambda,L+2\|Q^2\|\lambda,L\rangle =
\sqrt{\frac{6(\lambda-L)(\lambda+L+3)(L+2)(L+1)}{2L+3}}$$
$$  \times\left\{1 + \left(\lambda-\frac{1}{2}\right)\tau +
\left(\frac{2}{3}\lambda^2 + 
\frac{1}{2}L^2+\frac{3}{2}L+\frac{65}{24}\right)
\tau^2 + {\rm O}(\tau^3)\right\}, \eqno(29.6)$$
$$  \langle\lambda,L\|Q^2\|\lambda,L\rangle =
-(2\lambda+3)\sqrt{\frac{L(L+1)(2L+1)}{(2L-1)(2L+3)}}
\biggl\{1+\frac{2}{2\lambda+3}\{\lambda(\lambda+1)-L(L+1)\}\tau$$
$$  + \frac{1}{3(2\lambda+3)}\{(2\lambda+15)L(L+1)+(2\lambda+1)
(2\lambda^2+2\lambda+3)\}\tau^2 + {\rm O}(\tau^3)\biggr\}.\eqno(29.7)$$
 In the limit  $\tau=0$ one obtains the classical expressions for the 
corresponding reduced matrix elements.

From a physical point of view of great interest are the $E2$ transition
probabilities ($B(E2)$ factors), which in the classical case are 
expressed by means of the reduced matrix elements of the u(3) quadrupole 
operator. By analogy the $B(E2)$ factors corresponding to the chain
u$_q$(3)$\supset$so$_q$(3) can be assumed to be of the form
$$ B(E2; (\lambda,L+2) \rightarrow (\lambda,L))_q =
\frac{1}{[2L+5]}{|\langle\lambda,L+2\|Q^2\|\lambda,L\rangle|}^2, \eqno(29.8)$$
where $Q^2$ is the $q$-deformed quadrupole operator of eq. (29.2). 
Again in analogy with the classical case, 
the reduced matrix element of eq. (29.4) is related to the 
deformation of
the physical system in the state with angular momentum $L$.

It should be remembered  that the results obtained here
are valid only for real values of the deformation parameter $q$. On the 
other hand the comparison of the predictions of various models 
based on the su$_q$(2) algebra with the experimental data
(sections 19--23) shows that one
can achieve good agreement between theory and experiment in the case of  
$q$ being a phase factor ($q = e^{i\tau}$, with $\tau$ real). 
It is therefore desirable to repeat the steps followed in this section 
for the case of $q$ being a phase factor. 

\subsection{Matrix elements of the usual quadrupole operator in the 
u$_q$(3) $\supset$ so$_q$(3) basis} 

In order to be able to calculate matrix elements of the usual quadrupole 
operator in the $q$-deformed basis of eq. (28.11) we should first express 
the basis vectors in terms of usual (undeformed) bosons $a_i^\dagger$, 
$a_i$, which are connected to the $q$-deformed bosons $b_i^\dagger$, $b_i$
through \cite{Bie873,Mac4581,KD415}
$$ b^\dagger_i = \sqrt{\frac{[N_i]}{N_i}} a^\dagger_i,\qquad
b_i = a_i \sqrt{\frac{[N_i]}{N_i}},\qquad i = +,0,-,\eqno(29.9)$$
where $N_i$ is the number operator of the undeformed bosons. 
In this way the operators $B^\dagger_i$, $B_i$ of eq. (28.2)
can be written in the form
$$ B_0 = q^{-\frac{1}{2}N_0} a_0 \sqrt{\frac{[N_0]}{N_0}}, \qquad
B^\dagger_0 = \sqrt{\frac{[N_0]}{N_0}} a^\dagger_0 q^{-\frac{1}{2}N_0} ,
\eqno(29.10)$$ 
$$ B_i = q^{N_i+\frac{1}{2}} a_i \sqrt{\frac{[2N_i]}{N_i}}, \qquad
B^\dagger_i = \sqrt{\frac{[2N_i]}{N_i}} a^\dagger_i q^{N_i+\frac{1}{2}} , 
\qquad i=+,- . \eqno(29.11)$$
As a next step we can replace $B^\dagger_i$ in eq. (28.11)  with
their expressions from eq. (29.11).  Then, using the following
identity, which holds in the Fock space,
$$ \frac{(B^\dagger_+)^x}{\sqrt{[2x]!!}}
\frac{(B^\dagger_0)^y}{\sqrt{[y]!}}
\frac{(B^\dagger_-)^z}{\sqrt{[2z]!!}}|0\rangle =
q^{\frac{1}{2}(x^2+z^2)-\frac{1}{4}y(y-1)}
\frac{(a^\dagger_+)^x}{\sqrt{x!}}
\frac{(a^\dagger_0)^y}{\sqrt{y!}}
\frac{(a^\dagger_-)^z}{\sqrt{z!}}|0\rangle, \eqno(29.12)$$
we obtain (in a slightly different notation) \cite{VdJ1799,VdJ948}
$$ \left|\begin{array}{l}\lambda\\L,M\end{array}\right\rangle_q=
q^{-\frac{1}{4}\{(2\lambda+M)(M-1)+L(L+1)\}} N^{(q)}_{\lambda L M}
\sum_{t=0}^{(\lambda-L)/2}
\sum_{x=\max(0,M)}^{\lfloor(L+M)/2\rfloor}
\frac{(-1)^t q^{(\lambda-\frac{1}{2})x-(L+\frac{3}{2})t}}
{[2t]!![\lambda-L-2t]!!} $$
$$ \times 
\frac{\sqrt{[2x+2t]!![\lambda+M-2x-2t]![2x+2t-2M]!!}}
{[2x]!![L+M-2x]![2x-2M]!!} 
\frac{(a^\dagger_+)^{x+t}}{\sqrt{(x+t)!}}
\frac{(a^\dagger_0)^{\lambda+M-2x-2t}}{\sqrt{(\lambda+M-2x-2t)!}}
\frac{(a^\dagger_-)^{x+t-M}}{\sqrt{(x+t-M)!}}|0\rangle, \eqno(29.13)$$
where the normalization factor $N^{(q)}_{\lambda L M}$ is
$$ N^{(q)}_{\lambda L M} =
\sqrt{\frac{[\lambda-L]!![L+M]![L-M]![2L+1]}{[\lambda+L+1]!!}}. \eqno(29.14)$$

The expression of eq. (29.13) can be used for the calculation
of the matrix elements of ``standard'' (or classical)
tensor operators (which are built up from standard boson operators)
between the states with given $q$-deformed angular momentum.

Let us consider now the classical quadrupole operator $Q^{(c)}_m$.
Its zero component is of the form
$$ Q^{(c)}_0 = \sqrt{6} [a^\dagger\otimes \widetilde{a}]_{20}
= a^\dagger_{+}\widetilde{a}_{-} + 2 a^\dagger_{0}\widetilde{a}_{0}
+a^\dagger_{-}\widetilde{a}_{+} 
= -a^\dagger_{+}a_{+} + 2 a^\dagger_{0}a_{0}
-a^\dagger_{-}a_{-} = 3N_{0} - N, \eqno(29.15)$$
where $a^\dagger_m, a_m (m=+,0,-)$ are classical boson operators and
$\widetilde{a}_m=(-1)^m a_{-m}$.

The explicit expression for the matrix elements of $Q^{(c)}_0$
between the $q$-deformed states of eq. (29.13)  then becomes
$$ {\atop{\atop}}_q\!\left\langle\begin{array}{l}\lambda\\ J,M'
\end{array}\right| Q^{(c)}_0
\!\left|\begin{array}{l}\lambda\\ L,M\end{array}\right\rangle_q =
\delta_{M,M'}
{\atop{\atop}}_q\!\left\langle\begin{array}{l}\lambda\\ J,M
\end{array}\right| Q^{(c)}_0
\!\left|\begin{array}{l}\lambda\\ L,M\end{array}\right\rangle_q, \eqno(29.16)$$
and
$$  {\atop{\atop}}_q\!\left\langle\begin{array}{l}\lambda\\ J,M
\end{array}\right| Q^{(c)}_0
\!\left|\begin{array}{l}\lambda\\ L,M\end{array}\right\rangle_q =
(2\lambda + 3M) \delta_{L,J}
-6\,  q^{-\frac{1}{2}(2\lambda+M)(M-1) -\frac{1}{4}\{ L(L+1)+J(J+1) \}}
  N^{(q)}_{\lambda L M} N^{(q)}_{\lambda J M} $$
$$ \times \sum_{t=0}^{(\lambda-L)/2}\sum_{r=0}^{(\lambda-J)/2}
\frac{(-1)^{t+r} q^{-(\lambda+L+1)t - (\lambda+J+1)r} }{[2t]!!
[\lambda-L-2t]!![2r]!![\lambda-J-2r]!!} 
\sum_{x}\frac{[2x]!![\lambda+M-2x]![2x-2M]!!}
{[2x-2t]!![L+M-2x+2t]![2x-2t-2M]!!} $$ 
$$\times \frac{x~ q^{(2\lambda-1)x} }
{[2x-2r]!![J+M-2x+2r]![2x-2r-2M]!!}, \eqno(29.17)$$
where:
$$\max\{ t,r,M+t,M+r \} \leq x \leq
\min\{ \lfloor(L+M)/2\rfloor + t, \lfloor(J+M)/2\rfloor + r \}.\eqno(29.18)$$

Clearly the recipe described above can be applied for the
calculation of  matrix elements between the $q$-deformed
angular momentum states of any classical operator. The disadvanage
of this method is that one must perform special calculations for any
specific operator. This can be avoided by the use of the transformation
matrix between the ``standard'' and $q$-deformed states, which will be 
described in the next subsection. 

\subsection{Transformation between so$_q$(3) and so(3) basis states}

For fixed $\lambda$ one can expand the
$q$-deformed basis states of eq. (29.13) 
in terms of classical ones (with $q= 1$),
since both sets form orthonormal bases in the
eigenspace ${\cal H}_\lambda$ of the number operator
$N=N_+ + N_0 + N_-$, corresponding to the eigenvalue $\lambda$.
In particular,
$\dim {\cal H}_\lambda = \frac{1}{2}(\lambda+1)(\lambda+2)$
and the operator $L_0$ has the same form $L_0=N_+ - N_-$
in both classical and $q$-deformed cases. In this way
$$ \left|\begin{array}{l}\lambda\\ L,M\end{array}\right\rangle_q
= \sum_{J,M_J}
\left|\begin{array}{l}\lambda\\ J,M_J\end{array}\right\rangle_c
{\atop{\atop}}_c\!\left\langle\begin{array}{l}\lambda\\J,M_J
\end{array}\right.
\!\left|\begin{array}{l}\lambda\\ L,M\end{array}\right\rangle_q
= \sum_{J}
\left|\begin{array}{l}\lambda\\ J,M\end{array}\right\rangle_c
{\atop{\atop}}_c\!\left\langle\begin{array}{l}\lambda\\ J,M
\end{array}\right.
\!\left|\begin{array}{l}\lambda\\ L,M\end{array}
\right\rangle_q, \eqno(29.19)$$
where $J = \lambda, \lambda-2,\ldots,|M|\ \mbox{or}\ |M|+1$
and the subscript $c$ denotes the classical (undeformed) basis states.
Taking the scalar product between $q$-deformed basis states of eq. (29.13)
and classical basis states, for
the transformation matrix in eq. (29.19) one obtains
$$ {\atop{\atop}}_c\!\left\langle\begin{array}{l}\lambda\\
J,M'\end{array}\right.
\!\left|\begin{array}{l}\lambda\\ L,M\end{array}\right\rangle_q
= \delta_{M,M'}\ q^{-\frac{1}{4}\{(2\lambda+M)(M-1)+L(L+1)\}}
\ N^{(c)}_{\lambda J M}\ N^{(q)}_{\lambda L M} $$
$$\times \sum_{t=0}^{(\lambda-L)/2}\sum_{r=0}^{(\lambda-J)/2}
\frac{(-1)^{t+r}\ q^{-(\lambda+L+1)t}}
{[2t]!![\lambda-L-2t]!!(2r)!!(\lambda-J-2r)!!} 
 \sum_{x}
\frac{q^{(\lambda-\frac{1}{2})x} \sqrt{[2x]!![\lambda+M-2x]![2x-2M]!!}}
{[2x-2t]!![L+M-2x+2t]![2x-2t-2M]!!} $$
$$ \times \frac{\sqrt{(2x)!!(\lambda+M-2x)!(2x-2M)!!}}
{(2x-2r)!!(J+M-2x+2r)!(2x-2r-2M)!!}, \eqno(29.20)$$
where
 $$\max\{t,r,M+t,M+r\} \leq x
\leq \min\{\lfloor (L+M)/2 \rfloor + t,
\lfloor (J+M)/2 \rfloor + r \}. \eqno(29.21)$$
It should be noted that the scalar product given in eq. (29.20)
is real and has the properties
$$ {\atop{\atop}}_c\!\left\langle\begin{array}{l}\lambda\\ J,M'
\end{array}\right.
\!\left|\begin{array}{l}\lambda\\ L,M\end{array}\right\rangle_q =
\delta_{M,M'}
{\atop{\atop}}_c\!\left\langle\begin{array}{l}\lambda\\ J,M
\end{array}\right.
\!\left|\begin{array}{l}\lambda\\ L,M\end{array}\right\rangle_q =
{\atop{\atop}}_q\!\left\langle\begin{array}{l}\lambda\\ L,M
\end{array}\right.
\!\left|\begin{array}{l}\lambda\\ J,M'\end{array}\right\rangle_c,\eqno(29.22)$$
which have been used in eq. (29.19).
Though looking very involved, eq. (29.20) is quite
handy for computational purposes. Indeed, only few terms are
contained in the sum.
The same equation shows that the transformation matrix is diagonal in
the third projections of $J$ and $L$,
and is nonvanishing even for high values
of the difference $\Delta(J) = |L-J|$. However,
the values are peaked around the classical value $\Delta(J) = 0$,
if the parameter of deformation $q$ tends to unity.

Let us consider now a classical tensor operator $A^{(c)}_{jm}$
of rank $j$ according to the algebra so(3),
which conserves the number of particles.
Using the Wigner-Eckart theorem
in the classical case we can express the matrix elements of
$A^{(c)}_{jm}$ between the classical so(3) basis states
in the form
$$ {\atop{\atop}}_c\!\left\langle\begin{array}{l}\lambda\\ J,M'
\end{array}\right|
A^{(c)}_{jm}\left|\begin{array}{l}\lambda\\ L,M
\end{array}\right\rangle_c
= (-1)^{2j} \frac{C^{JM'}_{LM, jm}}{\sqrt{2J+1}}
\langle\lambda J\| A^{(c)}_j \|\lambda L\rangle, \eqno(29.23)$$
where $C^{JM'}_{LM, jm}$ are the classical Clebsch-Gordan coefficients
of the so(3) algebra, and we assume that the
classical reduced matrix elements of $A^{(c)}_j$ in
eq. (29.23)  are known. In this way, from eq. (29.19)
follows that
$$  A^{(c)}_{jm} \left|\begin{array}{l}\lambda\\ L,M
\end{array}\right\rangle_q =
\sum_{R}
A^{(c)}_{jm} \left|\begin{array}{l}\lambda\\ R,M
\end{array}\right\rangle_c
{\atop{\atop}}_c\!\left\langle\begin{array}{l}\lambda\\ R,M
\end{array}\right.\left|\begin{array}{l}\lambda\\ L,M
\end{array}\right\rangle_q $$
$$  = \sum_{R}\sum_{P,M_P}
\left|\begin{array}{l}\lambda\\ P,M_P\end{array}\right\rangle_c
{\atop{\atop}}_c\!\left\langle\begin{array}{l}\lambda\\ P,M_P
\end{array}\right|
A^{(c)}_{jm}\left|\begin{array}{l}\lambda\\ R,M\end{array}\right\rangle_c
{\atop{\atop}}_c\!\left\langle\begin{array}{l}\lambda\\ R,M
\end{array}\right.
\left|\begin{array}{l}\lambda\\ L,M\end{array}\right\rangle_q, \eqno(29.24)$$
and using eq. (29.23)  we have the expansion
$$  {\atop{\atop}}_q\!\left\langle\begin{array}{l}\lambda\\ J,M'
\end{array}\right|
A^{(c)}_{jm}\left|\begin{array}{l}\lambda\\ L,M
\end{array}\right\rangle_q =
\sum_{R}\sum_{P} (-1)^{2j}
\frac{C^{PM'}_{RM, jm}}{\sqrt{2P+1}} $$
$$ \times
{\atop{\atop}}_q\!\left\langle\begin{array}{l}\lambda\\ J,M'
\end{array}\right.
\left|\begin{array}{l}\lambda\\ P,M'\end{array}\right\rangle_c
{\langle\lambda P\| A^{(c)}_j \|\lambda R\rangle}
{\atop{\atop}}_c\!\left\langle\begin{array}{l}\lambda\\ R,M
\end{array}\right.
\left|\begin{array}{l}\lambda\\ L,M\end{array}\right\rangle_q.\eqno(29.25)$$
In particular, eq. (29.25) shows that the matrix elements
of the classical tensor operator $A^{(c)}_{jm}$ between $q$-deformed
so$_q$(3) states are nonvanishing if $M'=M+m$, as in the classical
basis. It is also worth to remark that the summation expression
of eq. (29.25) gives an universal way for (numerical) computation
of matrix elements of different types of classical operators
between the deformed basis states of eq. (29.13). For instance, it is
straightforward to show that the numerical values of the quadrupole
operator calculated by eq. (29.17) of the previous subsection  and
the method used in this subsection (involving transformation matrices)
coincide. However, the method descibed in this subsection is universal
and can be applied to any classical operator.

Examples of transformation brackets can be found in \cite{RRIT4383}.

\section{ The question of complete breaking of symmetries and some 
applications}

In the cases examined so far the $q$-deformed symmetries considered were close
to their classical counterparts, to which they reduce for $\tau=0$ ($q=1$), 
since the values of $\tau$ were relatively small. One can then argue that 
results similar to the ones provided by the quantum symmetries can also
be obtained from the usual Lie symmetries through the addition of suitable
perturbations. What can be  very useful is to start with one limiting 
symmetry and, through large deformations, reach another limiting symmetry. 
We shall refer to this as the complete breaking of the symmetry. 
In this way one could hope to ``bridge'' different Lie symmetries through
the use of $q$-deformations, providing in addition new symmetries in the 
regions intermediate between the existing Lie ones. 

The question of complete breaking of the symmetry in the framework of the toy
IBM of sec. 27 has been studied by Cseh \cite{Cseh1225}, Gupta 
\cite{Gup1067}, and Del Sol Mesa {\it et al.} \cite{Del1147}. 

Cseh \cite{Cseh1225} started with the su$_q$(2) (vibrational) limit (in a 
form different from  the one used in sec. 27) and tried to 
reach the so$_q$(3) (rotational) limit. He noted that for $q$ being a 
phase factor this is not possible, while for real $q$ some rotational features
are obtained, but without all of the requirements for rotational behaviour
being satisfied simultaneously.

Gupta \cite{Gup1067} started with the su$_q$(2) limit, in the form given 
in sec. 27. He 
noted that for $q$ being a phase factor a recovery of the su(3) symmetry occurs
(see also the next paragraph),
while for real $q$, and even better
for $q$ complex ($q=e^s$ with $s=a+ib$), the so$_q$(3) limit is indeed 
reached. 

Del Sol Mesa {\it et al.} \cite{Del1147} considered the o$_q$(3) limit of the 
model of sec. 27, since it corresponds to the symmetry of a $q$-deformed 
version of the spherical Nilsson Hamiltonian with spin-orbit coupling term. 
They found that for $q$ being a phase factor ($q=e^{i\tau}$) and for $\tau$ 
obtaining values in the region $0.5\leq \tau\leq 2$ the u(3) symmetry, 
which is broken in the initial model because of the presence of the spin-orbit
term, is recovered. This offers a way of recovering the u(3) symmetry 
alternative to the one developed for the spherical Nilsson model 
\cite{BCM2841,CMQ238} and the deformed Nilsson model \cite{MDel146,CVHH303}
through the use of appropriate unitary operators.  

Complex deformations have also been used in \cite{GCLG73} in 
the framework of a deformed u(2) model, possessing the u(2)$\supset$u(1)
and u(2)$\supset$o(2) chains. Again it has been found that complex 
deformations can bridge the two limiting symmetries. 

Possible complete breaking of the symmetry has also been studied in the 
framework of the $q$-deformed version of the full Interacting Boson Model
(see sec. 31). 

A problem associated with complex deformations as the ones considered 
above is that the energy eigenvalues become complex as well. A way to 
avoid this problem has been introduced recently by Jannussis and 
collaborators \cite{FJMS94,ABF605}.

Finally, the o$_q$(3) limit of the model of sec. 27 has been used for 
describing the $^{16}$O + $\alpha$ cluster states in $^{20}$Ne \cite{Cseh63}. 
It turns out that an 
improved description of the energy spectrum and the $\alpha$-particle 
spectroscopic factors occurs for $q=e^{0.124}$. 

\section{ $q$-deformation of the Interacting Boson Model (IBM)}

The Interacting Boson Model \cite{AI201,AI253,AI468} 
(see \cite{IA1987,Bon88} for recent overviews) is the 
most popular algebraic model of nuclear structure. It describes the 
collective properties of medium-mass and heavy nuclei away from closed 
shells in terms of bosons, which correspond to correlated valence fermion
pairs. In its simplest form, called IBM-1, only  $s$ ($J=0$) and $d$ ($J=2$)
bosons are used. The overall symmetry of the model is u(6), possessing 
three limiting symmetries: the u(5) (vibrational) limit, corresponding 
to the chain of subalgebras
$$ {\rm u}(6) \supset {\rm u}(5) \supset {\rm o}(5) \supset {\rm o}(3),
\eqno(31.1)$$
the su(3) (rotational) limit, characterized by the chain
$$ {\rm u}(6) \supset {\rm su}(3) \supset {\rm o}(3),\eqno(31.2)$$
and the o(6) ($\gamma$-unstable) limit, for which the relevant chain is
$$ {\rm u}(6) \supset {\rm o}(6) \supset {\rm o}(5) \supset {\rm o}(3).
\eqno(31.3)$$

If one of these dynamical symmetries is present, the Hamiltonian can be 
written in terms of the Casimir operators of the algebras appearing in 
the relevant chain. Thus the Hamiltonian can be analytically diagonalized
in the corresponding basis. This is a great advantage of IBM and its 
numerous generalizations: they provide us with a large number of exactly 
soluble models, the predictions of which can be directly compared to 
experiment, without any need for lengthy numerical calculations. 

From what we have already seen in sec. 27, it is worth examining if 
a $q$-deformed version of the IBM has any advantages in comparison to the
standard version. In order to accomplish this, one has to construct 
the $q$-analogues of the three chains mentioned above. The difficulties 
associated with the su(3) chain have already been discussed in subsec. 27.3. 
In what follows  we are going to focus attention on the o(6) chain, for
which the relevant construction has been carried out \cite{WY492}. 
The technique used is based on the notion of complementary subalgebras, 
which is explained in detail in \cite{ABS1088}, while here only final results
will be reported. We only mention here that the notion of complementary 
subalgebras was introduced by Moshinsky and Quesne 
\cite{MQ1631,CDQ779,Que2675}. Two subalgebras 
A$_1$ and A$_2$ of a larger algebra A are complementary within a definite 
irrep of A, if there is an one-to-one correspondence between all the irreps 
of A$_1$ and A$_2$ contained in this irrep of A.  

In the o(6) limit of IBM the Hamiltonian is
$$ H= E_0 + \beta C_2({\rm o}(5))+\gamma C_2({\rm o}(3)) + \eta C_2({\rm o}
(6)).\eqno(31.4)$$
The eigenvalues of the energy in the relevant basis have already been given 
in subsec. 26.1. 

Using the notion of complementarity it turns out that, instead of using
the o(6) chain mentioned above, it suffices to study the chain 
$$ {\rm su}^{sd}(1,1) \otimes {\rm so}(6) \supset {\rm su}^d(1,1) 
\otimes {\rm so}(5) \supset {\rm so}(3), \eqno(31.5)$$
where su$^{sd}$(1,1) is the algebra closed by the pair operators formed 
out of the $s$ and $d$ bosons, while su$^d$(1,1) is the algebra closed
by the pair operators formed out of $d$ bosons alone. 
(Details on the basis for symmetric irreps of su(1,1)$\otimes$o(5)
can be found in \cite{SG76}.)  The irreps of 
su$^{sd}$(1,1) are characterized by the same quantum numbers as the 
irreps of o(6) in the o(6) chain of the IBM, while the irreps of 
su$^d$(1,1) are characterized by the same quantum numbers as the irreps
of o(5) in the o(6) limit of IBM. 
Therefore in the Hamiltonian one can use the Casimir operators of the 
su$^{sd}$(1,1), su$^d$(1,1) and su(2) subalgebras (the deformed versions
of which are well known, as seen in secs 14--16) instead of the 
Casimir operators of o(6), o(5), o(3) respectively. Keeping the same 
notation as in eq. (26.3) the final result reads 
$$E(N,\sigma,\tau,\nu_{\Delta},J,M_J) = E_0 + \beta 8 \left[{\tau\over 2}
\right]_q \left[{\tau+3\over 2}\right]_q + \gamma 2 [J]_q [J+1]_q +\eta 8
 \left[{\sigma\over 2}\right]_q  \left[ {\sigma+4\over 2}\right]_q, 
\eqno(31.6)$$
where the free parameters have been chosen so that the present equation 
reduces to its classical counterpart for $q\rightarrow 1$. 
For the ground state band then the analog of eq. (26.4) is 
$$ E(J) = E_0+ \beta' 8 [J]_{q^{1/4}} [J+6]_{q^{1/4}} + \gamma 2 [J]_q [J+1]_q
+\eta' 8 [N]_{q^{1/2}} [N+4]_{q^{1/2}},\eqno(31.7)$$
where the identities
$$ \left[ {x \over 2} \right]_q = [ x]_{q^{1/2}} (q^{1/2}+q^{-1/2})^{-1}, 
\eqno(31.8)$$
$$   \left[ {x \over 4} \right]_q = [ x]_{q^{1/4}} (q^{1/2}+q^{-1/2})^{-1}
                        (q^{1/4}+q^{-1/4})^{-1}, \eqno(31.9)$$
have been used and $\beta'$, $\eta'$ are related to $\beta$, $\eta$ and $q$
in an obvious way. 
We remark that the Casimir operator of su$_q^d$(1,1), 
which is complementary to o(5) in the undeformed case, leads to 
a term of the form $[J]_{q'} [J+6]_{q'}$ with $q'=q^{1/4}$. 

In refs \cite{GL593,Gup117}
 the question has been studied if large values of the 
deformation parameter can lead us from the o(6) limit to the su(3) 
(rotational) or u(5) (vibrational) limits, so that complete breaking 
of the symmetry, in the sense of sec. 30, could be obtained. It turns out
that for $q$ real the spectrum of the ground state band goes towards the 
rotational limit, while $q$ being a phase factor leads towards the vibrational 
limit. Many more detailed studies, of both spectra and electromagnetic 
transition probabilities, are required before such a claim can be made. 

A different method for constructing the $q$-deformed versions of the u(5)
and o(6) limits of the IBM has been used by Pan \cite{Pan1876}. 
The method is based on the use of $q$-deforming functionals (see secs 10, 14). 
The same method has been used in \cite{Pan257} for studying the $q$-deformed
version of the su(3)$\supset$so(3) decomposition. The final  result 
for the energy eigenvalues in the o(6) case is similar to the one reported in 
eq. (31.6), the main difference being that different deformation parameters 
are allowed in each of the three deformed terms in the rhs. Some comparisons 
of the model predictions for spectra and B(E2) values to the experimental data
have been performed \cite{Pan1876}.

It is clear that several deformed versions of the IBM can be constructed,
providing us with a large number of exactly soluble models.  
In order to demonstrate their usefulness, one has to show that by deforming 
the model one gets some advantages over the classical (non-deformed) 
version. One way to achieve this is the use of parameter-independent 
tests based on systematics of the data, like the ones used in 
\cite{Bon2001,BSR63,BSR865} for the usual IBM. It is also
desirable for the deformation parameter to be associated with some 
physical quantity, as in the case of the su$_q$(2) model. Much work is 
still required in these directions. Some mathematical results which 
can be useful in these efforts are reported below: 

i) Casimir operators for su$_q$(n) have been given in \cite{Bin1133}, 
while the quadratic Casimir of so$_q$(5) can be found in \cite{ZGB937}.

ii) Raising and lowering operators for u$_q$(n) have been given by 
Quesne \cite{Que357}. 

iii) Irreps of u$_q$(m+n) in the u$_q$(m)$\oplus$u$_q$(n) basis have been 
constructed in \cite{Pan1992}, while generalized $q$-bosonic operators acting 
in a tensor product of $m$ Fock spaces have been constructed as double
irreducible tensors with respect to u$_q$(m)$\oplus$u$_q$(n) in 
\cite{Que298,Que322}.

\section{Deformed versions of other collective models}

The Moszkowski model \cite{Moszk} is a schematic two-level model which 
provides a description of the phase transition from 
the vibrational regime to the rotational one. A $q$-deformed version 
of the model has been developed \cite{MAP6317}
and the RPA modes in it have been discussed \cite{BBMPP895}.
In addition the validity of the coherent states variational method 
has been checked in the framework of this model and the time evolution 
of the system has been studied by means of a time dependent variational 
principle \cite{FSM547}. 
Furthermore, the $q$-deformed Moszkowski model with cranking has been 
studied in the mean field approximation and the relation between 
$q$-deformation and temperature has been discussed \cite{PBPBM1209}.
It should be noticed 
here that quantum algebraic 
techniques have also been found useful in describing thermal effects in 
the framework of the $q$-deformed Thouless model for superconductivity
\cite{BAMP192}. 

The Lipkin--Meshkov--Glick (LMG) model \cite{LMG188} 
is an exactly soluble schematic shell 
model. $q$-deformed versions of the 2-level LMG model (in terms of an 
su$_q$(2) algebra) \cite{AB174,LS50,AMMC831,ACMM701,AEGPL4915}
and of the 3-level LMG model (in terms of an su$_q$(3)
algebra) \cite{Brito92} have been developed. 

\section{Fermion pairs as deformed bosons: approximate mapping}

We have seen so far that several quantum algebraic phenomenological models 
have been proposed for the description of nuclear collective properties.
These models make use of $q$-deformed bosons, which satisfy commutation 
relations differing from the standard boson commutation relations, to
which they reduce in the limit $q\rightarrow 1$. 

On the other hand, it is known that vibrational nuclear spectra, which are
described in the simplest way by a pairing Hamiltonian, show anharmonicities
(see also sec. 26), described, for example, by the {\sl Anharmonic Vibrator
 Model (AVM)} \cite{DDK632}
$$ E(J)=aJ+bJ(J-2).\eqno(33.1)$$
In the framework of the single-j shell model 
\cite{DK236,VD272,VKD2188,Klein39,BKL521},
which can be extended to several non-degenerate j-shells 
\cite{BK87,MBK381,KM441} these anharmonicities are related to the fact
that correlated fermion pairs satisfy commutation relations
which resemble boson commutation relations but in addition include corrections
due to the presence of the Pauli principle. This fact has been
the cause for the development of boson mapping techniques
(see the recent reviews by Klein and Marshalek \cite{KM375} and Hecht 
\cite{Hecht87}  and references therein), by
which the description of systems of fermions in terms of bosons
is achieved. In recent years boson mappings have attracted
additional attention in nuclear physics as a necessary tool
in providing a theoretical justification for the success of
the phenomenological Interacting  Boson Model \cite{IA1987}
 and its various extentions, in
which low lying collective states of medium and heavy mass
nuclei are described in terms of bosons.

From the above observations it is clear that both q-bosons
and correlated fermion pairs satisfy commutation relations
which resemble the standard boson commutation relations but
they deviate from them, due to the $q$-deformation in the former
case and to the Pauli principle in the latter.
A question is thus created: Are $q$-bosons
suitable for the approximate description of correlated fermion
pairs? In particular, is it possible to construct a boson mapping
in which correlated fermion pairs are mapped onto $q$-bosons, in a way
that the $q$-boson operators approximately satisfy the same
commutation relations as the correlated fermion pair operators?
In this section we show for the simple case of su(2) that such
a mapping is indeed possible.

\subsection{The single-j shell model}

Let us consider the single-j shell model 
\cite{DK236,VD272,VKD2188,Klein39,BKL521}. 
One can define fermion pair and multipole operators as
$$A_{JM}^\dagger = {1\over \sqrt{2}} \sum_{m m'} ( j m j m' | J M)
a^\dagger_{jm} a^\dagger_{jm'}, \eqno(33.2)$$
$$B_{JM}= {1\over \sqrt{2J+1}} \sum _{m m'} (j m j -m'| J M)
(-1)^{j-m'}  a^\dagger_{jm} a_{jm'}, \eqno(33.3)$$
with the following definitions
$$ A_{JM}= [ A^\dagger_{JM}]^\dagger , \quad B^\dagger_{JM}= 
[B_{JM} ]^\dagger.\eqno(33.4)$$
In the above $a_{jm}^\dagger$ ($a_{jm}$) are fermion creation (annihilation)
operators and $(jm jm'|JM)$ are the usual Clebsch--Gordan
coefficients.

The pair and multipole operators given above satisfy the following
commutation relations:
$$[ A^\dagger_{JM}, A^\dagger_{J'M'} ] =0, \eqno(33.5)$$
$$[A_{JM}, A^\dagger_{J'M'}]=\delta_{JJ'} \delta_{M M'}
-2\sum_{J''} (-1)^{2j+M} \sqrt{(2J+1)(2J'+1)(2J''+1)} $$
$$(J -M J' M' | J'' M'-M) \left\{\matrix{J & J' & J''\cr
j & j & j \cr}\right\} B_{J'', M'-M}, \eqno(33.6)$$
$$[ B_{JM}^\dagger, A^\dagger_{J'M'}]= \sum_{J''} 2 \sqrt{2J'+1} (-1)^{2j-M}
(J -M J' M' | J'' M'-M)                             $$
$$\left\{\matrix{J & J' & J''\cr j & j & j\cr}\right\}
A^\dagger_{J'', M'-M} {1+(-1)^{J''} \over 2}, \eqno(33.7)$$
$$[B_{JM}, B_{J'M'}] = \sum_{J''} (-1)^{2j-J''}
[1-(-1)^{J+J'+J''}] \sqrt{2J''+1}$$ $$ (J M J' M' | J'' M+M')
\left\{\matrix{J & J' & J''\cr j & j & j \cr}\right\}
B_{J'', M+M'}, \eqno(33.8)$$
in which the curly brackets are the usual 6-j symbols.
These are the commutation relations of the so(2(2j+1)) algebra.

\subsection{Fermion pairs of zero angular momentum} 

In the present subsection we will restrict ourselves to fermion pairs
coupled to angular momentum zero. The relevant commutation relations take
the form
$$[ A_0, A_0^\dagger] = 1-{N_F\over\Omega}, \qquad
  [{N_F\over 2}, A^\dagger_0]= A^\dagger_0, \qquad
  [{N_F\over 2}, A_0]= -A_0, \eqno(33.9)$$
where $N_F$ is the number of fermions, $2\Omega=2j+1$ is the
size of the shell, and $$B_0=N_F/\sqrt{2\Omega}. \eqno(33.10)$$
 With the identifications
$$J_+= \sqrt{\Omega} A^\dagger_0, \quad J_-=\sqrt{\Omega} A_0, \quad
J_0= {N_F-\Omega\over 2}, \eqno(33.11)$$
eqs (33.9) take the form of the usual su(2) commutation relations
$$[J_+, J_-]= 2J_0, \quad [J_0, J_+]=J_+, \quad [J_0, J_-]=-J_-.
\eqno(33.12)$$
An exact boson mapping of the su(2) algebra is given in \cite{KM375,BKL521}
$$A_0^\dagger= a_0^\dagger \sqrt{1-{n_0\over \Omega}}, \quad
  A_0=\sqrt{1-{n_0\over \Omega}} a_0, \quad
N_F=2 n_0, \eqno(33.13)$$
where $a_0^\dagger$ ($a_0$) are boson creation (annihilation) operators
carrying angular momentum zero and $n_0$ is the number of these
bosons.

The simplest pairing Hamiltonian one can consider has the form
$$H=-G\Omega A^\dagger_0 A_0. \eqno(33.14)$$
The Casimir operator of su(2) can be written as  
$$\{A^\dagger_0, A_0\} + {\Omega\over 2} \left(1-{N_F\over \Omega}\right)^2 =
{\Omega\over 2}+1, \eqno(33.15)$$
while the pairing energy takes the form
$${E\over (-G\Omega)} = {N_F\over 2} -{N_F^2\over 4\Omega} +
{N_F\over 2\Omega}.\eqno(33.16) $$

Our aim is to check if there is a boson mapping
for the operators $A^\dagger_0$, $A_0$ and $N_F$
in terms of $q$-deformed bosons, having the following
properties:

i) The mapping is simpler than the one of eq. (33.13), i.e. to
each fermion pair operator $A^\dagger_0$, $A_0$ corresponds a bare
$q$-boson operator and not a boson operator accompanied by
a square root (the Pauli reduction factor).

ii) The commutation relations (33.9) are satisfied up to a
certain order.

ii) The pairing energies of eq. (33.16) are reproduced up to the
same order.

\subsection{Mapping using the $q$-deformed oscillator}

In the case of the $q$-deformed harmonic oscillator (sec. 10),
the commutation relation
$$ [a, a^\dagger] = [N+1] -[N] \eqno(33.17)$$
for $q$ being a phase can be written as
$$[a, a^\dagger] = {\cos{(2N+1)\tau\over 2}\over\cos{\tau\over 2}}.
\eqno(33.18)$$

In physical situations $\tau$ is expected to be small (i.e.
of the order of 0.01).
Therefore in eq. (33.18) one can take Taylor expansions
of the functions appearing there and thus find an expansion of the
form
$$[a, a^\dagger] = 1 -{\tau^2\over 2} (N^2+N) +{\tau^4\over 24}
(N^4+2 N^3-N)-\ldots .\eqno(33.19)$$
We remark that the first order corrections contain not only a term
proportional to $N$, but in addition a term proportional to $N^2$,
which is larger than $N$. Thus one cannot make the simple mapping
$$A_0\rightarrow a,\quad A^\dagger_0\rightarrow a^\dagger,\quad N_F\rightarrow
2N, \eqno(33.20)$$ 
because then one cannot get the first of the commutation 
relations (33.9) correctly up to
the first order of the corrections. The same problem appears
in the case that $q$ is real as well. In addition, by making the
simple mapping of eq. (33.20) the pairing Hamiltonian can be written
as
$${H\over -G \Omega} = a^\dagger a = [N].\eqno(33.21)$$
In the case of small $\tau$, one can take Taylor expansions
of the functions appearing in the definition of the $q$-numbers
(eq. (2.2) or (2.3)) and thus obtain the following expansion
$$ [N]= N \pm {\tau^2\over 6} (N-N^3)+{\tau^4\over 360}
(7N-10N^3+3 N^5) \pm {\tau^6\over 15120} (31N-49N^3+21N^5-3N^7)+
\ldots, \eqno(33.22)$$
where the upper (lower) sign corresponds to $q$ being a phase factor (real).
We remark that while the first order corrections in eq. (33.16)
are proportional to $N_F^2$ and $N_F$, here the first order
corrections are proportional to $N$ and $N^3$. Thus neither the
pairing energies can be reproduced correctly by this mapping.

\subsection{Mapping using the $Q$-deformed oscillator}

In the case of the $Q$-oscillator of sec. 11, however, 
the commutation relation among the bosons is 
$$[b, b^\dagger] = Q^N.\eqno(33.23)$$
Defining $Q=e^T$ this can be written as
$$[b, b^\dagger]= 1 + TN + {T^2  N^2\over 2} +{T^3 N^3\over 6} +\ldots.
\eqno(33.24)$$
We remark that the first order correction is proportional to $N$.
Thus, by making the boson mapping
$$A_0^\dagger\rightarrow b^\dagger, \quad
A_0\rightarrow b, \quad N_F\rightarrow 2N ,\eqno(33.25)$$ 
one can satisfy the first commutation relation of eq. (33.9) up to the first 
order of the corrections by determining $T=-2/\Omega$.

We should now check if the pairing energies (eq. (33.16)) can be found
correctly up to the same order of approximation when this
mapping is employed. The pairing Hamiltonian in this case takes
the form
$${ H\over -G\Omega}= b^\dagger b = [N]_Q.\eqno(33.26)$$

Defining $Q=e^T$ it is instructive to construct the expansion of
the $Q$-number of eq. (6.1) in powers of $T$. Assuming that $T$ is
small and taking Taylor expansions in eq. (6.1) one finally has
$$[N]_Q= N+{T\over 2} (N^2 -N) +{T^2\over 12} (2N^3-3N^2+1) +
{T^3\over 24} (N^4-2N^3+N^2) +\ldots\eqno(33.27)$$
 Using the
 value of the deformation
parameter $T=-2/\Omega$, determined above from the requirement
that the commutation relations are satisfied up to first order
corrections, the pairing energies become
$${E\over -G \Omega}= N- {N^2-N\over \Omega} +
{2N^3-3N^2+1\over 3\Omega^2} - {N^4-2N^3+N^2\over 3\Omega^3} +
\ldots. \eqno(33.28)$$
The first two terms in the rhs of eq. (33.28), which correspond to the
leading term plus the first order corrections, are exactly equal to
the pairing energies of eq. (33.16), since $N_F\rightarrow 2N$.
 We therefore conclude that
through the boson mapping of eq. (33.25) one can both satisfy the
fermion pair commutation relations of eq. (33.9) and reproduce the
pairing energies of eq. (33.16) up to the first order corrections.

The following comments are also in place:

i) By studying the spectra of the two versions of the $q$-deformed
harmonic oscillator, given in eqs.  (10.10) and (11.10), one can easily
draw the following conclusions: when compared to the usual oscillator
spectrum, which is equidistant, the spectrum of the $q$-oscillator
is getting shrunk for $q$ being a phase, while the spectrum of the
$Q$-oscillator gets shrunk when $T<0$. In a similar way, the spectrum
of the $q$-oscillator gets expanded for $q$ real, while the spectrum of
the $Q$-oscillator gets expanded for $T>0$. In physical situations
(secs 19--23)  it has been found that the physically interesting results
are gotten with $q$ being a phase. Thus in the case of the
$Q$-oscillator it is the case $T<0$ the one which corresponds to the
physically
interesting case. As we have already seen, it is exactly for
$T=-2/\Omega<0$ that the present mapping gives the fermion pair
results.

ii) It should be recalled that the pairing model under discussion
is studied under the assumptions  that the degeneracy of
the shell is large ($\Omega >>1$), that the number of particles
is large ($N>>1$), and that one stays away from the center of the
shell ($\Omega -N=O(N)$). The accuracy of the
present mapping in reproducing the pairing energies has been checked 
in \cite{Bon101}, where results for $\Omega=11$
(the size of the nuclear fp major shell), $\Omega=16$
(the size of the nuclear sdg major shell) and $\Omega=22$ (the
size of the nuclear pfh major shell) are reported, along with
results for the case $\Omega=50$ (as an
example of a large shell). In all cases good agreement
between the classical pairing model results and the
$Q$-Hamiltonian of eq. (33.26) is obtained up to the point
at which about 1/4 of the shell is filled. The deviations
observed near the middle of the shell are expected, since there the
expansion used  breaks down.

 We have thus shown that an approximate  mapping of the fermion
pairs coupled to angular momentum zero in a single-j shell
onto suitably defined $q$-bosons (the $Q$-bosons) is possible. The su(2) 
commutation relations are satisfied up to the first order corrections, while 
at the same time the eigenvalues of a simple pairing Hamiltonian are correctly
reproduced up to the same order. The small parameter of the
expansion, which is $T$ (where $Q=e^T$), turns out to be negative
and inversely proportional to the size of the shell.

The present results are an indication that suitably defined $q$-bosons
could be used for approximately describing systems of correlated fer\-mi\-ons
under certain conditions in a simplified way. 
The construction of $q$-bosons
which would {\sl exactly} satisfy the fermion pair su(2) commutation
relations  will be undertaken in the following section. 

\section { Fermion pairs as deformed bosons: exact mapping} 

From the contents of the previous section, 
the following question is created: Is it possible to construct
a generalized deformed oscillator (as in sec. 12) using deformed bosons in such
a way that the spectrum of the oscillator will exactly correspond to the
pairing energy in the single-j shell model, while the
commutation relations of the deformed bosons will exactly correspond to
the commutation relations of the correlated fermion pairs in the
single-j shell under discussion? In this section  we show that such an 
oscillator can indeed be constructed \cite{BD2781}
by using the method of sec. 12.

\vfill\eject
\subsection{An appropriate generalized deformed oscillator}

We apply the procedure of sec. 12 in the case of the pairing in a
single-j shell mentioned before. The boson number is half
the fermion number, i.e. $N=N_F/2$. Then eq. (33.16) can be
written as
$$ {E\over -G \Omega} = N -{N^2\over \Omega} +{N\over
\Omega}.\eqno(34.1)$$
One can then use a generalized deformed  oscillator with structure function 
$$ F(N) = a^\dagger a =N-{N^2\over \Omega}+{N\over \Omega}.\eqno(34.2)$$
In addition one has
$$ F(N+1) = a a^\dagger  = N+1 -{(N+1)^2\over \Omega}+{N+1\over\Omega}.
\eqno(34.3)$$
What we have constructed is a boson mapping for the operators
$A_0$, $A_0^\dagger$, $N_F$:
$$A_0\rightarrow a, \quad A_0^\dagger\rightarrow a^\dagger, \quad
N_F\rightarrow 2N. \eqno(34.4)$$
From eq. (34.2) it is clear that this mapping gives the
correct pairing energy. In addition one has
$$[a, a^\dagger]= F(N+1)-F(N)=1-{2N\over \Omega}, \eqno(34.5)$$
in agreement to the first commutation relation of eq. (33.9). Thus the 
correct commutation relations
are also obeyed. (The last two commutation relations of eq. (33.9) 
are satisfied because of (12.1).)

As we have already seen in the previous section, 
an exact hermitian boson mapping for the su(2) algebra is known
to have the form of eq. (33.13). In this mapping the Pauli principle effects 
are carried by the square roots accompanying the ordinary boson
operators, while in the mapping of eq. (34.4) the Pauli
principle effects are ``built in'' the deformed bosons.

The generalized oscillator obtained here has energy spectrum
$$E_N = {1\over 2} (F(N)+F(N+1))=N+{1\over 2}-{N^2\over \Omega},
\eqno(34.6)$$
which is the spectrum of an anharmonic oscillator.

\subsection{Related potentials}

The classical potential giving the same spectrum, up to first
order perturbation theory, can be easily determined (see also 
subsecs 13.1, 13.2). The potential
$$V(x)=\kappa x^2 +\lambda x^4 , \eqno(34.7)$$
is known to give in first order perturbation theory the
spectrum
$$E_n = \kappa (2n+1) +\lambda (6n^2 +6 n +3)
= (2\kappa+6\lambda) (n+{1\over 2}) + 6\lambda n^2. \eqno(34.8)$$
Comparing eqs. (34.6) and (34.8) one finds
$$\kappa = {1\over 2} (1+{1\over \Omega}), \quad \lambda=
-{1\over 6\Omega}.\eqno(34.9)$$
Then the classical potential giving the same spectrum, up to
first order perturbation theory, as the generalized oscillator
determined here, is
$$V(x)={1\over 2} (1+{1\over \Omega}) x^2 -{1\over 6\Omega}
x^4. \eqno(34.10)$$
It is therefore demonstrated that the Pauli principle effects
in a single-j shell with pairing interaction are equivalent
to an $x^4$ anharmonicity.

The generalization of the results obtained in this section  for the
pairing hamiltonian to any anharmonic oscillator is
straightforward. For example, the potential
$$V(x)=\kappa x^2 + \lambda x^4 + \mu x^6 +\xi x^8, \eqno(34.11)$$
is known to give up to first order perturbation theory the
spectrum
$$E_n =\kappa (2n+1)+\lambda (6n^2 +6n+3)   +\mu
(20 n^3+30 n^2 + 40 n+15) +\xi (70 n^4+140 n^3+350 n^2 + 280 n
+ 105), \eqno(34.12)$$
which can be rewritten in the form
$$E_n =(n+(n+1)) (\kappa +5\mu) + (n^2+(n+1)^2) (3\lambda+70\xi)
 + (n^3+(n+1)^3) (10\mu) + (n^4+(n+1)^4) (35 \xi).
\eqno(34.13)$$
Taking into account eq. (12.11),  from eq. (34.13) one gets
$${F(N)\over 2} = (\kappa +5\mu) n + (3\lambda+70 \xi) n^2
+(10\mu) n^3 + (35\xi) n^4.\eqno(34.14)$$
For $\mu=\xi=0$ and $\kappa$, $\lambda$ given from eq. (34.9),
the results for the pairing problem are regained.

It is worth mentioning at this point that the energy spectrum
of the generalized oscillator corresponding to the pairing
correlations (eq. (34.6))  can be rewritten as
$$E_N= {2\over \Omega} \left(-{1\over 8} +{\Omega+1\over 2} 
\left(N+{1\over 2}\right)
-{1\over 2} \left(N+{1\over 2}\right)^2\right). \eqno(34.15)$$
On the other hand, it is known that for the modified P\"oschl--Teller
potential (see also subsec. 13.2)
$$ V(x)= D \tanh^2(x/R), \eqno(34.16)$$
the energy spectrum is given by \cite{Haar75}
$$ E_N = {\hbar^2\over m R^2} \left(-{1\over 8} + 
{1\over 2}\sqrt{8mDR^2/\hbar^2
+1}\quad \left(N+{1\over 2}\right) -{1\over 2} \left(N+{1\over 2}\right)^2 
\right).\eqno(34.17)$$
It is thus clear that the energy spectrum of the generalized oscillator
studied here
can be obtained from the modified P\"oschl--Teller potential for
special values of the potential depth $D$.

It is also worth remarking that the ``structure function'' $F(N)$
of the generalized oscillator obtained here (eq. (34.2)) can be
written as
$$F(N)= {N\over \Omega} (\Omega+1-N), \eqno(34.18)$$
which is similar to the one of the para-fermionic oscillator
of Ohnuki and Kamefuchi \cite{OK82} (see also secs 12, 18). 

In summary, we have constructed a generalized deformed oscillator
 which satisfies the same commutation relations as fermion
pair and multipole operators of zero angular momentum in a
single-j shell, and, in addition, reproduces the pairing
energy of this shell exactly. We have thus demonstrated that
an exact hermitian boson mapping of a system of angular-momentum-zero
fermion pairs in terms of bare deformed bosons can be constructed,
while in the usual case the ordinary bosons  are accompanied
by square roots due to the Pauli principle effects.
The oscillator corresponding to the pairing problem has a spectrum
which can be reproduced up to first order perturbation theory by a
harmonic oscillator with an $x^4$ anharmonicity. The construction
of a generalized deformed oscillator corresponding to any
anharmonic oscillator has also been achieved.

The results obtained in this section indicate that deformed bosons
might be a convenient tool for describing systems of fermion
pairs under certain conditions. The generalisation
of the results obtained here to fermion pairs of nonzero
angular momentum, which will allow for a fuller treatment of
the single-j shell in terms of deformed bosons, is a very interesting problem. 

\section{The seniority scheme}

In the previous two sections we have seen how correlated fermion pairs of zero
angular momentum can be mapped onto deformed bosons. It is however known that 
pairs of non-zero angular momentum play an important role in the formation of 
nuclear properties. 
In the present section a first step in the direction of describing the 
$J\neq 0$ pairs in terms of deformed bosons is taken. 

\subsection{Uncovering a dynamical symmetry}

In the usual formulation of the theory of pairing in a single-j 
shell \cite{Heyde90}, fermion pairs of angular momentum $J=0$ are created
by the pair creation operators 
$$ S^\dagger = {1\over \sqrt{\Omega}} \sum_{m>0} (-1)^{j+m}  a^\dagger_{jm}
a^\dagger_{j-m}, \eqno(35.1)$$
where $a^\dagger_{jm}$ are fermion creation operators and $2\Omega=2j+1$
is the degeneracy of the shell. In addition, pairs of nonzero 
angular momentum are created by the $\Omega-1$ operators
$$ B^\dagger_J=\sum_{m>0} (-1)^{j+m} (j m j -m | J 0) a^\dagger_{jm} 
a^\dagger_{j-m}, \eqno(35.2)$$
where $( j m j -m | J 0)$ are the usual Clebsch Gordan coefficients. 
The fermion number operator is defined as 
$$ N_F = \sum_m a^\dagger_{jm} a_{jm} = \sum_{m>0} (a^\dagger_{jm} a_{jm} +
a^\dagger_{j-m} a_{j-m} ). \eqno(35.3)$$
As we have already seen, the $J=0$ pair creation and annihilation operators 
satisfy the  commutation relation 
$$ [S, S^\dagger] = 1-{N_F\over \Omega}, \eqno(35.4)$$
while the pairing Hamiltonian is 
$$ H = -G \Omega S^\dagger S.\eqno(35.5)$$
The seniority $V_F$ is defined as the number of fermions not
coupled  to $J=0$. If only pairs of $J=0$ are present (i.e. $V_F=0$), 
the eigenvalues of the Hamiltonian are (as already seen in eq. (33.16))
$$ E(N_F,V_F=0) = -G \Omega \left( {N_F\over 2} + {N_F\over 2\Omega}
-{N_F^2\over 4\Omega}\right). \eqno(35.6)$$
For non-zero seniority the eigenvalues of the Hamiltonian are
$$ E(N_F,V_F)= -{G\over 4} (N_F-V_F) (2\Omega-N_F-V_F+2).\eqno(35.7)$$
We denote the operators $N_F$, $V_F$ and their eigenvalues by the
same symbol for simplicity. 

In subsec. 33.4 it has been proved that the behaviour of the $J=0$ pairs
can be described, up to first order corrections, in terms of
$Q$-bosons. In particular, making the mapping 
$$ S^\dagger \rightarrow b^\dagger, \quad\quad S\rightarrow b, \quad\quad
N_F\rightarrow 2 N, \eqno(35.8)$$
the relevant pairing Hamiltonian of eq. (35.5)  becomes
$$ H(N,V=0) = -G \Omega b^\dagger b = -G\Omega [N]_Q,\eqno(35.9)$$
which coincides with eq. (35.6) up to 
first order corrections in the small parameter, which is identified
as $T=-2/\Omega$. Furthermore, the $Q$-bosons satisfy the commutation
relation of eq. (33.24),
which coincides with eq. (35.4) up to first order corrections in the
small parameter, which is, consistently with the above finding, 
identified as $T=-2/\Omega$. Therefore the fermion pairs of $J=0$
can be approximately described as $Q$-bosons, which correctly reproduce
both the pairing energies and the commutation relations up to
first order corrections in the small parameter. 

For the case of nonzero seniority, one observes that eq. (35.7) 
can be written as 
$$ E(N_F,V_F)= G\Omega \left({V_F\over 2} + {V_F\over 2\Omega} -
{V_F^2\over 4\Omega} \right) -G \Omega \left({N_F\over 2} +
 {N_F\over 2\Omega} -
{N_F^2\over 4\Omega} \right), \eqno(35.10)$$
i.e. it can be separated into two parts, formally identical to 
each other. Since the second part (which corresponds to the $J=0$
pairs) can be adequately described by the $Q$-bosons $b$, $b^\dagger$,
and their number operator $N$, 
 as we have already seen, it is reasonable to  assume that the first
 part can also be described in terms of some $Q$-bosons $d$,
$d^\dagger$, and their number operator $V$ (with $V_F\rightarrow 2V$),
satisfying commutation relations similar to eqs (11.2) and (11.3):
$$ [V,d^\dagger]= d^\dagger, \quad\quad [V,d]=-d,\quad\quad d d^\dagger
-Q d^\dagger d = 1. \eqno(35.11)$$
 From the
physical point of view this description means that a set of $Q$-bosons
is used for the $J=0$ pairs and another set for the $J\neq 0$ pairs.
The latter is reasonable, since in the context of this theory
the angular momentum value of the $J\neq 0$ pairs is not used explicitly.
The $J\neq 0$ pairs are just counted separately from the $J=0$  
pairs. A Hamiltonian giving the same spectrum as in eq. (35.10),
up to first order corrections in the small parameter,  
 can then be written as
$$ H(N,V) =G \Omega ([V]_Q -[N]_Q). \eqno(35.12)$$
Using eq. (33.27) it is easy to see that this expression agrees
to eq. (35.10) up to first order corrections in the small parameter 
$T=-2/\Omega$. 

Two comments concerning eq. (35.12) are in place:

i) In the classical theory states of maximum seniority (i.e.
states with $N=V$) have zero energy. This is also holding for the 
Hamiltonian of eq. (35.12) to all orders in the deformation
parameter.   

ii) A landmark of the classical theory is that $E(N,V)-E(N,V=0)$
is independent of $N$. This also holds for eq. (35.12) to all orders
in the deformation parameter. 

Knowing the Schwinger realization of the su$_q$(2) algebra in terms
of $q$-bosons (sec. 15), one may wonder if the 
operators used here close an algebra. It is easy to see that
the operators $b^+d$, $d^+ b$ and $N-V$ do not close an algebra.
Considering, however, the operators (see \cite{CJ711} with $p=1$) 
$$ J_+ = b^\dagger Q^{-V/2} d, \quad\quad J_-= d^\dagger Q^{-V/2} b, 
\quad\quad
J_0= {1\over 2} (N-V), \eqno(35.13)$$
one can easily see that they satisfy the commutation relations 
\cite{CJ711,Jan91}
$$ [J_0, J_{\pm}] = \pm J_{\pm} , \quad\quad J_+ J_- - Q^{-1} J_- J_+
= [2 J_0] _Q .\eqno(35.14)$$
Using the transformation 
$$ J_0 = \tilde J_0, \quad\quad J_+ = Q^{(1/2)(J_0-1/2)}
\tilde J_+, \quad\quad J_-= \tilde J_- Q^{(1/2)(J_0-1/2)}, \eqno(35.15)$$
one goes to the usual su$_q$(2) commutation relations
$$ [\tilde J_0, \tilde J_{\pm}] = \pm \tilde J_{\pm}, \quad\quad
[\tilde J_+, \tilde J_-] = [2 \tilde J_0], \eqno(35.16)$$
where $q^2=Q$. One can thus consider eq. (35.14) as a rewriting of the 
algebra su$_q$(2), suitable for boson realization in terms of $Q$-bosons.

It is clear that $N+V$ is the first order Casimir operator of
the  u$_Q$(2) algebra formed above (since it commutes with all the
generators given in eq. (35.13)),
 while $N-V$ is the first order
Casimir operator of its u$_Q$(1) subalgebra, which is generated
by $J_0$ alone. Therefore the Hamiltonian of eq. (35.12) can be
expressed in terms of the Casimir operators of the algebras
appearing in the chain u$_Q$(2) $\supset$ u$_Q$(1) as
$$ E(N,V) = G\Omega ( \left[{C_1({\rm u}_Q(2))-C_1({\rm u}_Q(1))\over 2}
\right]_Q  -  \left[{C_1({\rm u}_Q(2))+C_1({\rm u}_Q(1))\over 2}\right]_Q),
\eqno(35.17)$$
  i.e. the Hamiltonian has a u$_Q$(2) $\supset$ u$_Q$(1) dynamical symmetry.  

\subsection{Comparison to experiment}

In the construction given above we have shown that $Q$-bosons can be
used for the approximate description of correlated fermion pairs in a
single-j shell.  The results obtained in the $Q$-formalism agree
to  the classical (non-deformed) results up to first order corrections
in the small parameter. However, the $Q$-formalism contains in
addition  higher order terms. The question is then born if these
additional terms are useful or not. For answering this question,
the simplest comparison with experimental data which can be made
concerns the classic example of the neutron pair separation energies
of the Sn isotopes, used by Talmi \cite{Talmi1,Talmi85}. 

In Talmi's formulation  of the pairing theory, the energy of the
states with zero seniority is given by 
$$ E(N)_{cl} = N V_0 + {N(N-1)\over 2} \Delta, \eqno(35.18)$$
where $N$ is the number of fermion pairs and $V_0$, $\Delta$
are constants. We  remark that this expression is the same
as the one in eq. (35.6), with the identifications 
$$\Delta/(2 V_0) =-1/\Omega, \quad\quad \Delta=2G,
 \quad\quad N_F=2N. \eqno(35.19)$$
 The neutron pair separation energies are given by 
$$ \Delta E(N+1)_{cl} =E(N+1)_{cl}-E(N)_{cl}=
V_0 \left( 1+ {\Delta\over V_0} N\right).\eqno(35.20)$$
Thus the neutron pair separation energies are expected to decrease
linearly with increasing $N$. (Notice from eq. (35.19) that $\Delta/V_0<0$,
since $\Omega>0$.) A similar  linear decrease is predicted also by the 
Interacting Boson Model \cite{IA1987}. 

In our formalism the neutron pair separation energies are given by
$$ \Delta E(N+1)_Q= -G \Omega ([N+1]_Q -[N]_Q) = -G \Omega Q^N =
-G \Omega e^{T N}.\eqno(35.21)$$
Since, as we have seen, $T$ is expected to be $-2/\Omega$, i.e. 
negative and small, the neutron pair separation energies are 
expected to fall exponentially with increasing $N$, but the small
value of $T$ can bring this exponential fall very close to a linear
one. 

The neutron pair separation energies of the even Sn isotopes
from $^{104}$Sn to $^{130}$Sn (i.e. across the whole sdg neutron
shell) have been fitted in \cite{BDF1299}  using both theories. 
 Furthermore, in \cite{BDF1299} a fit  
 of the logarithms of the energies has been performed, since eq. (35.21)
predicts a linear decrease of the logarithm of the energies with
increasing $N$. 
Both fits give almost identical results. Eq. (35.21) (in which the 
free parameters are $G\Omega$ and $T$),
gives a better result than eq. (35.20) (in which the free
parameters are $V_0$ and $\Delta/V_0$)
for every single isotope, 
without introducing any additional parameter, indicating that
the higher order terms can be useful.

One should, however, remark that $^{116}$Sn lies in the middle
of the sdg neutron shell. Fitting the isotopes in the lower half 
of the shell ($^{104}$Sn to $^{116}$Sn) and the isotopes in the 
upper half of the shell ($^{118}$Sn to $^{130}$Sn) separately, 
one finds that both theories give indistinguishably good results
in both regions. Therefore $Q$-deformation can be understood as 
expressing higher order correlations which manifest themselves 
in the form of particle-hole asymmetry.  It is also known that a strong 
subshell closure exists at N=64 (which corresponds to $^{114}$Sn). 
The presence of this subshell closure can also affect the neutron pair 
separation energies in a way similar to the one shown by the data.  

In \cite{BDF1299} a fit of the neutron pair 
separation energies of the Pb isotopes from $^{186}$Pb to $^{202}$Pb
has also been attempted. In
this case both theories give indistinguishably good fits. This 
result is in agreement with the Sn  findings, 
since all of these Pb
isotopes lie in the upper half of the pfh neutron shell. Unfortunately,
no neutron pair separation energy data exist for Pb isotopes in
the lower part of the pfh neutron shell. 

Concerning the values of $T$ obtained in the case of the Sn isotopes
($T=-0.0454$, $T=-0.0447$), 
one observes that they are slightly
smaller than the value ($T=-0.0488$) which would have been obtained
by considering the neutrons up to the end of the sdg shell as lying
in a single-j shell. This is, of course, a very gross approximation
which should not be taken too seriously, since it ignores the fact that most
properties of nuclei can be well accounted for by the valence nucleons alone,
without being affected by the closed core. In the case of the Pb isotopes
mentioned above, however, the best fit was obtained with
$T=-0.0276$, which is again slightly smaller than the value of
$T=-0.0317$ which corresponds to considering all the neutrons up to
the end of the pfh shell as lying in a single-j shell. 

In summary, we have shown that pairing in a single-j shell
can be described, up to first order corrections, by two $Q$-oscillators,
one describing the $J=0$ pairs and the other corresponding
to the $J\neq 0$ pairs, the deformation parameter $T=\ln Q$ being
related to the inverse of the size of the shell. 
 These two oscillators can be used for 
forming an su$_Q$(2) algebra. A Hamiltonian giving the correct 
pairing energies up to first order corrections in the small 
parameter can be  
written in terms of the Casimir operators of the algebras appearing
in the u$_Q$(2) $\supset$ u$_Q$(1) chain, thus exhibiting a
quantum algebraic dynamical symmetry. The additional terms introduced
by the $Q$-oscillators serve in improving the description of the 
neutron pair separation energies of the Sn isotopes, with no extra
parameter introduced. 

In the previous section a generalized deformed oscillator describing the
correlated fermion pairs of $J=0$ {\sl exactly} has been
introduced. This generalized deformed oscillator is the same as
the one giving the same spectrum as the Morse potential (sec. 37), up 
to a shift in the energy spectrum. The use of two generalized deformed
oscillators for the description of $J=0$ pairs and $J\neq 0$ pairs
in a way similar to the one of the present section is a straightforward task, 
while the construction out of them 
of a closed algebra analogous to the su$_Q$(2)
obtained here is an open problem. The extension of the ideas presented
here to the case of the BCS theory is an interesting open problem. 

\subsection{Other approaches}

A $q$-deformed version of the pairing theory was {\sl assumed} by 
Petrova \cite{Petr92} and Shelly Sharma \cite{Sha904},
with satisfactory results when compared to experimental
data. The present construction offers some justification for this 
assumption, since in both cases the basic ingredient is the modification
of eq. (35.4).  It should be noticed, however, that the deformed version of 
eq. (35.4)  considered in \cite{Petr92,Sha904} is different from 
the one obtained  here (eq. (33.24)). A basic difference is that the deformed 
theory of \cite{Petr92,Sha904}  reduces to the classical 
theory for $q\rightarrow 1$, so that $q$-deformation is introduced in order 
to describe additional correlations, while in the present formalism
the $Q$-oscillators involved for $Q\rightarrow 1$ reduce to usual 
harmonic oscillators, so that $Q$-deformation is introduced in order
to attach to the oscillators the anharmonicity needed by the energy
expression (eq. (35.6)).  

Continuing along the same line Shelly Sharma and Sharma \cite{SSS2323}
derived Random Phase Approximation (RPA) equations for the pairing 
vibrations of nuclei differing by two nucleons in comparison to the initial
one and applied their method to the study of the $0^+$ states of the
Pb isotopes, which offer a good example of pairing vibrations in 
nonsuperconducting nuclei. Furthermore, using deformed quasi-particle pairs 
coupled to zero angular momentum they developed a deformed version of the 
quasi-boson approximation for $0^+$ states in superconducting nuclei and 
tested it against a schematic two-level shell model \cite{SSS2323}. Another 
deformed two-level shell model has been developed by Avancini and 
Menezes \cite{AM6261}. 
$q$-deformed boson mappings have been developed in \cite{ACM006,AMM061}. 

In addition a $q$-deformed version of the many-body BCS approximation 
for a pure pairing force 
has been developed \cite{TL4127}, using $q$-deformed fermions satisfying 
su$_q$(n) anticommutation relations.  
A set of quantum BCS equations and a 
$q$-analog of the gap equation have been derived \cite{TL4127}. 

\section{Anisotropic quantum harmonic oscillators with rational ratios of 
frequencies}

3-dim anisotropic harmonic oscillators \cite{Barut}
with rational ratios of frequencies (RHOs) are of 
current interest because of their relevance as possible underlying symmetries
of superdeformed and hyperdeformed nuclei \cite{Mot522,Rae1343}. 
In particular, it is thought \cite{NT533,JK321}
that superdeformed nuclei correspond to a ratio of frequencies of 2:1, while
hyperdeformed nuclei correspond to a 3:1 ratio. In addition they have been
recently connected \cite{RZ599,ZRM61} 
to the underlying geometrical structure in the Bloch--Brink $\alpha$-cluster
model \cite{BB247}, 
and possibly to the interpretation of the observed shell structure 
in atomic clusters (see sec. 38), 
especially after the realization that large deformations
can occur in such systems \cite{BL4130}. 
The 2-dim RHO is also of interest, since its 
single particle level spectrum characterizes the underlying symmetry of 
``pancake'' nuclei \cite{Rae1343}. 

RHOs are examples of maximally superintegrable systems \cite{Hiet87,Eva5666}
in N dimensions. {\sl Superintegrable systems} in N dimensions have more than 
N independent integrals (constants of motion). {\sl Maximally superintegrable 
systems} in N dimensions have 2N$-1$ independent integrals. 

The two-dim \cite{JH641,Dem1349,Con273,Con1297,Cis870,CLP6685,GKM925}
 and three-dim 
\cite{DZ1203,Mai1004,Ven190,MV57,RD1323,BCD1401,Naz154} anisotropic harmonic
oscillators have been the subject of several investigations, both at the
classical and the quantum mechanical level. 
The special cases with frequency
ratios 1:2 \cite{Holt1037,BW2215} and 1:3 \cite{FL325} have also been 
considered. While
at the classical level it is clear that the su(N) or sp(2N,R) algebras can
be used for the description of the N-dimensional anisotropic oscillator, the
situation at the quantum level, even in the two-dimensional case, is not as
simple.

In this section we are going to prove that a generalized deformed u(2)
algebra is the symmetry algebra of the two-dimensional anisotropic quantum
harmonic oscillator, which is the oscillator describing the single-particle
level spectrum of ``pancake'' nuclei, i.e. of triaxially deformed nuclei
with $\omega_x >> \omega_y$, $\omega_z$ \cite{Rae1343}. The method can be 
extended to the 3-dim RHO in a rather straightforward way. 

\subsection{ A deformed u(2) algebra}

Let us consider the system described by the Hamiltonian: 
$$H=\frac{1}{2}\left( {p_x}^2 + {p_y}^2 + \frac{x^2}{m^2}
+ \frac{y^2}{n^2} \right), \eqno(36.1)$$
where $m$ and $n$ are two natural numbers mutually prime ones, i.e. their
great common divisor is $\gcd (m,n)=1$.

We define the creation and annihilation operators \cite{JH641} 
$$ a^\dagger=\frac{x/m - i p_x}{\sqrt{2}}, \qquad a =
\frac{x/m + i p_x}{\sqrt{2}}, \eqno(36.2)$$
$$ b^\dagger=\frac{y/n - i p_y}{\sqrt{2}}, \qquad b=\frac{y/n + i p_y}
{\sqrt{2}}.\eqno(36.3)$$ 
These operators satisfy the commutation relations: 
$$ \left[ a,a^\dagger \right] = \frac{1}{m}, \quad \left[
b,b^\dagger \right] = \frac{1}{n}, \quad \mbox{other commutators}=0. 
\eqno(36.4)$$

Further defining
$$ U=\frac{1}{2} \left\{ a, a^\dagger \right\}, \qquad W=\frac{1}{2} \left\{ b,
b^\dagger \right\},\eqno(36.5)$$
one can consider the enveloping algebra generated by
the operators: 
$$ S_+= \left(a^\dagger\right)^m \left(b\right)^n,\qquad S_-= \left(a\right)^m
\left(b^\dagger\right)^n, \eqno(36.6)$$
$$ S_0= \frac{1}{2}\left( U - W \right), \qquad H=U+W. \eqno(36.7)$$ 
These genarators satisfy the following relations: 
$$\left[ S_0,S_\pm \right]=\pm S_\pm, \quad \left[H,S_i\right]=0,
\quad \mbox{for}\quad i=0,\pm,\eqno(36.8) $$
and 
$$ S_+S_- = \prod\limits_{k=1}^{m}\left( U - \frac{2k-1}{2m} \right)
\prod\limits_{\ell=1}^{n}\left( W + \frac{2\ell-1}{2n} \right),\eqno(36.9) $$
$$ S_-S_+ = \prod\limits_{k=1}^{m}\left( U + \frac{2k-1}{2m} \right)
\prod\limits_{\ell=1}^{n}\left( W - \frac{2\ell-1}{2n} \right).\eqno(36.10) $$
The fact that the operators $S_i$, $i=0, \pm$ are integrals of motion has
been already realized in \cite{JH641}.

The above relations mean that the harmonic oscillator of eq. (36.1)
is described by the enveloping algebra of the
generalization of the u(2) algebra formed by the generators $S_0$, $S_+$, $%
S_-$ and $H$, satisfying the commutation relations of eq. (36.8) and 
$$\left[S_-,S_+\right] = F_{m,n} (H,S_0+1) - F_{m,n} (H,S_0), \eqno(36.11)$$ 
where
$$\quad F_{m,n}(H,S_0)= \prod\limits_{k=1}^{m}\left(
H/2+S_0 - \frac{2k-1}{2m} \right) \prod\limits_{\ell=1}^{n}\left( H/2-S_0 + 
\frac{2\ell-1}{2n} \right). \eqno(36.12)$$
In the case of $m=1$, $n=1$ this algebra is the usual u(2) algebra, and the
operators $S_0,S_\pm$ satisfy the commutation relations of the ordinary u(2)
algebra, since in this case one easily finds that 
$$ [S_-, S_+]=-2 S_0.\eqno(36.13)$$
In the rest of the cases, the algebra is a deformed version of u(2), in
which the commutator $[S_-,S_+]$ is a polynomial of $S_0$ of order $m+n-1$.
In the case with $m=1$, $n=2$ one has 
$$[S_-,S_+]= 3 S_0^2 - H S_0 -{\frac{H^2}{4}} +{\frac{3}{16}},\eqno(36.14)$$
i.e. a polynomial quadratic in $S_0$ occurs, while in the case of $m=1$, $n=3
$ one finds 
$$[S_-, S_+]= -4 S_0^3 + 3 H S_0^2 -{\frac{7}{9}} S_0 -{\frac{H^3}{4}} + {%
\frac{H}{4}},\eqno(36.15)$$
i.e. a polynomial cubic in $S_0$ is obtained.

\subsection{ The representations}

The finite dimensional representation modules  of this algebra can be found
using the concept of the generalized deformed oscillator (sec. 12), in a
method similar to the one used in \cite{BDK3407,BDK3700} for the study of 
quantum
superintegrable systems. The operators: 
$$ {\cal A}^\dagger= S_+, \quad {\cal A}= S_-, \quad {\cal N}%
= S_0-u, \quad u=\mbox{ constant}, \eqno(36.16)$$
where $u$ is a constant to be determined, are the generators of a deformed
oscillator algebra: 
$$ \left[ {\cal N} , {\cal A}^\dagger \right] = {\cal A}^\dagger, \quad \left[ 
{\cal N} , {\cal A} \right] = -{\cal A}, \quad {\cal A}^\dagger{\cal A}
=\Phi( H, {\cal N} ), \quad {\cal A}{\cal A}^\dagger =\Phi( H, {\cal N}+1 ). 
\eqno(36.17)$$
The structure function $\Phi$ of this algebra is determined by the function $%
F_{m,n}$ in eq. (36.12): 
$$ \Phi( H, 
{\cal N} )= F_{m,n} (H,{\cal N} +u )
= \prod\limits_{k=1}^{m}\left( H/2+%
{\cal N} +u - \frac{2k-1}{2m} \right) \prod\limits_{\ell=1}^{n}\left( H/2-%
{\cal N} - u + \frac{2\ell-1}{2n} \right).\eqno(36.18$$ 
The deformed oscillator corresponding to the structure function of eq.  
(36.18) has an energy dependent Fock space of dimension $N+1$ if 
$$\Phi(E,0)=0, \quad \Phi(E, N+1)=0, \quad \Phi(E,k)>0,
\quad \mbox{for} \quad k=1,2,\ldots,N. \eqno(36.19)$$
The Fock space is defined by: 
$$H\vert E, k > =E \vert E, k >, \quad {\cal N} \vert E, k >= k \vert E, k
>,\quad a\vert E, 0 >=0, \eqno(36.20)$$
$${\cal A}^\dagger \vert E, k> = \sqrt{\Phi(E,k+1)} \vert E, k+1>, 
\quad {\cal %
A} \vert E, k> = \sqrt{\Phi(E,k)} \vert E, k-1>. \eqno(36.21)$$
The basis of the Fock space is given by: 
$$\vert E, k >= \frac{1}{\sqrt{[k]!}} \left({\cal A}^\dagger\right)^k\vert E,
0 >, \quad k=0,1,\ldots N,\eqno(36.22) $$
where the ``factorial'' $[k]!$ is defined by the recurrence relation: 
$$[0]!=1, \quad [k]!=\Phi(E,k)[k-1]! \quad . \eqno(36.23)$$
Using the Fock basis we can find the matrix representation of the deformed
oscillator and then the matrix representation of the algebra of eqs (36.8), 
(36.12).
The solution of eqs (36.19) implies
the following pairs of permitted values for the energy eigenvalue $E$ and
the constant $u$: 
$$E=N+\frac{2p-1}{2m}+\frac{2q-1}{2n} , \eqno(36.24)$$
where $p=1,2,\ldots,m$, $\qquad q=1,2,\ldots,n$, and 
$$u=\frac{1}{2}\left( \frac{2p-1}{2m}-\frac{2q-1}{2n} -N \right), 
\eqno(36.25)$$
the corresponding structure function being given by: 
$$ \Phi(E,x)=\Phi^{N}_{(p,q)}(x)=  
\prod\limits_{k=1}^{m}\left( x + 
\displaystyle \frac{2p-1}{2m}- \frac{2k-1}{2m} \right)
\prod\limits_{\ell=1}^{n}\left( N-x+ \displaystyle \frac{2q-1}{2n} + \frac{%
2\ell-1}{2n}\right) $$
$$ =\displaystyle\frac{1}{m^m n^n} \displaystyle\frac{
\Gamma\left(mx+p\right) }{\Gamma\left(mx+p-m\right)} \displaystyle \frac{%
\Gamma\left( (N-x)n + q + n \right)} {\Gamma\left( (N-x)n + q \right)}, 
\eqno(36.26)$$
where $\Gamma(x)$ denotes the usual Gamma-function. 
In all these equations one has $N=0,1,2,\ldots$, while the dimensionality of
the representation is given by $N+1$. Eq. (36.24)  means that there
are $m\cdot n$ energy eigenvalues corresponding to each $N$ value, each
eigenvalue having degeneracy $N+1$. (Later we shall see that the degenerate
states corresponding to the same eigenvalue can be labelled by an ``angular
momentum''.) 

It is useful to show at this point that a few special cases are in agreement
with results already existing in the literature. 

i) In the case $m=1$, $n=1$ eq. (36.26) gives 
$$ \Phi(E,x)= x(N+1-x),\eqno(36.27)$$
while eq. (36.24) gives 
$$ E=N+1,\eqno(36.28)$$
in agreement with Sec. IV.A of \cite{BDK3700}. 

ii) In the case $m=1$, $n=2$ one obtains for $q=2$
$$\Phi(E,x)= x(N+1-x)\left(N+{3\over 2}-x\right), \qquad E=N+{5\over 4},
\eqno(36.29)$$
while for $q=1$ one has 
$$\Phi(E,x)= x(N+1-x)\left(N+{1\over 2}-x\right), \qquad E=N+{3\over 4}.
\eqno(36.30)$$
These are in agreement with the results obtained in Sec. IV.F of \cite{BDK3700}
for the Holt potential (for $\delta =0$). 

iii) In the case $m=1$, $n=3$ one has for $q=1$
$$\Phi(E,x)=x(N+1-x)\left(N+{1\over 3}-x\right) \left(N+{2\over 3}-x\right),
\qquad E=N+{2\over 3},\eqno(36.31)$$
while for $q=2$ one obtains 
$$\Phi(E,x)=x(N+1-x)\left(N+{2\over 3}-x\right) \left(N+{4\over 3}-x\right),
\qquad E=N+1,\eqno(36.32)$$
and for $q=3$ one gets 
$$\Phi(E,x)=x(N+1-x)\left(N+{4\over 3}-x\right) \left(N+{5\over 3}-x\right),
\qquad E=N+{4\over 3}.\eqno(36.33)$$
These are in agreement with the results obtained in Sec. IV.D of \cite{BDK3700}
for the Fokas--Lagerstrom potential. 

In all of the above cases we remark that the structure function has 
 forms  corresponding to various versions of the  generalized deformed 
parafermionic algebra of eq. (18.1), the relevant conditions of eq. (18.2) 
being satisfied in all cases. 
It is easy to see that the obtained algebra corresponds to this of the
generalized parafermionic oscillator in all cases with frequency 
ratios $1:n$. 

The energy formula can be corroborated by using the
corresponding Schr\"{o}dinger equation. For the Hamiltonian of eq. (36.1)
the eigenvalues of the Schr\"{o}dinger equation are given
by: 
$$E=\frac{1}{m}\left(n_x+\frac{1}{2}\right)+ \frac{1}{n}\left(n_y+%
\frac{1}{2}\right),\eqno(36.34) $$
where $n_x=0,1,\ldots$ and $n_y=0,1,\ldots$. Comparing eqs (36.24) and
(36.34) one concludes that: 
$$N= \left[n_x/m\right]+\left[n_y/n\right],\eqno(36.35)$$
where $[x]$ is the integer part of the number $x$, and 
$$p=\mbox{mod}(n_x,m)+1, \quad q=\mbox{mod}(n_y,n)+1. \eqno(36.36)$$

The eigenvectors of the Hamiltonian can be parametrized by the
dimensionality of the representation $N$, the numbers $p,q$, and the number $%
k=0,1,\ldots,N$. $k$ can be identified as $[n_x/m]$. One then has:
$$H\left\vert 
\begin{array}{c}
N \\ 
(p,q) 
\end{array}
, k \right>= \left(N+\displaystyle
\frac{2p-1}{2m}+\frac{2q-1}{2n} \right)\left\vert 
\begin{array}{c}
N \\ 
(p,q) 
\end{array}
, k \right>,\eqno(36.37) $$
$$S_0 \left\vert 
\begin{array}{c}
N \\ 
(p,q) 
\end{array}
, k \right>= \left( k+ \displaystyle
\frac{1}{2} \left( \frac{2p-1}{2m}- \frac{2q-1}{2n} -N \right) \right)
\left\vert 
\begin{array}{c}
N \\ 
(p,q) 
\end{array}
, k \right>, \eqno(36.38)$$
$$S_+\left\vert 
\begin{array}{c}
N \\ 
(p,q) 
\end{array}
, k \right> = \sqrt{ \Phi^N_{(p,q)}(k+1)} \left\vert 
\begin{array}{c}
N \\ 
(p,q) 
\end{array}
, k +1\right>,\eqno(36.39)$$ 
$$S_-\left\vert 
\begin{array}{c}
N \\ 
(p,q) 
\end{array}
, k \right> = \sqrt{ \Phi^N_{(p,q)}(k)} \left\vert 
\begin{array}{c}
N \\ 
(p,q) 
\end{array}
, k -1\right>. \eqno(36.40)$$

\subsection{ The ``angular momentum'' quantum number}

It is worth noticing that the operators $S_0,S_\pm$ do not correspond to a
generalization of the angular momentum, $S_0$ being the operator
corresponding to the Fradkin operator $S_{xx}-S_{yy}$ \cite{Hig309,Lee489}.
The corresponding ``angular momentum'' is defined by: 
$$L_0=-i\left(S_+-S_-\right).\eqno(36.41) $$
The ``angular momentum'' operator commutes with the Hamiltonian: 
$$\left[ H,L_0 \right]=0. \eqno(36.42)$$
Let $\vert \ell> $ be the eigenvector of the operator $L_0$ corresponding to
the eigenvalue $\ell$. The general form of this eigenvector can be given by: 
$$\vert \ell > = \sum\limits_{k=0}^N \frac{i^k c_k}{\sqrt{[k]!}} \left\vert 
\begin{array}{c}
N \\ 
(p,q) 
\end{array}
, k \right>. \eqno(36.43)$$

In order to find the eigenvalues of $L_0$ and the coefficients $c_k$ we use
the Lanczos algorithm \cite{Lan255}, as formulated in \cite{Flo331}. From 
eqs (36.39) and (36.40) we find 
$$L_0|\ell >=\ell |\ell >=\ell \sum\limits_{k=0}^N\frac{i^kc_k}{\sqrt{[k]!}}%
\left| 
\begin{array}{c}
N \\ 
(p,q)
\end{array}
,k\right\rangle = $$
$$=\frac 1i\sum\limits_{k=0}^{N-1}\frac{i^kc_k\sqrt{\Phi _{(p,q)}^N(k+1)}}{%
\sqrt{[k]!}}\left| 
\begin{array}{c}
N \\ 
(p,q)
\end{array}
,k+1\right\rangle -\frac 1i\sum\limits_{k=1}^N\frac{i^kc_k\sqrt{\Phi
_{(p,q)}^N(k)}}{\sqrt{[k]!}}\left| 
\begin{array}{c}
N \\ 
(p,q)
\end{array}
,k-1\right\rangle 
\eqno(36.44)$$
From this equation we find that: 
$$c_k=(-1)^k 2^{-k/2}H_k(\ell /\sqrt{2})/{\cal N},
\quad {\cal N}^2= \sum\limits_{n=0}^N 2^{-n}H_n^2(\ell /\sqrt{2})/n! 
\eqno(36.45)$$
where the function $H_k(x)$ is a generalization of the ``Hermite''
polynomials (see also \cite{BDEF150,KZ121}), satisfying the recurrence
relations: 
$$ H_{-1}(x)=0,\quad H_0(x)=1, \eqno(36.46)$$
$$ H_{k+1}(x)=2xH_k(x)-2\Phi _{(p,q)}^N(k)H_{k-1}(x), \eqno(36.47)$$
and the ``angular momentum'' eigenvalues $\ell $ are the roots of the
polynomial equation: 
$$ H_{N+1}(\ell /\sqrt{2})=0. \eqno(36.48)$$
Therefore for a given value of $N$ there are $N+1$ ``angular momentum''
eigenvalues $\ell $, symmetric around zero (i.e. if $\ell $ is an ``angular
momentum'' eigenvalue, then $-\ell $ is also an ``angular momentum''
eigenvalue). In the case of the symmetric harmonic oscillator ($m/n=1/1$)
these eigenvalues are uniformly distributed and differ by 2. In the general
case the ``angular momentum'' eigenvalues are non-uniformly distributed. For
small values of $N$ analytical formulae for the ``angular momentum''
eigenvalues can be found \cite{BDEF150}. Remember that to each value of $N$
correspond $m\cdot n$ energy levels, each with degeneracy $N+1$.

In order to have a formalism corresponding to the one of the isotropic  
oscillator, let us introduce  for every $N$ and 
$(m,n,p,q)$ an ordering of the ``angular momentum'' eigenvalues  
$$ \ell_\mu ^{L,m,n,p,q}, \quad \mbox{where} \quad L=N
\quad \mbox{and} \quad \mu=-L,-L+2,\ldots,L-2,L,\eqno(36.49)$$
by assuming that:
$$ \ell_\mu ^{L,m,n,p,q} \le \ell_\nu^{L,m,n,p,q} \quad \mbox{if} \quad 
\mu < \nu, \eqno(36.50)$$
the corresponding eigenstates being given by:
$$\AVector{L}{\mu}{m,n,p,q}=
\sum\limits_{k=0}^L
\frac{(-i)^k H_k(\ell_\mu^{L,m,n,p,q)} /\sqrt{2}) }
{{\cal N} \sqrt{2^{k/2}[k]! }}  
\CVector{N}{(p,q)}{k}
=
\sum\limits_{k=0}^L d_{k+1} \CVector{N}{(p,q)}{k}
\eqno(36.51)$$
The above vector elements constitute  the analogue corresponding
to the  basis of ``sphe\-rical harmonic'' functions of the usual oscillator.
The calculation of the ``angular momentum'' eigenvalues 
of eq. (36.49)
and the coefficients $d_1,d_2,\ldots,d_{L+1}$
 in the expansion of eq. (36.51) is a
quite difficult task. The existence of  general analytic expressions for
these quantities is not obvious. The first few ``angular momentum'' eigenvalues
are given by:
$$\ell^{1,m,n,p,q}_{\pm 1}=
\pm
\sqrt{ 
\frac{1}{m^m n^n} \frac{ \Gamma (m+p) }{\Gamma (p) } 
\frac{ \Gamma(n+q) }{\Gamma (q) }  }, \eqno(36.52)$$
and
$$\ell^{2,m,n,p,q}_{0}=0,\eqno(36.53)$$
$$ \ell^{2,m,n,p,q}_{\pm 2}=
\pm
\sqrt{ 
\frac{1}{m^m n^n}
\left( \frac{ \Gamma (m+p) }{\Gamma (p) } 
\frac{ \Gamma(2n+q) }{\Gamma (n+q) }
+\frac{ \Gamma (2m+p) }{\Gamma (m+p) } 
\frac{ \Gamma(n+q) }{\Gamma (q) } \right)
  }
\eqno(36.54)$$
For $L>2$ the analytic expressions of the angular momentum eigenvalues and
the coefficients $d_k$ are longer, but their calculation
is a straightforward task. Numerical results for these quantities in the cases 
of frequency ratios 1:2 and 1:3 are given in \cite{BDKLu2,BKLD3335}.

After working out a few examples (see \cite{BDKLu2,BKLD3335,Predeal95} for 
details) one finds out the following points:

i) In the basis described by eqs. (36.16)-(36.19) it is a trivial matter to
distinguish the states belonging to the same irrep for any $m:n$ ratio,
while in the Cartesian basis this is true only in the 1:1 case.

ii) In the 1:2 case the irreps have degeneracies 1, 1, 2, 2, 3,
3, 4, 4, \dots, i.e. ``two copies'' of the u(2) degeneracies 1, 2, 3, 4,
\dots are obtained.

iii) In the 1:3 cases the degeneracies are 1, 1, 1, 2, 2, 2, 3, 3, 3, \dots,
i.e. ``three copies'' of the u(2) degeneracies are obtained.

iv) It can be easily seen that the 1:n case corresponds to ``n copies'' of
the u(2) degeneracies.

v) Cases with both $m$, $n$ different from unity show more complicated
degeneracy patterns, also correctly reproduced by the above formalism. In
the 2:3 case, for example, the degeneracy pattern is 1, 1, 1, 1, 1, 2, 1, 2,
2, 2, 2, 3, 2, 3, 3, \dots.

vi) The only requirement for each energy eigenvalue to correspond to one
irrep of the algebra is that $m$ and $n$ have to be mutually prime numbers.
If $m$ and $n$ possess a common divisor other than 1, then some energy
eigenvalues will correspond to sums of irreps, i.e. to reducible
representations.

vii) The difference between the formalism used in 
\cite{DZ1203,Ven190,MV57,Naz154} 
and the one used here is that in the former case for given $m$ and $n$ 
appropriate
operators have to be introduced separately for each set of $(p,q)$ values,
while in the present case only one set of operators is introduced.

\subsection{Multisections of the isotropic oscillator}

In \cite{RSP1950} the concept of bisection of an isotropic harmonic oscillator
has been introduced. One can easily see that multisections (trisections, 
tetrasections, \dots) can be introduced in a similar way \cite{TEI96}. 
The degeneracies 
of the various anisotropic oscillators can then be obtained from these 
of the isotropic oscillator by using appropriate multisections. 

Using the Cartesian notation $(n_x, n_y)$ for the states of the isotropic 
harmonic oscillator we have the following list:

N=0: (00)

N=1: (10) (01)

N=2: (20) (02) (11)

N=3: (30) (03) (21) (12)

N=4: (40) (04) (31) (13) (22)

N=5: (50) (05) (41) (14) (32) (23), 
\hfill\break
where $N=n_x+n_y$.  The corresponding degeneracies are 1, 2, 3, 4, 5, 6, \dots,
i.e. these of u(2). 

A bisection can be made by choosing only the states with $n_y$=even. Then the
following list is obtained:  

N=0: (00)

N=1: (10)

N=2: (20) (02)

N=3: (30) (12)

N=4: (40) (04) (22) 

N=5: (50) (14) (32). 

The degeneracies are 1, 1, 2, 2, 3, 3, \dots, i.e. these of the anisotropic 
oscillator with ratio of frequencies 1:2. The same degeneracies are obtained 
by choosing the states with $n_y$=odd. Therefore a {\bf bisection} of the 
isotropic oscillator, distinguishing states with mod$(n_y,2)=0$ and states 
with mod$(n_y,2)=1$, results in two interleaving sets of levels of the 
1:2 oscillator.  

By analogy, a {\bf trisection} can be made by distinguishing states with 
mod$(n_y,3)=0$, or mod$(n_y,3)=1$, or mod$(n_y,3)=2$. One can easily see 
that in this case three interleaving sets of states of the 1:3 oscillator,
having degeneracies 1, 1, 1, 2, 2, 2, 3, 3, 3, \dots, occur. 

Similarly a {\bf tetrasection} can be made by distinguishing states with 
mod$(n_y,4)=0$, or mod$(n_y,4)=1$, or mod$(n_y,4)=2$, or mod$(n_y,4)=3$. 
The result is four interleaving sets of states of the 1:4 oscillator, 
having degeneracies 1, 1, 1, 1, 2, 2, 2, 2, 3, 3, 3, 3, \dots. 

By bisecting $n_x$ and trisecting $n_y$ one is left  with six interleaving 
sets of states with degeneracies 1, 1, 1, 1, 1, 2, 1, 2, 2, 2, 2, 3, 2, 3, 
3, \dots, i.e. degeneracies of the 2:3 oscillator. 

By bisecting (or trisecting, tetresecting, etc) both $n_x$ and $n_y$ one is 
obtaining the original u(2) degeneracies of the isotropic oscillator.  

It is therefore clear that the degeneracies of all $m:n$ oscillators
can be obtained from these of the isotropic oscillator by appropriate 
multisections. In particular:

i) The degeneracies of the $1:n$ oscillator can be obtained from these 
of the 1:1 (isotropic) oscillator by $n$-secting $n_y$ or $n_x$. 

ii) The degeneracies of the $m:n$ oscillator can be obtained from these 
of the 1:1 oscillator by $m$-secting $n_x$ and $n$-secting $n_y$. 

\subsection{ Connection to W$_3^{(2)}$ }

For the special case $m = 1$, $n=2$ it should be noticed that the  deformed
algebra received here coincides with the finite W algebra  W$_3^{(2)}$ \cite
{Tj1,Tj2,Tj3,Tj4}. The commutation relations of the W$_3^{(2)}$ algebra are 
$$ [H_W, E_W]= 2 E_W, \qquad [H_W, F_W]= -2 F_W, \qquad [E_W,F_W]= H^2_W + 
C_W, \eqno(36.55)$$
$$[C_W,E_W]=[C_W,F_W]=[C_W,H_W]=0,\eqno(36.56)$$
while in the $m=1$, $n=2$ case one has the relations 
$$[{\cal N}, {\cal A}^{\dagger}]= {\cal A}^{\dagger}, \qquad [{\cal N}, 
{\cal A%
}]= -{\cal A}, \qquad [{\cal A},{\cal A}^{\dagger}]= 3 S_0^2 -{\frac{H^2}{4}}
- H S_0 +{\frac{3 }{16}},\eqno(36.57)$$
$$ [H, {\cal A}^{\dagger}]= [H, {\cal A}]= [H, S_0] =0,\eqno(36.58)$$
with $S_0= {\cal N}+u$ (where $u$ a constant). It is easy to see that the
two sets of commutation relations are equivalent by making the
identifications 
$$F_W= \sigma {\cal A}^{\dagger}, \qquad E_W= \rho {\cal A}, \qquad H_W= -2
S_0 + k H, \qquad C_W= f(H), \eqno(36.59)$$
with 
$$\rho \sigma = {\frac{4}{3}}, \qquad k={\frac{1}{3}}, \qquad f(H)=
-{\frac{4}{9%
}} H^2 +{\frac{1}{4}}.\eqno(36.60)$$

\subsection{Discussion}

In conclusion, the two-dimensional anisotropic quantum harmonic oscillator
with rational ratio of frequencies equal to $m/n$, is described dynamically
by a deformed version of the u(2) Lie algebra, the order of this algebra
being $m+n-1$. The representation modules of this algebra can be generated
by using the deformed oscillator algebra. The energy eigenvalues are
calculated by the requirement of the existence of finite dimensional
representation modules. An ``angular momentum'' operator useful for
labelling degenerate states has also been constructed.  The algebras 
obtained in the special cases with $1:n$ ratios are shown to correspond to 
generalized parafermionic oscillators. 
In the special case 
of $m:n=1:2$ the resulting algebra has been identified as the finite W 
algebra W$_3^{(2)}$. Finally, it is demonstrated how the degeneracies 
of the various $m:n$ oscillators can be obtained from these of the 
isotropic oscillator by appropriate multisections. 

The extension of the present method to the three-dimensional anisotropic
quantum harmonic oscillator is already receiving attention 
\cite{Patras95,Varna95}, 
since it is of
clear interest in the study of the symmetries underlying the structure of
superdeformed and hyperdeformed nuclei \cite{Mot522,Rae1343}.

\section{The use of quantum algebras in molecular structure}

The techniques developed in this article can be applied in very similar
ways in describing properties of diatomic and polytomic molecules. A brief
list will be given here. 

1) Rotational spectra of diatomic molecules have been described in terms of 
the su$_q$(2) model \cite{Iwao363,BRRS300,ZS17,CY254,Est5614,Capps}.
 As in the case of nuclei, $q$ is a phase factor 
($q=e^{i\tau}$). In molecules $\tau$ is of the order of 0.01. 
The use of the su$_q$(2) symmetry leads to a partial summation of the Dunham
expansion describing the rotational--vibrational spectra of diatomic 
molecules \cite{BRRS300}. Molecular backbending (bandcrossing) has also been
described in this framework \cite{MRM1115}. Rotational spectra of symmetric 
top molecules have also been considered \cite{Chang1400,KN221} in the 
framework of the su$_q$(2) symmetry. Furthermore, two $q$-deformed rotators 
with slightly different parameter values have been used \cite{RMD2759}
for the description of $\Delta I=1$ staggering effects in rotational bands 
of diatomic molecules (see also \cite{MDRTB}). 
(For a discussion of $\Delta I=2$ staggering effects in diatomic molecules see
\cite{BDDKMMRR,BDDLMRR}. For a discussion of staggering
effects in nuclei see sec. 22.)

2) Vibrational spectra of diatomic molecules have been described in terms of 
$q$-deformed anharmonic oscillators having the 
su$_q$(1,1) \cite{BAR403} or the u$_q$(2) $\supset$ o$_q$(2) 
\cite{BRF221,GJC2715}
symmetry, as well as in terms of generalized deformed oscillators
similar to the ones used in sec. 26 \cite{BD75,CGY192,CGY183}. 
These results, combined with 1), lead 
to the full summation of the Dunham expansion \cite{BAR403,BRF221}. 
A two-parameter deformed anharmonic oscillator with u$_{qp}$(2) $\supset$ 
o$_{qp}$(2) symmetry has also been considered \cite{ZZH1053}. 

3) The physical content of the anharmonic oscillators mentioned in 2) 
has been clarified by constructing WKB equivalent potentials (WKB-EPs) 
\cite{BDKJMP,BDK191} and 
classical equivalent potentials \cite{BDK6153}, similar to the 
ones of sec. 13. The results have been corroborated by the study of the 
relation between su$_q$(1,1) and the anharmonic oscillator with  $x^4$ 
anharminicities \cite{NQ1699}. 
The WKB-EP corresponding to the su$_q$(1,1) anharmonic 
oscillator has been connected to a class of Quasi-Exactly Soluble Potentials 
(QESPs) \cite{BDM199}.                                                         

4) Generalized deformed oscillators 
giving the same spectrum as the Morse potential \cite{BD150} and the modified 
P\"oschl--Teller potential \cite{Das2261},  as well as a deformed oscillator
containing them as special cases \cite{Jan233,Jan180} have also been 
constructed. 
In addition,  $q$-deformed versions of the Morse potential have been given, 
either by using the so$_q$(2,1) symmetry \cite{CG941} or by solving a 
$q$-deformed Schr\"odinger equation for the usual Morse potential 
\cite{DD015}. For the sake of completeness it should be mentioned that a
deformed oscillator giving the same spectrum as the Coulomb potential 
has also been constructed \cite{DY4157}. 

5) A $q$-deformed version of the vibron model for diatomic molecules 
\cite{IL95,FVI94} has been 
constructed \cite{ABS1088,CL317,Pan47,GC3123}, in a way similar to that 
described in sec. 31.  

6) For vibrational spectra of polyatomic molecules a model of $n$ coupled 
generalized deformed oscillators has been built \cite{BD3611}, containg the 
approach of Iachello and Oss \cite{IO2976,IO500} as a special case.  
Furthermore, a system of two $Q$-deformed oscillators coupled so that the 
total Hamiltonian has the su$_Q$(2) symmetry (in the way described in 
\cite{Flo4739}) has been proved to be appropriate for the description of 
vibrational spectra of triatomic molecules \cite{BDK605}. 
In addition, a 3-dimensional anisotropic $q$-deformed harmonic oscillator 
has been used \cite{GMK115} for the description of vibrational spectra 
of triatomic molecules. A description of a tetrahedral molecule (methane)
in terms of $q$-deformed oscillators has been given in \cite{XHM1}. 

7) Quasi-molecular resonances in the systems $^{12}$C+$^{12}$C and 
$^{12}$C+$^{16}$O have been described in terms of a $q$-deformed oscillator
plus a rigid rotator \cite{CY325}.

A review of several of the above topics, accompanied by a detailed and 
self-contained introduction to quantum algebras, has been given by Raychev
\cite{RayAQC}. A recent review on algebraic methods in molecular 
spectroscopy has been given by Kellman \cite{Kel46}, while more 
extensive presentations of algebraic techniques applied in molecular 
spectroscopy can be found in the books by Iachello and Levine \cite{IL95}
and by Frank and Van Isacker \cite{FVI94}. 

\section{The use of quantum algebras in the structure of atomic clusters}

Metal clusters have been recently the subject of many investigations
(see \cite{deHeer,Brack,Nester} for relevant reviews). One of the first 
fascinating findings 
in their study was the appearance of magic numbers 
\cite{Martin,Bjorn,Knight1,Knight2,Peder,Brec,Persson}, analogous to 
but different from the magic numbers appearing in the shell structure of 
atomic nuclei \cite{Mayer}. 
This analogy led to the early description of metal 
clusters in terms of the Nilsson--Clemenger model \cite{Clem},
which is a simplified version of the Nilsson model 
\cite{Nilsson29,GLNN613,BR14,NR95} 
of atomic 
nuclei, in which no spin-orbit interaction is included. Further theoretical
investigations in terms of the jellium model \cite{Ekardt,Beck} 
demonstrated that the mean field potential in the case of simple metal 
clusters bears great similarities to the Woods--Saxon potential 
of atomic nuclei, with a slight modification of the ``wine bottle''
type \cite{Kotsos}. 
The Woods--Saxon potential itself looks like a harmonic 
oscillator truncated at a certain energy value and flattened at the bottom. 
It should also be recalled that an early schematic explanation of the 
magic numbers of metallic clusters has been given in terms of a scheme 
intermediate between the level scheme of the 3-dimensional harmonic 
oscillator and the square well \cite{deHeer}. Again in this case the 
intermediate 
potential resembles a harmonic oscillator flattened at the bottom.  

On the other hand, as we have seen in sec. 28, 
 a $q$-deformed version of the 3-dimensional harmonic 
oscillator has been constructed \cite{RRIT241}, taking advantage of the 
u$_q$(3) $\supset$ so$_q$(3) symmetry. 
The spectrum of this 3-dimensional $q$-deformed harmonic oscillator 
has been found \cite{RRIT241} to reproduce very well the spectrum of the 
modified harmonic oscillator introduced by Nilsson 
\cite{Nilsson29,NR95}, without the 
spin-orbit interaction term. Since the Nilsson model without the 
spin orbit term is essentially the Nilsson--Clemenger model used 
for the description of metallic clusters \cite{Clem}, it is worth examining 
if the 3-dimensional $q$-deformed harmonic oscillator can reproduce 
the magic numbers of simple metallic clusters. 

It has indeed been found \cite{BDKRRT} that the 3-dimensional $q$-deformed 
harmonic oscillator with u$_q$(3) $\supset$ so$_q$(3) symmetry correctly 
predicts all experimentally observed magic numbers of alkali metal clusters  
up to 1500, which is the expected limit of validity for theories based on 
the filling of electronic shells \cite{Martin}. 
This indicates that u$_q$(3), which 
is a nonlinear deformation of the u(3) symmetry of the spherical
(3-dimensional isotropic) harmonic oscillator, is a good candidate for 
being the symmetry of systems of alkali metal clusters.  

\section{Outlook}

Nobody likes binding himself by statements concerning the future. However, 
we attempt to give here a partial list of open problems, roughly following
the order of the material in this review:

1) The list of physical systems which can be classified under a generalized 
deformed su(2) symmetry (sec. 17) or under a generalized deformed 
parafermionic oscillator scheme (sec. 18) can be enlarged. Self-similar
potentials and isospectral oscillator Hamiltonian systems could probably 
be related to these symmetries. 

2) The description of B(E2) values in terms of the su$_q$(2) model 
attempted in sec. 21 takes into account only the kinematical deformation
effects. In order to take into account dynamical deformation effects, one 
has to build a larger algebra, of which the quadrupole operators will be 
members. The case of the u$_q$(3)$\supset$so$_q$(3) algebra has been 
considered in secs 28, 29. A detailed study is needed in order to see
if the experimental data support the modifications predicted by these 
models. In particular the su$_q$(2) prediction about B(E2) values 
increasing with increasing 
angular momentum $J$, supported by the predictions of other models as well
(see sec. 21), requires further testing against detailed experimental data.

3) The construction of Clebsch-Gordan coefficients for the subclass of 
generalized deformed su(2) algebras for which this could be possible 
(see secs 17, 25) is an open problem. 

4) The su$_q$(3) $\supset$ so$_q$(3) decomposition for su$_q$(3) irreps 
other than the completely symmetric ones (see secs 27--29 for the current
state of the art) remains an open problem, the 
solution of which is necessary for developing a deformed version of the
su(3) limit of the Interacting Boson Model. 

5) Realizations of multi-level shell models in terms of deformed bosons 
(see secs 33--35 for some one- and two-level cases) should be further 
pursued. 

6) The symmetry algebras of the various 3-dim anisotropic harmonic oscillators
with rational
ratios of frequencies should be worked out, since they are of interest in 
relation to superdeformed and hyperdeformed nuclei, and possibly to 
deformed atomic clusters (see secs 36, 38 for references). A deformed u(3)
algebra should occur in this case, which could serve as the basis for 
building a deformed analog of the Elliott model 
\cite{Ell128,Ell562,Ell557,Har67} 
suitable for superdeformed nuclei.   

7) In molecular physics (sec. 37) the study of vibrations of highly symmetric 
polyatomic molecules (including fullerenes) by these techniques is of 
interest. 

8) In the structure of atomic clusters (sec. 38) the 3-dimensional 
$q$-deformed harmonic oscillator (sec. 28) appears to be an appropriate 
basis for further investigations. 

In addition the following final comments are also in place: 

a) The usefulness of any ``new theory'' in nuclear structure is usually judged
by its ability to predict some novel excitation mode. For example, the 
prediction of the low-lying (around 3 MeV) 1$^+$ mode in even nuclei has 
been a success of the Interacting Boson Model-2 \cite{IA1987}. In the present 
case, the su$_q$(2) model prediction that B(E2) values in deformed 
nuclei increase as a function of angular momentum, and do not reach a 
saturation value as the su(2) symmetry suggests, is an interesting point 
which requires further experimental testing, as already mentioned in sec. 21.

b) Any ``new theory'' in nuclear structure oughts to acquire a microscopic 
justification, by establishing links to the underlying fermionic degrees 
of freedom. The mappings of secs 33 and 34 are a first step in this 
direction, although they deal only with fermion pairs of zero angular 
momentum in a single-$j$ shell. The generalization of these mappings 
to pairs of nonzero angular momentum and to multi-$j$ level schemes 
is a formidable but challenging task. 

c) Quantum algebras, being nonlinear extensions of Lie algebras, are 
specially suited for describing small perturbations  in  systems 
characterized by Lie symmetries. Several examples in nuclei, molecules, 
and atomic clusters have already been mentioned. Future work should 
clarify if the success of quantum algebras is limited within these 
bounds or if there is some deeper physical reason behind their success.
For the time being, quantum algebras have been proved to be a useful 
tool for the description of small deviations from the usual Lie symmetries. 

d) An interesting question is if quantum algebras are not only suitable for 
describing small deviations from Lie symmetries, but in addition can bridge 
different Lie symmetries. Preliminary work in this direction has been reported
in sec. 30. 

\bigskip
{\bf Acknowledgements}

\medskip
The authors are grateful to P. P. Raychev, R. P. Roussev, S. B. Drenska, 
N. Minkov and P. A. Terziev of the Institute for Nuclear Research and 
Nuclear Energy of the Bulgarian Academy of Sciences, Sofia, Bulgaria,
for several helpful discussions during the development of many of the ideas 
discussed in the present work.

\end{document}